\documentclass[12pt,a4paper]{article}
\pdfoutput=1

\usepackage[utf8]{inputenc}
\usepackage[english]{babel}
\usepackage{geometry}
\usepackage{amsmath, amsthm, amssymb, amsfonts, cases}
\usepackage{bm, bbm}
\usepackage[colorlinks=true,linkcolor=blue, linktoc=page, urlcolor=blue,citecolor=magenta]{hyperref}
\usepackage[dvipsnames]{xcolor}
\usepackage{graphicx}
\usepackage{comment}
\usepackage{mathtools}
\usepackage[font=small, font+=it, format=hang, margin={1cm, 1cm}]{caption}
\usepackage{float}

\usepackage{cite}

\allowdisplaybreaks[1]

\numberwithin{equation}{section}

\usepackage[normalem]{ulem}

\newcommand{\br}[1]{\left[#1\right]}

\newcommand{\pa}[1]{\left(#1\right)}

\begin{document}

\begin{titlepage}
    %
    \begin{flushright}\vspace{-2cm}
        {\small \today }
    \end{flushright}
    \bigskip

    \begin{center}
        %
    %
         {\Large\textbf{Blandford--Znajek monopole expansion revisited:}}\\[1mm]
        {\Large{novel non-analytic contributions to the power emission}}\\
        \line(2,0){450}
        \bigskip
        {\textbf{
            Filippo~Camilloni$^{\dagger\,\ddagger}$
            ,
            Oscar~J.~C.~Dias$^{\#}$
            ,
            Gianluca~Grignani$^\dagger$
            ,
            Troels~Harmark$^\ddagger$
            ,
            Roberto~Oliveri$^*$
            ,
            Marta~Orselli$^{\dagger\,\ddagger}$
            ,
            Andrea~Placidi$^{\dagger\,\ddagger}$
            ,
            Jorge~E.~Santos$^{\&}$
            }}
            \vspace{2pt}\\
        \bigskip
        \medskip
        \textit{{}$^\dagger$ Dipartimento di Fisica e Geologia, Universit\`a di Perugia, I.N.F.N. Sezione di Perugia, \\ Via Pascoli, I-06123 Perugia, Italy }\\
        \medskip
        \textit{{}$^{\#}$ STAG Research Centre and Mathematical Sciences, Highfield Campus,\newline{}
        University of Southampton, Southampton SO17 1BJ, United Kingdom}\\
        \medskip
        \textit{{}$^\ddagger$ Niels Bohr Institute, Copenhagen University,\\ Blegdamsvej 17, DK-2100 Copenhagen \O{}, Denmark}\\ 
        \medskip
        \textit{{}$^*$ CEICO, Institute of Physics of the Czech Academy of Sciences,\\ Na Slovance 2, 182 21 Praha 8, Czech Republic}\\
        \medskip
        \textit{{}$^{\&}$ 
        DAMTP, University of Cambridge,\\
        Wilberforce Road, Cambridge, CB3 0WA, UK}  
        
        
        \line(2,0){450}
        \vspace{8mm}


\begin{abstract}
\noindent The Blandford and Znajek (BZ) split-monopole serves as an important theoretical example of the mechanism that can drive the electromagnetic extraction of energy from Kerr black holes. It is constructed as a perturbative low spin solution of Force Free Electrodynamics (FFE). Recently, Armas~\emph{et al.}~put this construction on a firmer footing by clearing up issues with apparent divergent asymptotics. This was accomplished by resolving the behavior around the outer light surface, a critical surface of the FFE equations.
Building on this, we revisit the BZ perturbative expansion, and extend the perturbative approach to higher orders in the spin parameter of the Kerr black hole. 
We employ matched-asymptotic-expansions and semi-analytic techniques to extend the split-monopole solution to the sixth-order in perturbation theory. The expansion necessarily includes novel logarithmic contributions in the spin parameter. 
We show that these higher order terms result in non-analytic contributions to the power and angular momentum output. In particular, we compute for the first time the perturbative contributions to the energy extraction at seventh- and eighth-order in the spin parameter.
The resulting formula for the energy extraction
improves the agreement with numerical simulations at finite spin. 
Moreover, we present a novel numerical procedure for resolving the FFE equations across the outer light surface, resulting in significantly faster convergence and greater accuracy, and extend this to higher orders as well. 
Finally, we include a general discussion of light surfaces as critical surfaces of the FFE equations.
\end{abstract}

\vfill

\line(2,0){450}
        
\footnotesize{\textit{Emails:} filippo.camilloni@nbi.ku.dk, o.j.campos-dias@soton.ac.uk, gianluca.grignani@unipg.it,}
\footnotesize{harmark@nbi.ku.dk, roliveri@fzu.cz, orselli@nbi.dk, andrea.placidi@nbi.ku.dk, jss55@cam.ac.uk}      
        
\end{center}

\setcounter{footnote}{0}
\end{titlepage}

\tableofcontents


\section{Introduction and summary of the results}
The seminal paper of Blandford and Znajek (BZ) \cite{1977MNRAS.179..433B}, demonstrating the existence of a Penrose-like mechanism to extract the rotational energy from a Kerr black hole (BH), opened up one of the most fascinating research endeavours in astrophysics. In the BZ mechanism, the plasma-filled magnetosphere surrounding the rotating BH is described by force-free electrodynamics (FFE), which is an effective approximation of ideal relativistic magnetohydrodynamics (MHD) in a regime where the inertia of the plasma can be neglected. The FFE approximation is supported by numerical simulations \cite{macdonald,Contopoulos:1999ga,Nathanail:2014aua,Komissarov,McKinney_2005,KomissarovMHD2,Penna_2013,Parfrey:2018dnc,Mahlmann_2020,Mahlmann:2020nwe,Mahlmann:2020yxn,Talbot_2021} in the region away from the accretion disc and especially in the funnel region around the jets, where the plasma density is several orders of magnitude smaller than the energy density of the electromagnetic field.
To demonstrate their energy extraction mechanism, BZ explored perturbatively the split-monopole and paraboloidal profiles of the magnetosphere around a slowly rotating Kerr BH. 
In recent years, there has been renewed interest in extending these perturbative constructions \cite{Tanabe_2008,Pan_2015,Pan:2015iaa,Pan:2015imp,Grignani_2018,Grignani:2019dqc,Armas:2020mio} and exploring other profiles \cite{Pan:2014bja,Yang:2015ata,Gralla:2015vta,Pan:2017npg,East:2018ayf}. The BZ mechanism has been investigated  in the intermediate-spin regime \cite{McKinney_2004,KomissarovMHD,McKinney_2007,Tchekhovskoy_2008,Tchekhovskoy:2009ba,Tchekhovskoy2011,Nakamura_2018} using numerical simulations.
Finally, the high spin regime has been examined with analytical techniques \cite{Lupsasca_2014,Zhang_2014,Lupsasca_2015,Comp_re_2016,Camilloni:2020hns,Camilloni:2020qah} exploiting the enhancement of symmetries in the near-horizon region of a near-extreme Kerr BH.%
\footnote{Interestingly, there exist also 
exact solutions of FFE in the Kerr background, valid for any value of the spin. These are the Menon and Dermer class \cite{Menon_2005,Menon_2007,Menon_2015} and the Brennan, Gralla and Jacobson class \cite{Brennan_2013}. However, these two classes of solutions have an electromagnetic field with vanishing Lorentz invariant $F_{\mu\nu}F^{\mu\nu} =0$, contrary to the physical requirement of having $F_{\mu\nu}F^{\mu\nu} >0$. }
An important feature of the FFE configurations around the Kerr black hole is the presence of critical surfaces, which are the regular singular point of the second-order differential equation for the magnetosphere. Understanding the behavior of the magnetosphere near these surfaces is pivotal to find the correct solution. These critical surfaces include the event horizon and a surface at asymptotic infinity where one obtains what is know as Znajek conditions. Furthermore, they include the inner light surface (ILS) and the outer light surface (OLS), on which a co-rotating observer would need to travel at the speed of light. These critical surfaces play a crucial role in this work.
For the particular case of the split-monopole profile, BZ originally solved the FFE equations perturbatively to second order in $\alpha = J/M^2$ in the low spin limit $\alpha \ll 1$, with $M$ and $J$ being respectively the mass and angular momentum of the Kerr black hole. 
Subsequently, in \cite{Tanabe_2008} it was noticed that the perturbation scheme to extend the split-monopole configuration up to fourth-order breaks down asymptotically. Despite this result, there were further attempts to push forward the perturbative BZ split-monopole solution to higher orders in $\alpha$ \cite{Pan_2015,Pan:2015iaa,Pan:2015imp}. More recently in \cite{Grignani_2018,Grignani:2019dqc}, further inconsistencies in the perturbation analysis were explored, pointing to apparent issues with imposing the correct asymptotic behavior even at second order in $\alpha$. 
In \cite{Armas:2020mio} a resolution was found to the apparent inconsistencies in the perturbative approach to the BZ split-monopole at low spin. The crucial observation is that one needs to provide a smooth solution across the OLS. For a slowly rotating Kerr black hole $\alpha \ll 1$, the OLS emerges from infinity at a Boyer-Lindquist radius proportional to $\alpha^{-1}$. Introducing a rescaled radial coordinate, that resolves the behavior near the OLS, one can numerically find a magnetosphere configuration that is continuous across the OLS and which behaves correctly in the asymptotic region. This was accomplished to third order in $\alpha$ in the region near the OLS \cite{Armas:2020mio}, being the first order for which these issues appear.
In this work we build on the approach of \cite{Armas:2020mio} and pay particular attention to the non-analytic behavior in $\alpha$ that it results in. Already in \cite{Armas:2020mio} non-analytic terms in $\alpha$ were found in the region near the OLS in the magnetic flux function $\psi$, since one needs to include terms proportional to $\alpha^{2n} |\alpha|$ if one should impose the physical requirement that $\psi$ is even under $\alpha \rightarrow -\alpha$.
However, this non-analytic behavior did not reveal any consequences for the physical quantities in \cite{Armas:2020mio}. 
In our work, we extend the perturbative analysis to include higher-order terms in $\alpha$. The space-time is divided into two regions, one being the region that includes the horizon, in this paper called the $r$-region after the Boyer-Lindquist radial coordinate $r$, and the other being the $\bar{r}$-region, in which the rescaled radial coordinate $\bar{r}=\alpha r$ is kept finite for $\alpha \rightarrow 0$. In addition, one has an overlap region between these, as well as an asymptotic region which is contained in the $\bar{r}$-region.

For the $r$-region we find that one has a new $|\alpha|^5$ contribution to $\psi$. In addition we consider $\alpha^6$ and $\alpha^6 \log |\alpha|$ contributions as well. This is the first time that non-analytical contributions to $\psi$ have been found in the $r$-region. 
The presence of logarithmic contributions in $\alpha$ is a new feature, which shows that the non-analytic behavior of the perturbative expansion plays an increasingly central role at higher orders. 

Turning to the $\bar{r}$-region we find new $\alpha^4$ and $\alpha^4 \log |\alpha|$ contributions to $\psi$, showing the presence of non-analytical terms also in the $\bar{r}$-region. Finding these new contributions involves an intricate matching between the two regions in the overlap region, where perturbative orders are mixed non-trivially between the two regions. 
To solve for $\psi$ in the $\bar{r}$-region one has to proceed numerically, imposing continuity across the OLS and a monopole-like profile in the asymptotic region.
One of our main results is that the non-analytic behavior also appears in the extraction of energy and angular momentum of the magnetosphere-equipped Kerr black hole in the perturbative expansion for $\alpha\ll 1$.
Opting for the angular velocity of the event horizon $\Omega_H$ as the perturbative parameter,\footnote{Usually in this paper we work with the perturbative parameter $\alpha$ but one can always recast any $\alpha$-expanded quantity in terms of $\Omega_H$ via the simple relation
\begin{equation*}
    \Omega_H = \frac{\alpha}{\left(\sqrt{1-\alpha ^2}+1\right) r_0}.
\end{equation*}} the result we get in the low spin regime $r_0|\Omega_H| \ll1$ for the energy flux at the horizon is
\begin{equation}
\begin{split}
    \label{Power_OmegaH}
    \dot{E}\big\vert_{r_+} = \frac{2\pi}{3} \Omega_H^2 \bigg\{1 &+ 0.34588327 ~ r_0^2\Omega_H^2 -0.70309718 ~r_0^4\Omega_H^4 + 0.0483269(2)~r_0^5|\Omega_H|^5\\
    & +\left[0.1837383(5) -0.0027081(2) \log(r_0 |\Omega_H|)\right]r_0^6 \Omega_H^6 + \cdots \bigg\} ,
\end{split}
\end{equation}
where $r_0=2M$ is a dimensional parameter proportional to the black hole mass and $r_+$ is the location of the event horizon. Equation~\eqref{Power_OmegaH} shows the high-order contributions with respect to the BZ power emission formula $\dot{E}^{BZ}\vert_{r_+} = (2\pi/3) \Omega_H^2$. 
The new terms in Eq.~\eqref{Power_OmegaH} are those proportional to $|\Omega_H|^7$, $\Omega_H^8$ and $\Omega_H^8 \log(r_0 |\Omega_H|)$. Clearly, the contributions proportional to $|\Omega_H|^7$ and $\Omega_H^8 \log(r_0 |\Omega_H|)$ are non-analytic in a perturbative expansion of $r_0 \Omega_H$ around $\Omega_H=0$. 
Interestingly, we find that in order to determine the coefficients of $|\Omega_H|^7$ and $\Omega_H^8$ one needs to extend smoothly the magnetic flux function $\psi$ across the OLS. To do so we have implemented a novel numerical procedure. Such method, which is explained in detail in Appendix \ref{App:Numerics}, improves significantly the convergence and accuracy with respect to previous attempts in the literature \cite{Armas:2020mio}. It involves expanding $\psi$ in a basis of mode functions that capture its angular dependence. The numerical procedure amounts to solve a system of second order ordinary differential equations with a spectral collocation scheme.  
Contrary to \cite{Armas:2020mio}, no minimization procedure is required. The same technique is subsequently employed to solve numerically for $\psi$ at order $\alpha^4$ and $\alpha^4 \log |\alpha|$ in the $\Bar{r}$-region.
The analytic expressions of all the coefficients in Eq.~\eqref{Power_OmegaH} are given in Sec.~\ref{Sec: ELoutflows}; here we provide their numerical values for the ease of the presentation. The coefficients are approximated to the 8th significant digit. The uncertainty due to the numeric procedure only affects the coefficients of $|\Omega_H|^7$ and $\Omega_H^8$. The first subleading correction (i.e. in $\Omega_H^4$) is in perfect agreement with the results of \cite{Tanabe_2008} and we discuss the agreement of the second subleading correction (i.e.  in $\Omega_H^6$) with \cite{Pan:2015iaa} in Sec.~\ref{Sec: ELoutflows}. 
Note that we are imposing the requirement that there is no difference in the physics of the BZ split-monopole solution if one reverses the direction of both the angular velocity of the black hole as well as the rotational direction of the magnetosphere. Thus, sending $\alpha$ to $-\alpha$, the magnetic flux $\psi$ should be the same, but spin-direction dependent quantities such as angular velocities and angular momenta should change sign. Clearly, the energy flux should instead be invariant, thus $\dot{E}(-\Omega_H) = \dot{E}(\Omega_H)$ which is saying that the energy flux is an even function of the angular velocity. In other words, the power emitted does not depend on the direction of the black hole rotation, meaning that clockwise and anti-clockwise rotations are physically equivalent because both configurations extract the same amount of energy per unit time.
For the flux of angular momentum we also find logarithmic contributions. Indeed, at the horizon we compute
\begin{equation}
\begin{split}
    \dot{L}\big\vert_{r_+}= 
    \frac{4\pi}{3}\Omega_H\bigg\{ 1 &+ 0.17294163 ~r_0^2\Omega_H^2-0.0694564(6)~r_0^3 |\Omega_H|^3 \\
    &+ \left[
    -0.3736222(9) + 0.0038095(2) \log{\left(r_0|\Omega_H|\right)}\right]r_0^4\Omega_H^4+\cdots \bigg\}.
\end{split}
\end{equation}
The novelty here is the presence of logarithmic contributions proportional to  $\Omega_H^5$. This is yet another manifestation of the non-analytic behavior of the observables in the low spin regime $r_0 |\Omega_H| \ll 1$. As for the energy flux, we approximated the numerical values at the 8th significant digit.
The paper is organized as follows. In Sec.~\ref{Sec:FFE_Kerr} we briefly review the Kerr geometry and provide the basics of stationary and axisymmetric force-free fields. Sec.~\ref{Subsec: SE_cs} introduces a new partially-covariant expression for the stream equation in the Kerr background, which turns out to highlight the presence of its critical surfaces and the relative regularity conditions which will be used throughout the entire paper. 
The reader familiar with FFE models of magnetospheres in the Kerr geometry can skip the first two sections. In Sec.~\ref{sec: MAE_BZ}, we review the main feature of the BZ perturbative approach 
and the method of the \emph{matched asymptotic expansions} (MAE) applied to the BZ split-monopole case. 
Secs.~\ref{Sec: Exp} and \ref{Sec: ELoutflows} are entirely devoted to presenting our new results outlined in the Introduction.
We conclude with a discussion of our results in Sec.~\ref{sec: conclusions}. We refer the reader to Appendix \ref{App: previous_orders} for a brief, nonetheless self-consistent, computation of the MAE method applied to lower orders in the spin parameter, as previously derived in the literature; Appendix \ref{App:Green} contains a discussion about the Green's function method in the context of the BZ expansion; Appendix \ref{App:Numerics} contains a detailed description of the numerical strategies exploited to solve and make the flux function continuous across the OLS in the $\bar{r}$-region.

\section{Force-free electrodynamics in Kerr background}
\label{Sec:FFE_Kerr}

We briefly review force-free magnetospheres around Kerr black holes. This section is intended to recap definitions and equations. For a detailed treatment of the subject, we refer the reader to the reviews \cite{Beskin_2003,Gralla:2014yja,Blandford_2019,Punsly}.

\subsection{Kerr geometry}
\label{Subsec:2.1}

%
The Kerr metric describes the spacetime geometry of a rotating black hole with mass $M$ and angular momentum $J$. In the Boyer-Lindquist (BL) coordinates and in geometric units ($G=1=c$), the Kerr metric reads 
\begin{equation} 
    \label{KerrBL}
    ds^2=-\Big(1-\frac{r_0r}{\Sigma}\Big)dt^2-\frac{2r_0r}{\Sigma}a\sin^2 \theta ~dt d\phi +\frac{(r^2+a^2)^2-a^2\Delta\sin^2 \theta}{\Sigma}\sin^2 \theta ~d\phi^2
    +\frac{\Sigma}{\Delta}dr^2+\Sigma d\theta^2,
\end{equation}
with $r_0=2M$, specific angular momentum $a=J/M$ and
\begin{equation}
    \Sigma = r^2+a^2 \cos^2 \theta, \quad \Delta=(r-r_+)(r-r_-),
    \quad r_\pm = \frac{r_0}{2}\left(1\pm \sqrt{1-\frac{4a^2}{r_0^2}}\right).
\end{equation}
The position of the event horizon is $r=r_+$.
The Kerr black hole \eqref{KerrBL} is stationary and axisymmetric, thus its isometry group is $\mathbb{R}\times U(1)$. 
In the BL coordinates the metric \eqref{KerrBL} can be rearranged in a block-diagonal form so as to separate the toroidal coordinates $(t,\phi)$ from the poloidal coordinates $(r,\theta)$. The metric takes the product form
\begin{equation}
    \label{metric_TP}
    ds^2 = ds_T^2+ ds_P^2, \quad\text{where}\quad ds_T^2 = g_{tt} dt^2 + 2g_{t\phi} dt d\phi + g_{\phi\phi} d\phi^2, \quad ds_P^2 = g_{rr} dr^2 + g_{\theta\theta} d\theta^2,
\end{equation}
and one can define the determinants
\begin{equation}
    g_T=g_{tt}\,g_{\phi\phi}-{g_{t\phi}}^2=-\Delta \sin^2\theta,\quad  g_P=g_{rr}\,g_{\theta\theta}=\frac{\Sigma^2}{\Delta},\quad g=g_T\,g_P=-\Sigma^2\sin^2\theta.
\end{equation}

\subsection{Stationary and axisymmetric force-free magnetospheres}
\label{Subsec:2.2}

%
FFE is the low-density limit of ideal magnetohydrodynamics (MHD), in which one assumes that the energy contribution of the magnetic field overwhelms the inertia of the plasma. Under such conditions the dynamics of the magnetosphere is governed by the Maxwell's equations 
    \begin{equation}
        \label{Maxwell_eq}
        D_\mu F^{\mu\nu}=-j^\nu, \quad D_{[\rho}F_{\mu\nu]}=0,
    \end{equation}
supplemented with the force-free constraints
    \begin{equation}
        \label{FF_constr}
        F_{\mu\nu}j^\nu=0, 
        \quad
        j^\mu\neq0.
    \end{equation}
Here $F_{\mu\nu}=\partial_\mu A_\nu-\partial_\nu A_\mu$ and $A_\mu$ are the electromagnetic field strength and the gauge potential, respectively, whereas $j^\mu$ is the current density of the plasma. 
Equations~\eqref{Maxwell_eq} and \eqref{FF_constr} can be used to eliminate the current density $j^\mu$ to get 
\begin{equation}
    \label{FFE_0}
    F_{\mu\nu}D_{\rho}F^{\nu\rho}=0.
\end{equation}
FFE, therefore, captures the non-linear dynamics of the plasma-filled magnetosphere around compact objects, such as black holes, pulsars and stars \cite{Punsly}.

We consider magnetospheres around a Kerr black hole that share the same symmetries of the background, \emph{i.e.} stationarity and axisymmetry, and our task is to solve \eqref{FFE_0} on a fixed Kerr background. In this case, it is possible to choose a gauge such that the gauge potential is independent of the time and azimuthal coordinates, namely $\partial_t A_\mu=\partial_\phi A_\mu=0$. One defines
\begin{equation}
    \label{psi_definition}
    \psi(r,\theta)\equiv A_\phi(r,\theta)
\end{equation}
as the \emph{magnetic flux} (divided by $2\pi$) through a circular loop of radius $r\sin\theta$ surrounding the rotational axis of the black hole. Combining the toroidal components of Eq.~\eqref{FFE_0}, one gets $\partial_\theta  A_t\partial_r \psi=\partial_\theta  \psi\partial_r A_t$, which implies that $A_t$ can be regarded as a function of $\psi$.\footnote{Indeed recall that the chain rule for a generic function $Q\big(\psi(r,\theta)\big)$ states that
\begin{equation}\label{Integrability} 
 \left\{
\begin{array}{ll}
\partial_r Q =\frac{dQ}{d\psi}\partial_r\psi\,,& \\
\partial_\theta Q =\frac{dQ}{d\psi}\partial_\theta \psi  \,,
\end{array}
\right.
\Rightarrow 
\frac{\partial_r Q}{\partial_r\psi}=\frac{dQ}{d\psi}=\frac{\partial_\theta Q}{\partial_\theta\psi}\,.
\end{equation}
Conversely, if the latter condition holds then $Q$ must be a function of $\psi$. We use this integrability property for the quantities $A_t$, $\Omega$ and $I$. 
 } 
 The \emph{angular velocity of magnetic field lines $\Omega$} can thus be defined as $\Omega=-\frac{\partial_r A_t}{\partial_r \psi}=-\frac{\partial_\theta A_t}{\partial_\theta \psi}$, \emph{i.e.} 
\begin{equation}\label{defOmega}
    \partial_\theta A_t=-\Omega \partial_\theta \psi,\quad  \partial_r A_t=-\Omega \partial_r \psi.
\end{equation}
Taking $\partial_r$ of the first condition in \eqref{defOmega} and subtracting $\partial_\theta$ of the second condition in \eqref{defOmega}, after using the integrability condition for $A_t(\psi)$ one finds that  $\Omega$ satisfies itself the integrability condition
\begin{equation}
    \label{intcond1}
    \partial_\theta \Omega\, \partial_r\psi=\partial_\theta\psi\, \partial_r\Omega,
\end{equation}
meaning that the angular velocity of field lines is a function of $\psi$ alone, $\Omega=\Omega(\psi)$.
Finally, one defines the \emph{poloidal current} $I$ as
\begin{equation}
    \label{I_def}
    I=\sqrt{-g}F^{\theta r}.
\end{equation}
The radial component of Eq.~\eqref{FFE_0} reads
\begin{equation}
    \label{prestr1}
    \partial_r \psi \left[ - \Omega\, \partial_\rho ( \sqrt{-g} F^{t\rho}) + \partial_\rho ( \sqrt{-g} F^{\phi\rho}) \right] + F_{r\theta} \partial_r I =0,
\end{equation}
while for the polar component of Eq.~\eqref{FFE_0} we obtain
\begin{equation}
    \label{prestr2}
    \partial_\theta \psi \left[ - \Omega\, \partial_\rho ( \sqrt{-g} F^{t\rho}) + \partial_\rho ( \sqrt{-g} F^{\phi\rho}) \right] + F_{r\theta} \partial_\theta I =0.
\end{equation}
For Eq.~\eqref{prestr1} to be consistent with Eq.~\eqref{prestr2} it is necessary that the poloidal current is a function of  $\psi$, $I=I(\psi)$, which is equivalent to the integrability condition
\begin{equation}
    \label{intcond2}
    \partial_r I \,\partial_\theta\psi = \partial_\theta I\, \partial_r \psi.
\end{equation}
Finally, from Eqs.~\eqref{prestr1}-\eqref{prestr2} we get 
\begin{equation}
\label{stream1}
\Omega \partial_\rho ( \sqrt{-g} F^{\rho t}) - \partial_\rho ( \sqrt{-g} F^{\rho\phi})  + F_{r\theta} \frac{dI}{d\psi} =0,
\end{equation}
which is the stream equation in a stationary and axisymmetric background.
The above equations all follow from the FFE conditions \eqref{FFE_0}. As a consistency check, we can now verify that they also imply that the associated Maxwell energy momentum tensor on the Kerr background is conserved: $\nabla_\mu T^{\mu\nu}=0$ with $T_{\mu\nu}=F_\mu^{\:\:\alpha}F_{\nu\alpha}-\frac{1}{4}g_{\mu\nu}F^2$.

\section{Stream equation and its critical surfaces}
\label{Subsec: SE_cs}

In this section we exhibit a new partially-covariant form for the stream equation \eqref{stream1}, useful for analyzing the critical surfaces.
Subsequently, we review the four possible critical surfaces for a magnetosphere in the background of the Kerr geometry.

\subsection{Partially covariant form of the stream equation}

In order to write the stream equation \eqref{stream1} in a more covariant fashion, it is useful to introduce the one-form \cite{Gralla:2014yja}
\begin{equation}
    \eta=d\phi-\Omega(\psi) dt.
\end{equation}
By means of the block-diagonal decomposition \eqref{metric_TP}, any stationary and axisymmetric force-free field can be written as\cite{Gralla:2014yja}
\begin{equation}
    \label{FF_field}
    F=-I(\psi)\sqrt{-\frac{g_P}{g_T}}dr\wedge d\theta+d\psi\wedge \eta.
\end{equation}
 Therefore, any stationary and axisymmetric force-free magnetosphere is fully determined once the explicit functional expressions for $\psi$, $I$ and $\Omega$ are known. The stream equation \eqref{stream1} can be conveniently recast as
\begin{equation}
    \label{stream2}
    \eta_\mu \partial_\nu \Big(  \eta^\mu \sqrt{-g}\, g^{\nu\rho} \partial_\rho \psi \Big) = F_{r\theta} \frac{dI}{d\psi}.
\end{equation}
Specializing to the Kerr metric, the stream equation takes the form
\begin{equation}
    \label{stream3}
    \eta_\mu \partial_r \Big( \eta^\mu  \Delta \sin \theta \,  \partial_r \psi \Big) + \eta_\mu \partial_\theta \Big( \eta^\mu  \sin \theta \,  \partial_\theta \psi \Big) + \frac{\Sigma}{\Delta \sin \theta} I \frac{dI}{d\psi} =0.
\end{equation}
As shown below, such partially covariant expression for the stream equation makes the analysis of the critical surfaces of the force-free magnetosphere around rotating black holes easier to perform.

Equation~\eqref{stream3} constitutes the general relativistic version of the force-free Grad-Shafranov equation in the Kerr geometry. It follows from the full MHD Grad-Shafranov equation for stationary and axisymmetric flows \cite{Beskin_2003} after taking the limit of vanishing density for the plasma filling the magnetosphere. From a pure mathematical perspective, Eq.~\eqref{stream3} is a second-order quasi-linear partial differential equation (PDE) for the magnetic flux $\psi$ and the difficulties one encounters in solving this equation are, in first place, due to the presence of the two integrals of motion $I(\psi)$ and $\Omega(\psi)$ in the form of free functions that have to be determined.
The problem of providing prescriptions for these two functions is not only related to the nature of the astrophysical object immersed in the magnetosphere -- whether it is a pulsar or a black hole, say -- but also to the topology of the magnetic field lines that one aims to study \cite{Gralla:2014yja,Uzdensky:2003cy,Uzdensky:2004qu}.\footnote{See \cite{Uzdensky:2004qu} for configurations of closed magnetic field lines connecting the black hole with a thin accretion disc, or \cite{Nathanail:2014aua} for the case of a black hole immersed in a vertical magnetic field.}

By considering the canonical form for a second-order PDE, namely,
\begin{equation}\label{PDEcan}
    \mathcal{A}\,\partial^2_r\psi+2\mathcal{B}\,\partial_r\partial_\theta\psi+\mathcal{C}\,\partial^2_\theta\psi+\dots=0,
\end{equation} 
it is possible to study the character of Eq.~\eqref{stream3}. When the discriminant is non-negative
\begin{equation}\label{discriminant}
    \mathcal{A}\mathcal{C}-\mathcal{B}^2=\sin^2\theta(\eta^\mu\eta_\mu)^2\Delta\geq 0,
\end{equation}
the stream equation is elliptic everywhere in the exterior region of the Kerr black hole $(r>r_+,\theta)$, with the exception of critical surfaces \cite{Beskin_2003}. In the Kerr background, one identifies four critical surfaces which are regular singular points
(or surfaces) of the stream equation \eqref{stream3}: the event horizon, the ILS, the OLS, and the asymptotic region. 

%
\subsection{Light surfaces and regularity conditions}
\label{Subsec: 2.3.1}

The position of the light surfaces can be determined by looking at those surfaces where the velocity vector field $\chi=\partial_t+\Omega(\psi)\partial_\phi$ of an observer co-rotating with the magnetosphere becomes null:
    \begin{equation}
        \label{chi2}
        \chi^\mu\chi_\mu=g_{tt}+2\Omega g_{t\phi}+\Omega^2g_{\phi\phi}=0.
    \end{equation}
Equation~\eqref{chi2}, which implicitly defines the position $r_{\rm LS}(\theta)$ of the light surfaces, admits two solutions in the Kerr geometry,\footnote{There are two solutions under the assumption $\Omega\neq0$. A non-trivial interesting case in which there is only one light surface is the uniform vertical configuration, where outside the cylindrical separatrix the field lines are assumed not to rotate \cite{Nathanail:2014aua,Pan:2014bja,Yang:2015ata,Pan:2017npg,East:2018ayf}.} corresponding to the ILS and the OLS. While the OLS represents the black hole analogue of the light cylinder in pulsar magnetospheres, the presence of the ILS is a feature of General Relativity; in particular, the ILS originates at the pole of the event horizon and it is always located inside the ergosphere \cite{Kom2004}.
Light surfaces are also present in MHD flows as the critical surfaces of the full Grad-Shafranov equation for which the electric field equals in magnitude the poloidal component of the magnetic field. In the force-free limit, where the inertia of the plasma is negligible, both the ILS and OLS also coincide with Alfvén surfaces\cite{Kom2004,Beskin_2003}.

Notice that, whenever the toroidal sector of the metric is non-degenerate, the one-form $\eta$ is null at $r_{LS}(\theta)$ as well; in this case indeed
\begin{equation}
\label{eta2}
\eta_\mu\eta^\mu=\frac{1}{g_T}\chi^\mu\chi_\mu=-\frac{1}{\Delta\sin^2\theta}\chi^\mu\chi_\mu=0.
\end{equation}
Accordingly, the discriminant \eqref{discriminant} vanishes, $\mathcal{A}\mathcal{C}-\mathcal{B}^2=0$. At the light surfaces both the second derivatives vanish, and Eq.~\eqref{stream2} becomes
    \begin{equation}
        \label{stream_LS1}
        \sqrt{-g} \, g^{\nu\rho} \partial_\rho  \psi \, \eta_\mu \partial_\nu  \eta^\mu  = F_{r\theta} \frac{dI}{d\psi}.
    \end{equation}
For the Kerr metric this reads
    \begin{equation}
        \label{stream_LS2}
        \Delta\,   \eta_\mu \partial_r \eta^\mu  \partial_r \psi + \eta_\mu \partial_\theta  \eta^\mu  \partial_\theta \psi  + \frac{\Sigma}{\Delta \sin^2 \theta} I \frac{dI}{d\psi} =0.
    \end{equation}
Since the second-order terms in 
Eq.~\eqref{stream3} vanish at $r_{\rm LS}(\theta)$, the stream equation can admit solutions for which the second derivatives of $\psi$ diverge as $\partial^2\psi\propto (\eta^\mu\eta_\mu)^{-1}$. For such solutions, the first derivative of $\psi$ has a discontinuity across the light surface, which reflects a discontinuity of poloidal magnetic fields according to Eq.~\eqref{FF_field}. This jump can be supported only if an infinitesimally-thin current sheet is present at the light surface, which in turn constitutes a layer where the force-free assumption is violated.

Equation~\eqref{stream_LS1}, or equivalently Eq.~\eqref{stream_LS2}, serves to discard such discontinuous solutions by constraining the first derivative of $\psi$, and acts as a \emph{regularity condition} across the light surface. Since regularity at $r_{LS}(\theta)$ is not automatically implied by the stream equation, in order to construct realistic models of force-free magnetospheres, such regularity conditions have to be imposed as additional \emph{physical constraints} on the FFE system of equations. This was first noticed in \cite{Contopoulos:1999ga} and applied in other magnetospheric configurations~\cite{Uzdensky:2004qu,Nathanail:2014aua}, where Eq.~\eqref{stream_LS1} was employed in an iterative numerical optimization procedure to derive expressions for the unknown functions $I(\psi)$ and $\Omega(\psi)$ under the assumption of a smooth matching for $\psi$ across the light surface.

Finally, notice that the ``classical" perspective about the BZ mechanism \cite{1977MNRAS.179..433B} regards open field lines extending from the horizon to infinity as responsible for the extraction of energy and angular momentum from the black hole.\footnote{Although for the extraction to take place what is really crucial is to consider magnetic field lines connecting the ergosphere with infinity, irrespective of whether they penetrate the horizon or not \cite{Kom2004}.} Magnetic field lines in this configuration necessarily have to intersect both the ILS and the OLS \cite{Gralla:2014yja}, so that for open field lines threading the horizon of a rotating black hole the regularity condition \eqref{stream_LS2} has to be imposed at $r_{\rm OLS}(\theta)$ and at $r_{\rm ILS}(\theta)$.
    
As a final remark, from Eq.~\eqref{FF_field}, one has 
\begin{equation}
F^2=-\frac{I^2}{g_T}+|d\psi|^2|\eta|^2.
\end{equation}
Hence at the light surface, when $d\psi$ is regular, $F^2|_{\rm LS}=-I^2/g_T\geq0$. This means that the regularity condition at the light surface also imposes the field to be magnetically dominated or null on that surface.

%
\subsection{Event horizon and the Znajek condition} 

 The event horizon of a Kerr black hole is a regular singular surface as well. This conclusion can be drawn by observing that in Eq.~\eqref{stream3} the quantity $\eta^\mu=g^{\mu\nu}\eta_\nu$ appears and it explicitly reads 
    \begin{equation}
        \label{eta_h}
        \eta^\mu=\frac{1}{g_T}\left[-\left(g_{t\phi}+g_{\phi\phi}\Omega\right)\,{\delta^\mu}_t+\left(g_{tt}+g_{t\phi}\Omega\right)\,{\delta^\mu}_\phi\right] \equiv \frac{1}{\Delta}h^\mu.
    \end{equation}
As a consequence, all the terms containing radial derivatives in $\psi$ disappear from the stream equation~\eqref{stream3}, when this is evaluated at the horizon. The stream equation in this case reduces to the following ODE
    \begin{equation}
        \label{stream_EH}
        \left.\left[\eta_\mu\partial_\theta\left(h^\mu\sin\theta\partial_\theta\psi\right)+\frac{\Sigma}{\sin\theta}I\frac{dI}{d\psi}\right]\right|_{r_+}=0.
    \end{equation}
    It is useful to notice that
    \begin{equation}
        \label{h+}
        \left. h^\mu \right\vert_{r_+}=-\left.\left[\frac{r_0^2r_+^2}{\Sigma}(\Omega_H-\Omega)\left({\delta^\mu}_t+\Omega_{H}{\delta^\mu}_\phi\right)\right]\right\vert_{r_+},
    \end{equation}
    where $\Omega_H=-(g_{t\phi}/g_{\phi\phi})\vert_{r_+}=a(r_0r_+)^{-1}$ is the ZAMO (i.e. zero-angular-momentum-observer) angular velocity at $r=r_+$. Since the event horizon is a surface of constant $r$, all the derivatives in Eq.~\eqref{stream_EH} are performed along the $\theta$-direction. By means of Eq.~\eqref{h+}, and with a few algebraic steps, it is possible to integrate the reduced stream equation~\eqref{stream_EH} as follows
    \begin{equation}
        \left.\partial_\theta\left[\left(\frac{r_0r_+}{\Sigma}\sin\theta\right)^2(\Omega_H-\Omega)^2\left(\partial_\theta\psi\right)^2-I^2\right]\right\vert_{r_+}=0.
    \end{equation}
    The integration constant can be set to zero by requiring the current $I$ to vanish along the rotation axis of the black hole at $\theta=0$. The solution of $I_+$ with the positive sign ensures regularity on the future event horizon of the black hole and it is known as the \emph{Znajek condition} \cite{10.1093/mnras/179.3.457} or frozen-in condition at the horizon
    \begin{equation}
        \label{ZC}
        I(r_+,\theta)=\left.\left[\left(\frac{r_0r_+}{\Sigma}\sin\theta\right)\left(\Omega_H-\Omega\right)\partial_\theta\psi\right]\right\vert_{r_+}.
    \end{equation}

    In general, the Znajek condition at the horizon should not be regarded as a boundary condition: since it directly descends from the stream equation \eqref{stream3}, it cannot be independently imposed to select a particular solution to the magnetosphere problem \cite{Nathanail:2014aua}. As a matter of fact, Eq.~\eqref{ZC} is a true regularity condition which guarantees finiteness of the electromagnetic field \eqref{FF_field} when measured by a freely-falling observer crossing the horizon.
    This is also related to the fact that the origin of the Znajek condition can be traced back into the more general context of MHD as a regularity condition imposed at the fast magnetosonic critical surface which, in the low-density limit of FFE, coincides with the horizon \cite{Komissarov:2002my,Beskin_2003,Kom2004,Uzdensky:2003cy,Uzdensky:2004qu}.

Numerical resolutions of the Grad-Shafranov equation \cite{Mahlmann:2018ukr}, as well as time-dependent numerical simulations \cite{Kom2004,Nathanail:2014aua}, revealed that solutions with a smooth matching across the light surfaces automatically satisfy the horizon regularity condition \eqref{ZC}. 
In the present paper we give an analytical confirmation of this fact at the level of the ILS, for each one of the perturbative orders taken into account.
The redundancy of the condition at the ILS is understood as a consequence of the small spin expansion. As we shall discuss later, this feature can be seen by identifying the critical points in the differential operators associated to the stream equation within the MAE scheme; see comments below Eqs.~\eqref{Lr} and \eqref{Lbarr}. 

%
\subsection{Asymptotic infinity}
\label{subsec:2.3.3}
    
As mentioned earlier, open magnetic field lines connecting the system to infinity are crucial for the energy and angular momentum to be carried away from the black hole. However, $r=\infty$ is a regular singular point of the stream equation \eqref{stream3} and, as such, it should be equipped with its own regularity condition. Labelling the asymptotic value of a function with a superscript $\infty$, we can notice that the vector $h^\mu$, defined in Eq.~\eqref{eta_h}, asymptotes to
\begin{equation}
       \label{h_inf}
    (h^\mu)^\infty\approx -r^2\Omega^\infty{\delta^\mu}_t.
\end{equation}
Analogously to what discussed previously about the horizon, in the limit $r\to\infty$ one observes that radial-derivative terms are negligible as compared to angular derivatives; the reduced stream equation can be easily integrated to derive the following regularity condition in the asymptotic region
\begin{equation}
     \label{inf_ZC}
    I^\infty(\theta)=\sin\theta\,\Omega^\infty(\partial_\theta\psi)^\infty.
\end{equation}
We refer to this as the Znajek condition at infinity and, as for the Znajek condition at the horizon \eqref{ZC}, this should not be regarded as a boundary condition, but rather as an outgoing radiation condition \cite{Nathanail:2014aua,PhysRevD.92.084017}.


\section{Matched asymptotic expansions I: past results}
\label{sec: MAE_BZ}

In this section we briefly review the BZ perturbative approach and the MAE method, while recalling all the results in the literature that are relevant to our analysis. 

In addition to reviewing past results we also provide novel analytical expressions for two coefficients $W_0$ and $V_0$ (Eqs.~\eqref{Ws} and \eqref{Vs}) that previously were only known numerically, as well as present the complete analytic expression for $\psi_4(r,\theta)$ in Appendix \ref{app:explicit}.

\subsection{Blandford--Znajek split-monopole in the slow-spin regime}
\label{Subsec:3.1}
The perturbative parameter of the BZ approach is related to the specific angular momentum of a Kerr black hole as follows
\begin{equation}
    \alpha=\frac{J}{M^2}=\frac{2a}{r_0},
\end{equation}
with the slow-spin regime defined by the condition $\alpha\ll1$. Under this assumption, it is possible to expand perturbatively in the parameter $\alpha$ all the relevant quantities of the Kerr background.

Note that we assume $\alpha > 0$ in the following, unless otherwise noted. As explained in the Introduction, the physics of magnetospheres of clockwise $\alpha > 0$ and anti-clockwise $\alpha < 0$ rotations are equivalent. Thus, one should have a symmetry of the magnetosphere if one sends $\alpha \rightarrow -\alpha$ along with $\Omega \rightarrow -\Omega$ and $I \rightarrow -I$ while keeping $\psi$ the same.

At the leading order in $\alpha$, the Kerr metric reduces to the Schwarzschild geometry
\begin{equation}
    ds^2=-\left(1-\frac{r_0}{r}\right)dt^2+\left(1-\frac{r_0}{r}\right)^{-1}dr^2+r^2(d\theta^2+\sin^2\theta\, d\phi^2)+\mathcal{O}(\alpha).
\end{equation}
The seed solution for this construction is therefore a static force-free field, which can be derived by solving the source-free stream equation in the Schwarzschild spacetime,\footnote{Notice that the stream equation in the Schwarzschild spacetime only has two critical surfaces coincident with the horizon $r=r_0$ and infinity $r=+\infty$.} namely
\begin{equation}
    \label{L_schw}
    \frac{1}{\sin\theta}\partial_r\left[\left(1-\frac{r_0}{r}\right)\partial_r\psi\right]+\frac{1}{r^2}\partial_\theta\left(\frac{1}{\sin\theta}\partial_\theta\psi\right)=0.
\end{equation}

Among the static magnetospheres, there are configurations with monopole \cite{1977MNRAS.179..433B}, paraboloidal\cite{1977MNRAS.179..433B}, hyperbolic\cite{Gralla:2015vta} and vertical \cite{Nathanail:2014aua,Pan:2014bja} magnetic field line profiles. We refer the reader to \cite{Grignani:2019dqc} for a classification of the asymptotic behaviour.
Typically, in solving Eq.~\eqref{L_schw}, one restricts to consider only the northern hemisphere $0\leq\theta<\pi/2$, and then patches that solution with the one in the southern hemisphere. These \emph{split-field configurations} are, by construction, discontinuous across the equatorial plane, $\theta=\pi/2$, onto which an infinitesimal current-sheet develops. One can regard this current-sheet as a very crude picture of a thin accretion disc, whose presence has to be invoked in order to sustain the magnetic field surrounding the black hole, as a consequence of the no-hair theorem.\footnote{
Notice that the force-free assumptions are not valid in the current-sheet region. Furthermore, it is important to stress that approximating the accretion disc with a current sheet misses some crucial aspects of the physics of black hole accretion; for example the equatorial current-sheet extends from infinity all the way down to the event horizon, which contrasts with the presence of an ISCO, or more generally an inner edge, in accretion disc models.}

In this work the main focus will be devoted to the split-monopole configuration
\begin{equation}
\label{monopole_config}
    \psi_0=1-\cos\theta,\quad I_0=0, \quad \Omega_0=0.
\end{equation}
This is the seed solution of the BZ perturbative approach  which consists in constructing corrections in $\alpha$ for $\psi$, $I$, and $\Omega$ and, by means of these, turning-on a slow rotation in the system both for the black hole and for the force-free field. These corrections are derived in such a manner as to be consistent with the boundary conditions of the split-monopole field \cite{Armas:2020mio, Grignani:2019dqc}, namely
\begin{equation}
\label{BC_split-monopole}
    \psi\big\vert_{\theta=0}=0,\quad \psi\big\vert_{\theta=\pi/2}=1,\quad \partial_\theta\psi\big\vert_{\theta=0}=0, \quad \lim_{r\to\infty}\psi=\psi^\infty(\theta).
\end{equation}

The positions of the relevant surfaces for the black hole magnetosphere are expanded in $\alpha$ as well. The  event horizon position admits an analytic expansion around $\alpha=0$ 
\begin{equation}
    \label{rplus_exp}
    \frac{r_+}{r_0}=\frac{1}{2}+\frac{1}{2}\sqrt{1-\alpha^2}=1-\frac{1}{4}\alpha^2-\frac{1}{16}\alpha^4+\mathcal{O}(\alpha^6).
\end{equation}
Similarly, concerning the position of the static limit
\begin{equation}
    \label{rergo_exp}
    \frac{r_{\rm ergo}(\theta)}{r_0}=\frac{1}{2}+\frac{1}{2}\sqrt{1-\alpha^2\cos^2\theta}=1-\frac{1}{4}\alpha^2\cos^2\theta-\frac{1}{16}\alpha^4\cos^4\theta+\mathcal{O}(\alpha^6),
\end{equation}
one observes that the ergosphere detaches from the event-horizon as the rotation increases.

When one does not know the expression of $\Omega$, the position of the light surfaces cannot be determined a priori. Nevertheless it is natural to assume that, as for the other critical surfaces involved in this problem, the position of the inner and outer light surfaces can be expanded in $\alpha$. From this, together with Eqs.~\eqref{rplus_exp} and \eqref{rergo_exp}, we can gather already 
\begin{equation}
    \label{rILS_leading}
    \frac{r_{\rm ILS}(\theta)}{r_0}=1+\mathcal{O}(\alpha^2),
\end{equation}
whereas $r_{\rm OLS}$ at this point is only known to approach infinity as $\alpha$ goes to zero and the Schwarzschild spacetime is recovered.
All the missing information has to be determined consistently from Eq.~\eqref{chi2} as part of the resolution of the relative perturbative order. For instance, we anticipate that, for the monopole configuration, at order $\alpha$ in the field variables one gets the first subleading $\mathcal{O}(\alpha^2)$ term in $r_{\rm ILS}$ and the leading order of $r_{\rm OLS}$, namely \cite{Armas:2020mio}
\begin{equation}
    \label{rOLS_leading}
    \frac{r_{\rm OLS}(\theta)}{r_0}= \frac{4}{\sin\theta}\frac{1}{\alpha}+\mathcal{O}(\alpha^0).
\end{equation}
\subsection{Matched Asymptotic Expansions}
\label{Sec: MAE}

As correctly pointed out in \cite{Armas:2020mio}, since the OLS scales as $r/r_0\sim\alpha^{-1}$, it is necessary to introduce a new radial coordinate in order to resolve its position and to distinguish the OLS from the asymptotic region. We therefore consider the coordinate
\begin{equation}
    \bar{r}=\alpha r,
\end{equation} 
which is designed to keep the position of the OLS fixed when performing a small $\alpha$-expansion. 
This allows one to distinguish two different regions in which perturbation theory should be applied: the \emph{$r$-region} and the \emph{$\bar{r}$-region}.

The $r$-region is defined by 
\begin{equation}
    \frac{r}{r_0}\ll\frac{1}{\alpha} \quad \Leftrightarrow \quad \frac{\bar{r}}{r_0}\ll1.
\end{equation}
In the $r$-region one solves the stream equation in terms of the coordinates $(r,\theta)$ which are both kept finite in the $\alpha$ expansion.
The force-free field variables are expanded in this region as \cite{Armas:2020mio}~\footnote{See Appendix~\ref{App: previous_orders} for a detailed explanation about the form of these expressions. Note in particular that the terms proportional to $\alpha^2$ in the functions $I(\psi)$ and $\Omega(\psi)$ are zero as well as those proportional to $\alpha$ and $\alpha^3$ in $\psi(r,\theta)$. As for the notation, we use $i_n(\psi_0)$ to highlight the dependence of the poloidal current on the flux function. In this sense $i_n'(\psi_0)=\partial_\theta i_n(\theta)/\partial_\theta \psi_0(\theta)$. The same considerations apply for $\omega_n(\psi_0)$. }
\begin{equation}
    \label{Exp_r_Jay}
    \begin{split}
        \psi(r,\theta)&=\psi_0(\theta)+\alpha^2\,\psi_2(r,\theta)+\alpha^4\,\psi_4(r,\theta)+\mathcal{O}(\alpha^5),
        \\
        r_0I(\psi)&=\alpha\, i_1(\psi_0)+\alpha^3\left[i_1'(\psi_0)\psi_2(r,\theta)+i_3(\psi_0)\right]+\alpha^4\,i_4(\psi_0)+\mathcal{O}(\alpha^5),
        \\
        r_0\Omega(\psi)&=\alpha\, \omega_1(\psi_0)+\alpha^3\,\omega_3(\psi_0)+\alpha^4\,\omega_4(\psi_0)+\mathcal{O}(\alpha^5).
    \end{split}
\end{equation}
At the $n$-th order in perturbation theory, in the $r$-region the stream equation can be formally written as 
\begin{equation}
    \label{SE_r_reg}
    \mathcal{L}\psi_n(r,\theta)=S_n(r,\theta;\psi_{k<n},i_{k<n},\omega_{k<n}), 
\end{equation}
where the operator $\mathcal{L}$, in the $r$-region, is defined as 
\begin{equation}
    \label{Lr}
    \mathcal{L}=\frac{1}{\sin\theta}\partial_r\left[\left(1-\frac{r_0}{r}\right)\partial_r\;\right]+\frac{1}{r^2}\partial_\theta\left(\frac{1}{\sin\theta}\partial_\theta\;\right).
\end{equation}
Notice that this operator is characterized by a critical surface at $r=r_0$, corresponding to the horizon and to the leading order term of the ILS location.

\begin{figure}[ht!]
    \centering
    \includegraphics[width=0.8\textwidth]{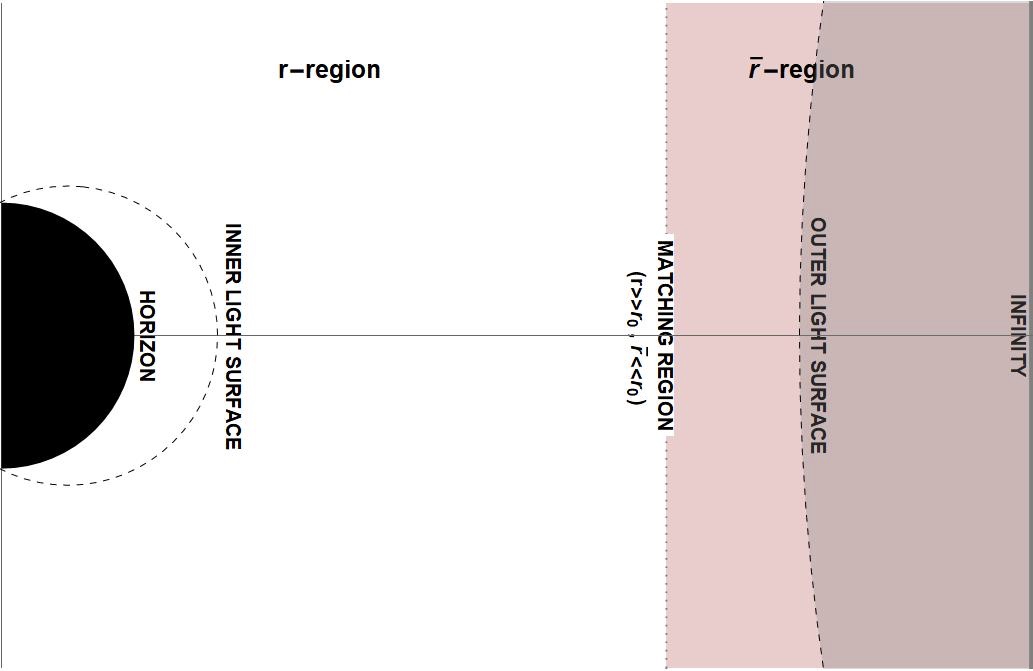}
    \caption{
    In the figure we sketched the magnetosphere structure around a Kerr black hole with the division in $r$-region and $\bar{r}$-region as introduced by the MAE scheme. 
    We highlighted with colors that the $\bar{r}$-region is further splitted by the OLS.}
    \label{fig:LS_Illustration}
\end{figure}

The $\bar{r}$-region is defined by
\begin{equation}
    \frac{r}{r_0}\gg1 \quad \Leftrightarrow \quad \frac{\bar{r}}{r_0}\gg\alpha.
\end{equation}
In the $\bar{r}$-region one solves the stream equation in terms of the coordinates $(\bar{r},\theta)$ which are both kept finite in the $\alpha$ expansion.
The force-free field variables are expanded in this region as \cite{Armas:2020mio}
\begin{equation}
\label{Exp_rbar_Jay}
    \begin{split}
        \psi(\bar{r},\theta)&=\psi_0(\theta)+\alpha^3\,\bar{\psi}_3(\bar{r},\theta)+\mathcal{O}(\alpha^4),
        \\
        r_0I(\psi)&=\alpha\, i_1(\psi_0)+\alpha^3\,i_3(\psi_0)+\alpha^4\left[i'_1(\psi_0)\bar{\psi}_3(\bar{r},\theta)+i_4(\psi_0)\right]+\mathcal{O}(\alpha^5),
        \\
        r_0\Omega(\psi)&=\alpha\, \omega_1(\psi_0)+\alpha^3\,\omega_3(\psi_0)+\alpha^4\,\omega_4(\psi_0)+\mathcal{O}(\alpha^5).
    \end{split}
\end{equation}
This expansion is reviewed and discussed in detail in  Appendix \ref{App: previous_orders}. In particular, we have shown why one cannot have corrections of order $\alpha$ and $\alpha^2$ to $\psi(\bar{r},\theta)$.
At the order $\alpha^n$ in perturbation theory the stream equation can be formally written as
\begin{equation}
    \label{SE_rbar_reg}
     \bar{\mathcal{L}}\bar{\psi}_n(\bar{r},\theta)=\bar{S}_n(\bar{r},\theta;\bar{\psi}_{k<n},i_{k\leq n+1},\omega_{k \leq n+1}), 
\end{equation}
where the operator $\bar{\mathcal{L}}$, in the $\bar{r}$-region, is defined as
\begin{equation}
\label{Lbarr}
    \bar{\mathcal{L}}= \frac{\sin{\theta}}{r_0^2}\partial_{\theta}\left[\sin{\theta}\left(\frac{r_0^2}{\bar{r}^2\sin^2{\theta}}-\frac{1}{16}\right)\partial_{\theta}\right] + \sin^2{\theta}\partial_{\bar{r}}\left[\frac{\bar{r}^2}{r_0^2}\left(\frac{r_0^2}{\bar{r}^2\sin^2{\theta}}-\frac{1}{16} \right)\partial_{\bar{r}}\right]+\frac{\left( 2-3\sin^2{\theta}\right)}{8r_0^2}.
\end{equation}
It is clear from this equation that the OLS at $\bar{r}_{\rm OLS}=4r_0/\sin\theta$ constitutes a critical surface in the bulk of the $\bar{r}$-region. 
The $\bar{r}$-region is therefore divided by the OLS in two subdomains, that we dub \emph{inner} and \emph{outer} regions. The stream equation is elliptic in both of these subdomains. 
 The inner and outer domains are depicted in two different colors in Fig.~\ref{fig:LS_Illustration}.

One proceeds to construct an expansion in $\alpha$ of the force-free fields $\psi$, $\Omega$ and $I$ for each of the two regions.
According to the {\sl Matched Asymptotic Expansions} (MAE)
scheme,  the two expansions have to match where the $r$-region and the $\bar{r}$-region overlap. This  \emph{overlap region}  is defined by the limits
\begin{equation}
    \label{match_reg}
    1\ll\frac{r}{r_0}\ll \frac{1}{\alpha} \quad \Leftrightarrow \quad \alpha \ll \frac{\bar{r}}{r_0} \ll 1.
\end{equation}
In Ref.~\cite{Armas:2020mio} the expansions \eqref{Exp_r_Jay} and \eqref{Exp_rbar_Jay} have been proved to match and all of their coefficients have been successfully computed up to $\mathcal{O}(\alpha^4)$. A brief review of their computations is given in Appendix \ref{App: previous_orders}, while a list of the results is given in the next subsection. The main purpose of this work is to explore higher order terms both in the $r$-region and in the $\bar{r}$-region.

Notice that what we call $r$-region also includes the ILS and the event horizon $r=r_+$. This is due to the fact that the regularity condition at the ILS does not play any role in our construction since it is equivalent to the Znajek condition at the horizon for the split-monopole configuration.

\subsection{Summary of past relevant results}
\label{Sec: summary}
What follows is a list of all the relevant results already found in the literature \cite{Armas:2020mio,1977MNRAS.179..433B,Tanabe_2008,Pan_2015,Pan:2015iaa,Pan:2015imp,Grignani_2018,Grignani:2019dqc} and that will be recalled in the next sections. For a more detailed discussion about the computations that lead to these results we refer the reader to Appendix \ref{App: previous_orders}.

When solving Eq. \eqref{SE_r_reg} by means of the expansion \eqref{Exp_r_Jay}, one obtains the following results at order $\alpha^2$ and $\alpha^4$, respectively,
\begin{equation}
    \label{psi2psi4}
    \psi_2(r,\theta)=R_2^{(2)}(r)\Theta_2(\theta),\quad \psi_4(r,\theta)=R_2^{(4)}(r) \Theta_2(\theta)+R_4^{(4)}(r)\Theta_4(\theta),
\end{equation}
where we adopted the notation for which $\psi_n=\sum_{k}R_{2k}^{(n)}\Theta_{2k}(\theta)$ and the functions $\Theta_k(\theta)$ are defined in Appendix \ref{App:Green}. The structure of the solutions \eqref{psi2psi4} in terms of the even harmonics $\Theta_{2k}$ is dictated by the form of the corresponding sources $S_2$ and $S_4$ in Eq.~\eqref{SE_r_reg}, explicitly written in Eqs.~\eqref{stream_2nd} and \eqref{stream_r_4}.

In order to compute the energy and angular momentum flux through the horizon (see Sec. \ref{Sec: ELoutflows}), it is useful to recall the expansion for small $r$ of the radial functions. For $r/r_0\ll 1$, one has
\begin{subequations}
    \begin{align}
    \label{R2smallr}
        & R_2^{(2)}(r) = U_0 + \frac{r-r_0}{r_0} U_1 +\left(\frac{r-r_0}{r_0}\right)^2 U_2 + \mathcal{O}\left( (r-r_0)^3\right),
        \\
     \label{R42smallr}
        &R_2^{(4)}(r) = W_0 + \frac{r-r_0}{r_0} W_1+\left( \frac{r-r_0}{r_0}\right)^2 W_2 + \mathcal{O}\left( (r-r_0)^3\right),
        \\
     \label{R44smallr}
        &R_4^{(4)}(r) = V_0 + \frac{r-r_0}{r_0} V_1 +\left( \frac{r-r_0}{r_0}\right)^2 V_2 + \mathcal{O}\left( (r-r_0)^3\right)
    \end{align}
\end{subequations}
with
\begin{subequations}
    \begin{align}
    \label{Us}
    U_0 &= \frac{6\pi^2 - 49}{72}, \qquad U_1 = \frac{6\pi^2 - 61}{12}, \qquad U_2 =\frac{3 \pi^2-29}{4},
    \\
    \label{Ws}
    W_0 &=\frac{39 \zeta (3)}{3920}+\frac{17929399}{2540160}-\frac{3877 \pi ^2}{12096}-\frac{19 \pi ^4}{480},  \qquad W_1 =\frac{6048 W_0-222 \pi ^2+1831}{1008} ,\cr \qquad W_2 &= \frac{4032 W_0-168 \pi ^2+1481}{448},
    \\
    \label{Vs}
    V_0 &=\frac{79 \pi ^4}{640}-\frac{963 \zeta (3)}{3920}-\frac{2012505017}{67737600}+\frac{9791 \pi ^2}{5376},   \qquad V_1 = \frac{8960 V_0+24 \pi ^2-169}{448}, \cr
    \qquad V_2 & =\frac{179200 V_0+756 \pi ^2-6295}{1792}.
    \end{align}
\end{subequations}
We obtained $W_0$ and $V_0$ using two independent but equivalent methods. In the first method, we determined $R_2^{(4)}(r)$ and $R_4^{(4)}(r)$ for all $r$. This is a novel result, that we explicitly present in Appendix \ref{app:explicit}. In the second method, which we outline in Appendix \ref{App:Green}, we determine $W_0$ and $V_0$ using Green functions. Both methods agree on the analytic expressions for $V_0$ and $W_0$. Note that previously these quantities were only known numerically.

As already explained, the matching with the $\bar{r}$-region, instead, exploits the expansion for $r/r_0\gg1$, which reads
\begin{subequations}
    \begin{align}
    \label{R2greatr}
        &R_2^{(2)}(r)=\frac{1}{8}\frac{r_0}{r}-\frac{11}{800}\frac{r_0^2}{r^2}+\frac{1}{40}\frac{r_0^2}{r^2}\log\frac{r}{r_0}+\mathcal{O}\left(\frac{r_0^3}{r^3}\log\frac{r}{r_0}\right),
        \\
    \label{R42greatr}
        &R_2^{(4)}(r)=\frac{1}{224}\frac{r}{r_0}+\frac{227}{100800}+\frac{1}{1680}\log\frac{r}{r_0}+\mathcal{O}\left(\frac{r_0}{r}\log\frac{r}{r_0}\right),
        \\
    \label{R44greatr}
        &R_4^{(4)}(r)=\frac{9}{8960}\frac{r}{r_0}+\frac{363}{896000}+\frac{3}{22400}\log\frac{r}{r_0}+\mathcal{O}\left(\frac{r_0}{r}\log\frac{r}{r_0}\right).
    \end{align}
\end{subequations}

Concerning the poloidal current $I(\psi)$ and the angular velocity of the field line $\Omega(\psi)$, these can be determined after expanding the Znajek condition at the horizon and at infinity (or, analogously, the ILS and the OLS condition) by means of Eq.~\eqref{Exp_r_Jay}, yielding
\begin{equation}
\label{znajek_i1_i4}
	\begin{split}
		i_1(\theta)&=\frac{\sin^2\theta}{4}, \\ i_3(\theta)&=-\omega_3(\theta)\sin^2\theta +\frac{1}{40}\left[\left(7-8U_0\right)\Theta_1(\theta)+\frac{\left(1-4U_0\right)}{2}\Theta_3(\theta)\right], \\ 
		i_4(\theta)&=-\omega_4(\theta) \sin^2\theta
	\end{split}
\end{equation}
and
\begin{equation}
\label{omega1and3}
	\omega_1 =\frac{1}{4},\quad
	\omega_3(\theta)=\frac{1}{16}\left[1+\left(1-4U_0\right)\frac{\sin^2\theta}{2}\right].
\end{equation}
To determine $\omega_4(\theta)$ and $i_4(\theta)$ one needs to solve for $\bar{\psi}_3(\bar{r},\theta)$ in the $\bar{r}$-region expansion \eqref{Exp_rbar_Jay}.
This requires solving the equation for $\bar{\psi}_3(\bar{r},\theta)$ 

\begin{equation}
    \label{eq_psibar3}
    \Bar{\mathcal{L}}\Bar{\psi}_3 (\Bar{r}, \theta) = -\frac{r_0 \sin ^2{\theta}  \cos{\theta} }{2 \Bar{r}^3}+\frac{\partial_\theta\left\{\sin^2{\theta} \left[\sin^2\theta\omega_{4}(\theta)-i_{4}(\theta)\right]\right\}}{4r_0^2\sin{\theta}}.
\end{equation}
Here one needs to take into account the Znajek condition at the horizon \eqref{znajek_i1_i4} to eliminate $i_4(\theta)$ and the Znajek condition at infinity
\begin{equation}
\label{zc_inf_4}
    \sin\theta\partial_\theta \bar{\psi}_3^\infty(\theta)-2\cos\theta \bar{\psi}_3^\infty(\theta)=-8\omega_4(\theta)\sin^2\theta,
\end{equation}
where $\bar{\psi}^\infty_3(\theta)$ is the asymptotic value of $\bar{\psi}_3(\bar{r},\theta)$ for $\bar{r}\rightarrow \infty$. That $\bar{\psi}^\infty_3(\theta)$ is finite is in accordance with the asymptotic behavior of a split-monopole configuration. 
Equation \eqref{eq_psibar3} for $\bar{\psi}_3(\bar{r},\theta)$ is quite non-trivial as it includes the OLS. In \cite{Armas:2020mio} it was solved numerically, taking into account that one needs to demand continuity of $\bar{\psi}_3(\bar{r},\theta)$ at the OLS.

One can always decompose $\bar{\psi}_3^\infty(\theta)$ as
\begin{equation}
    \label{psibar3_num_ansatz}
    \bar{\psi}_{3}^\infty(\theta)=\bar{c}^{(3)}_2\Theta_2(\theta)+\bar{c}^{(3)}_4\Theta_4(\theta)+\bar{c}^{(3)}_6\Theta_6(\theta)+\bar{c}^{(3)}_8\Theta_8(\theta)+\dots\;,
\end{equation}
where $\bar{c}_{2k}^{(3)}$ are coefficients to be determined and the basis functions $\Theta_{2k}(\theta)$ are defined in Eq.~\eqref{eq:ThetaEven}. The presence of only even harmonics in \eqref{psibar3_num_ansatz}  is required by the split-monopole boundary conditions \eqref{BC_split-monopole}: only the $\Theta_{2k}(\theta)$ vanish at the equator.
In \cite{Armas:2020mio} the  coefficients $\bar{c}_{2k}^{(3)}$ were found numerically as part of solving Eq.~\eqref{eq_psibar3} and demanding continuity at the OLS. 
In Appendix \ref{App:Numerics} we have implemented a new numerical procedure to solve Eq.~\eqref{eq_psibar3} that provides greater accuracy and faster convergence. See Fig.~\ref{fig:plot_barpsi4} for the plot of the numerical solution. Using this we found the coefficients $\bar{c}_{2k}^{(3)}$ with increased precision. The explicit values of the first four are
\begin{equation}
\begin{split}
    \label{c_coeff3}
    \bar{c}^{(3)}_2 &=0.02170516(3), \qquad \bar{c}^{(3)}_4=0.002706130(5),  \\ 
    \bar{c}^{(3)}_6 &=-0.000283909(5), \quad \bar{c}^{(3)}_8=0.000072528(9),
\end{split}
\end{equation}
where the round bracket is placed on the first uncertain digit.

Plugging Eq.~\eqref{psibar3_num_ansatz} into Eq.~\eqref{zc_inf_4} one sees that
$\omega_4(\theta)$ has the following structure in terms of odd harmonics, $\Theta_{2k+1}$,
\begin{equation}
    \label{w4_num_ansatz}
    \omega_4(\theta)=b^{(4)}_1\Theta_1(\theta)+b^{(4)}_3\Theta_3(\theta)+b^{(4)}_5\Theta_5(\theta)+b^{(4)}_7\Theta_5(\theta)+\dots\;,
\end{equation}
we can use Eqs.~\eqref{psibar3_num_ansatz}-\eqref{c_coeff3} to find%
\footnote{Notice that, using the Znajek condition at infinity \eqref{zc_inf_4}, the expression for each $b_n^{(4)}$ contains infinite numbers of $\bar{c}_l^{(3)}$. To get enough accuracy for the results \eqref{b_coeff4} we computed the coefficients $\bar{c}_l^{(3)}$ up to $l=22$.}
\begin{equation}
\begin{split}
\label{b_coeff4}
  b^{(4)}_1&=0.00256778(0), \quad b^{(4)}_3=0.0005087(7), \\ 
  b^{(4)}_5&=-0.0000691(5), \quad b^{(4)}_7=0.0000212(9).
\end{split}
\end{equation}
We notice that the fact that one is able to impose the continuity of $\psi$ at the OLS by fixing an ansatz for $\psi^\infty$ is another hint of the equivalence between the OLS and the Znajek condition at infinity.
%

\section{Matched asymptotic expansions II: new results}
\label{Sec: Exp}

In this section we extend the perturbative construction of the BZ split-monopole to higher orders  in $\alpha$ both in the $r$-region and in the $\bar{r}$-region. We first present the expansions and show the matching. 
Building on these results, we show in Sec.~\ref{Sec: ELoutflows}
 how this perturbative construction allows us to derive the expressions for the energy and angular momentum fluxes to higher orders than previously known in the literature.

\subsection{Expansion in the \texorpdfstring{$r$}{}-region}

In the $r$-region the expansion of the field variables we consider is of the form 
\begin{align}\label{EXP_r_reg}
	\psi(r,\theta)&=\psi_0(\theta)+\alpha^2 \, \psi_2(r,\theta)+ \alpha^4 \, \psi_4(r,\theta)+\alpha^5 \, \psi_5(r,\theta)+\mathcal{O}(\alpha^6\log\alpha),
	\cr
	r_0 I(\psi)&=\alpha \, i_1(\psi_0)+\alpha^3 \left[i_1'(\psi_0)\psi_2(r,\theta)+i_3(\psi_0)\right]+ \alpha^4 \, i_4(\psi_0) \\ \nonumber
	&\quad+\alpha^5 \left[I_5(r,\theta)+\log\alpha \,I_{5_L}(r,\theta)\right]+\mathcal{O}(\alpha^6\log\alpha),
\\ 
	r_0 \Omega(\psi)&=\alpha \, \omega_1(\psi_0)+\alpha^3 \, \omega_3(\psi_0)+ \alpha^4 \,\omega_4(\psi_0)+\alpha^5\left[ \Omega_5(r,\theta)+\log\alpha \,\Omega_{5_L}(r,\theta)\right]+\mathcal{O}(\alpha^6\log\alpha). \nonumber
\end{align}
Notice that we introduced terms proportional to $\log\alpha$ in $I(\psi)$ and $\Omega(\psi)$.
As we discuss more extensively in Sec.~\ref{Sec: Matching2}, the inclusion of these terms is in fact unavoidable once the matching with  the $\bar{r}$-region is established: the reason lies in the appearance of $\log r/r_0$ in the expansion of $\psi_2$ and $\psi_4$ for $r/r_0\gg1$; see Eqs.~\eqref{R2greatr}-\eqref{R42greatr}-\eqref{R44greatr}.
Solving the integrability conditions \eqref{intcond1} and \eqref{intcond2} yields  
\begin{equation}
    \begin{split}
        I_5(r,\theta)&=i_5(\psi_0)+i'_1(\psi_0)\psi_4(r,\theta)+i'_3(\psi_0)\psi_2(r,\theta)+\frac{i_1''(\psi_0)}{2}\psi_2^2(r,\theta), \quad I_{5_L}(r,\theta)=i_{5_L}(\theta),
        \\
        \Omega_5(r,\theta)&=\omega_5(\psi_0)+\omega'_3(\psi_0)\psi_2(r,\theta), \hspace{5.97cm} \Omega_{5_L}(r,\theta)=\omega_{5_L}(\theta).
    \end{split}
\end{equation}
From the Znajek condition at the horizon, Eq.~\eqref{ZC}, expanded at order $\alpha^5$, one can establish relations among $i_5$ and $\omega_5$ as well as $i_{5_L}$ and $\omega_{5_L}$. More specifically one has
\begin{equation}
    \label{zc_5}
    i_5(\theta )= -\omega_5(\theta ) \sin ^2(\theta )+\left[k_1\Theta_1(\theta)+k_3\Theta_3(\theta)+k_5\Theta_5(\theta)\right], \quad i_{5_L}(\theta) = -\omega_{5_L}(\theta ) \sin ^2(\theta ),
\end{equation}
where the constants $k_1, k_3$ and $k_5$ are expressed in terms of the horizon values of $\psi_2$ and $\psi_4$ as 
\begin{equation}
    \begin{split}
        &k_1=\frac{6+(33+8U_0)U_0-28W_0}{140},
        \\
        &k_3=\frac{-1+(17-28U_0)U_0-4(3W_0+10V_0)}{240},
        \\
        &k_5=\frac{1+16(1-11U_0)U_0-224V_0}{2688}.
    \end{split}
\end{equation}
Note that the same results follow by imposing that the solution of the reduced stream equation \eqref{stream_LS2} at the ILS is regular on the axis. 

Turning the attention to the flux function $\psi_5$, its stream equation reads
\begin{equation}
    \label{eqpsi5}
    \mathcal{L}\psi_5(r,\theta)=\frac{r^2+r_0 r+r_0^2}{2r_0^2 r^2\sin^2\theta}\partial_\theta\left[\omega_4(\theta)\sin^4\theta\right].
\end{equation}
It is possible to derive some important semi-analytical results for $\psi_5$. First, notice that in the source term in Eq.~\eqref{eqpsi5} there is the function $\omega_4$ which is given by the expression in Eq.~\eqref{w4_num_ansatz}. Therefore, in contrast to what happened for $\psi_2$ and $\psi_4$, the decomposition in harmonics  for  $\psi_5(r,\theta)$ has an infinite number of terms. In other words, the general solution is expressed as 
\begin{equation}
    \label{exp_psi5}
    \psi_5(r,\theta)=R^{(5)}_{2}(r)\Theta_2(\theta)+R^{(5)}_{4}(r)\Theta_4(\theta)+R^{(5)}_{6}(r)\Theta_6(\theta)+\dots\;
\end{equation}
in terms only of even harmonics, $\Theta_{2k}(\theta)$. This is again dictated by the split-monopole boundary conditions \eqref{BC_split-monopole}, see comment below Eq.~\eqref{psibar3_num_ansatz}.
Since the source term in Eq.~\eqref{eqpsi5} does not receive any contribution from $\psi_2$ and $\psi_4$, all the radial solutions in Eq.~\eqref{exp_psi5} which converge at the horizon are polynomial in $r$.
In particular, for the first two functions one gets
\begin{subequations}
    \begin{align}
        R^{(5)}_{2}(r)&=C_1\frac{r^2 (4 r-3 r_0)}{4r_0^3} -\frac{(99 b^{(4)}_1+44 b^{(4)}_3+8 b^{(4)}_5) \left(344 r^3-249 r^2 r_0+6 r r_0^2+7 r_0^3\right)}{2079 r_0^3},
        \\
        R^{(5)}_{4}(r)&=C_2\frac{r^2\left(56 r^3-105 r^2 r_0+60 r r_0^2-10 r_0^3\right)}{56r_0^5}+ \cr
        &\quad+(715 b^{(4)}_1+2340 b^{(4)}_3+1800 b^{(4)}_5+448 b^{(4)}_7)\times\cr
        &\quad\times\left(\frac{27}{286000}+\frac{11981 r^5}{35750 r_0^5}-\frac{35943 r^4}{57200 r_0^4}+\frac{35943 r^3}{100100 r_0^3}-\frac{11961 r^2}{200200 r_0^2}+\frac{9 r}{100100 r_0} \right),
    \end{align}
\end{subequations}
with $C_1$ and $C_2$ integration constants. Similar but quite longer results follow for $R^{(5)}_{6}(r)$ and higher.
In general one has $R^{(5)}_{n}(r)\sim (r/r_0)^{n+1}$ for $r/r_0\gg 1$, a behaviour that would ruin the perturbation scheme in the matching with the $\bar{r}$-region.\footnote{For example, in the overlap region $R^{(5)}_{6}(r)$ generates non-perturbative terms like $\alpha^5(r/r_0)^7\sim \alpha^{-2}(\bar{r}/r_0)^7$. Other powers, for example in $R^{(5)}_{2}(r)$, cannot match with the $\bar{r}$-expansion \eqref{Exp_rbar_Jay} since they contribute as $\alpha^5(r/r_0)^3\sim \alpha^2 (\bar{r}/r_0)^3$.}
However, in a very non trivial manner, for each radial function one can choose the integration constant in such a way to get rid of all the troublesome powers in $r$. For the first three radial functions this procedure yields 
\begin{subequations}
\begin{align}
        R^{(5)}_{2}(r)&=-\left(99 b^{(4)}_1+44 b^{(4)}_3+8 b^{(4)}_5\right) \left(\frac{1}{231} \frac{r^2}{r_0^2}+\frac{2}{693} \frac{r}{r_0}+\frac{1}{297}\right),
        \\
        \label{R5sol}
        R^{(5)}_{4}(r)&=-\left(715 b^{(4)}_1+2340 b^{(4)}_3+1800 b^{(4)}_5+448 b^{(4)}_7\right) \left(\frac{1}{10010}\frac{r^2}{r_0^2}+\frac{9}{100100}\frac{r}{r_0}+\frac{27}{286000}\right),
        \\
        R^{(5)}_{6}(r)&=-\left(8075 b^{(4)}_3+266 (85 b^{(4)}_5+72 b^{(4)}_7)+5280 b^{(4)}_9\right) \left(\frac{1}{106590}\frac{r^2}{r_0^2}+\frac{2}{223839}\frac{r}{r_0}+\frac{43}{4700619}\right).
\end{align}
\end{subequations}
This allows one to express the value of $\psi_5$ at the horizon solely in terms of the coefficients $b^{(4)}_n$ of $\omega_4$, as follows
\begin{equation}
\begin{split}\label{psi5}
    \psi_5(r_0,\theta)=&-\frac{2\left(99b^{(4)}_1+44b^{(4)}_3+8b^{(4)}_5\right)}{189}\Theta_2(\theta)+\\
    &-\frac{569\left(715 b^{(4)}_1+2340 b^{(4)}_3+1800b^{(4)}_5+448 b^{(4)}_7\right)}{2002000}\Theta_4(\theta)+\\
    &-\frac{1291\left[8075b^{(4)}_3+266\left(85b^{(4)}_5+72b^{(4)}_7\right)+5280b^{(4)}_9\right]}{47006190}\Theta_6(\theta)+\dots\;\; .
\end{split}
\end{equation}
We recall that the values of the first coefficients $b^{(4)}_n$ are listed in Eq.~\eqref{b_coeff4}.
Conversely, in the regime $r/r_0\gg 1$, one observes that $\psi_5(r,\theta)=r^2/r_0^2 \mathcal{F}(\theta)+\mathcal{O}(r/r_0)$, with the function $\mathcal{F}$ constructed by means of all the leading contributions of $R^{(5)}_{2n}(r)\Theta_{2n}(\theta)$, and obeying the following ODE
\begin{equation}
    \label{eqF}
    \mathcal{F}''(\theta)-\cot\theta \mathcal{F}'(\theta)+2\mathcal{F}(\theta)=\frac{\partial_\theta\left(\sin^4\theta \omega_4(\theta)\right)}{\sin\theta}.
\end{equation}
%


\subsection{Expansion in the \texorpdfstring{$\bar{r}$}{}-region}

In the $\bar{r}$-region, which encompasses the OLS, the expansions we consider are of the kind 
	\begin{align}
\label{EXP_rbar_reg}
		\psi(\bar{r},\theta)&=\psi_0(\theta)+\alpha^3 \, \bar{\psi}_3(\bar{r}, \theta)+\alpha^4\left[\bar{\psi}_4(\bar{r},\theta)+\log\alpha \,\bar{\psi}_{4_L}(\bar{r},\theta)\right]+\mathcal{O}(\alpha^5\log\alpha), \nonumber
		\\ 
		r_0I(\psi)&=\alpha \, i_1(\psi_0)+\alpha^3 \, i_3(\psi_0)+\alpha^4 \left[i_1'(\psi_0)\bar{\psi}_3(\bar{r},\theta)+i_4(\psi_0)\right] \\
		&\quad+\alpha^5\left[\bar{I}_5(\bar{r},\theta)+\log\alpha\; \bar{I}_{5_L}(\bar{r},\theta)\right]+\mathcal{O}(\alpha^6\log\alpha),
	\nonumber	\\
		r_0\Omega(\psi)&=\alpha \, \omega_1(\psi_0)+\alpha^3 \, \omega_3(\psi_0)+\alpha^4 \, \omega_4(\psi_0)+\alpha^5\left[\bar{\Omega}_5(\bar{r},\theta)+\log\alpha\; \bar{\Omega}_{5_L}(\bar{r},\theta)\right]+\mathcal{O}(\alpha^6\log\alpha).\nonumber
\end{align}
Again, there are coefficients proportional to $\log\alpha$, this time not only on $I(\psi)$ and $\Omega(\psi)$ but also at the level of the flux function $\psi(\bar{r},\theta)$. 
This form of the expansion follows from our analysis in Section \ref{Sec: Matching2} of the matching between the $r$-region and $\bar{r}$-region.
The integrability conditions \eqref{intcond1} and \eqref{intcond2} fix the dependence on $\bar{r},\theta$ in terms of $\psi(\bar{r},\theta)$ as follows
\begin{equation}
    \begin{split}
        \bar{I}_5(\bar{r},\theta)&=i_1'(\psi_0)\bar{\psi}_{4}(\bar{r},\theta)+i_5(\psi_0),\quad \bar{I}_{5_L}(\bar{r},\theta)=i_1'(\psi_0)\bar{\psi}_{4_L}(\bar{r},\theta)+i_{5_L}(\psi_0),
        \\
        \bar{\Omega}_5(\bar{r},\theta)&=\omega_5(\psi_0),\quad \hspace{2.7cm} \bar{\Omega}_{5_L}(\bar{r},\theta)=\omega_{5_L}(\psi_0).
        \end{split}
\end{equation}
The stream equation for the flux function $\bar{\psi}_{4_L}$ reads
\begin{align}
     \label{eq_barpsi4L}
     \bar{\mathcal{L}} \bar{\psi}_{4_L}(\bar{r},\theta)
     & = \frac{\partial_\theta\left\{\sin^2{\theta} \left[\sin^2\theta\omega_{5_L}(\theta)-i_{5_L}(\theta)\right]\right\}}{4r_0^2\sin{\theta}}.
\end{align}
Here we can trade $i_{5_L}$ for $\omega_{5_L}$ via the Znajek condition at the horizon \eqref{zc_5}, while in turn  $\omega_{5_L}$ is related to the asymptotic value of $\bar{\psi}_{4L}$ by means of the Znajek condition at infinity \eqref{inf_ZC}, that yields
\begin{equation}
\label{zc_inf_5L}
    \sin\theta\partial_\theta \bar{\psi}_{4_L}^\infty(\theta)-2\cos\theta \bar{\psi}_{4_L}^\infty(\theta)=-8\sin^2\theta\omega_{5_L}(\theta).
\end{equation}
This results in the following equation
\begin{align}     
\label{neweq_barpsi4L}
     \bar{\mathcal{L}} \bar{\psi}_{4_L}(\bar{r},\theta)
     & =- \frac{\partial_\theta\left\{\sin^2 \theta \left[\sin\theta\partial_\theta \bar{\psi}_{4_L}^\infty(\theta)-2\cos\theta \bar{\psi}_{4_L}^\infty(\theta) \right]\right\}}{16r_0^2\sin{\theta}}.
\end{align}
Adopting an ansatz for $\bar{\psi}_{4_L}(\bar{r},\theta)$ for which the angular dependence is a linear combination of only $\Theta_2(\theta)$ and $\Theta_4(\theta)$ one finds the simple particular solution
\begin{equation}
    \label{presol_barpsi4L}
    \bar{\psi}_{4_L}(\bar{r},\theta) =42 K \frac{r_0^2}{\bar{r}^2}\Theta_2(\theta )+ K\left[\Theta_2(\theta )+\frac{9}{40} \Theta_4(\theta )\right],
\end{equation}
where $K$ is an arbitrary constant.
We recognise immediately that Eq.~\eqref{presol_barpsi4L} is well behaved at infinity and smooth across the OLS.
We shall see in Sec.~\ref{Sec: Matching2} that $K=-1/1680$ by comparing the large $r$ behavior of the $r$-region with the small $\bar{r}$ behavior of the $\bar{r}$-region. Thus, the solution for $\bar{\psi}_{4_L}(\bar{r},\theta)$ is
\begin{equation}
    \label{sol_barpsi4L}
    \bar{\psi}_{4_L}(\bar{r},\theta) =- \frac{1}{40}\frac{r_0^2}{\bar{r}^2}\Theta_2(\theta )-\left[\frac{1}{1680}\Theta_2(\theta )+\frac{3}{22400} \Theta_4(\theta )\right].
\end{equation}
In general, we require that solutions of Eq.~\eqref{eq_barpsi4L} should obey the boundary conditions inferred from Eq.~\eqref{greatr} at small $\bar{r}$ as well as 
$\bar{\psi}_{4_L} \rightarrow \bar{\psi}_{4_L}^\infty(\theta)$ for large $\bar{r}$. From the fact that \eqref{eq_barpsi4L} is elliptic in the inner and outer subdomains on either side of the OLS, and that $\bar{\psi}_{4_L}$ obeys a Robin type boundary condition at the OLS, one can show that the solution \eqref{sol_barpsi4L} is locally unique.

Coming to $\omega_{5_L}(\theta)$, from Eq.~\eqref{sol_barpsi4L} we can easily read $\bar{\psi}_{4_L}^\infty(\theta)$ and replace it in Eq.~\eqref{zc_inf_5L}, so as to compute
\begin{equation}
\label{sol_omgea5L}
\omega_{5_L}(\theta) = b^{(5_L)}_1 \Theta_1(\theta)+b^{(5_L)}_3\Theta_3(\theta),
\end{equation}
where
\begin{equation}
\label{sol_omgea5L_2}
    b^{(5_L)}_1 = - \frac{13}{192000} \;, \quad b^{(5_L)}_3 =-\frac{3}{128000}. 
\end{equation}
The stream equation for the flux function $\bar{\psi}_4$ is instead given by
\begin{align}
     \label{eq_barpsi4}
     \bar{\mathcal{L}} \bar{\psi}_{4}(\bar{r},\theta)
     & =\frac{\cos\theta \sin^4\theta}{32 \bar{r}^2}+\frac{\partial_\theta\left\{\sin^2{\theta} \left[\sin^2\theta\omega_{5}(\theta)-i_{5}(\theta)\right]\right\}}{4r_0^2\sin{\theta}}+\bar{\Sigma}_4(\bar{r},\theta;\bar{\psi}_3).
\end{align}
Notice the presence of the source term $\bar{\Sigma}_4(\bar{r},\theta;\bar{\psi}_3)$, containing the function $\bar{\psi}_3(\bar{r},\theta)$ discussed in Sec.~\ref{Sec: summary}.
The explicit expression for $\bar{\Sigma}_4(\bar{r},\theta;\bar{\psi}_3)$ is given by 
\begin{align}
    \bar{\Sigma}_4(\bar{r},\theta;\bar{\psi}_3)&=2\partial_{\bar{r}}\left[\frac{\bar{r}}{r_0}\left(\frac{r_0^2}{\bar{r}^2}-\frac{\sin^2\theta}{32}\right)\partial_{\bar{r}}\psi_{3}(\bar{r},\theta)\right]+\left(\frac{r_0}{\bar{r}^2}-\frac{\sin^2\theta}{16r_0}\right)\partial_{\bar{r}}\psi_{3}(\bar{r},\theta)\nonumber\\&\quad+\frac{r_0\sin\theta}{\bar{r}^3}\partial_\theta\left(\frac{\partial_\theta\psi_{3}(\bar{r},\theta)}{\sin\theta}\right).
\end{align}
Again we can resort to the Znajek condition at the horizon \eqref{zc_5} to express $i_5$ in terms of $\omega_5$ and relate the latter to the asymptotic value of the numerical solutions for $\bar{\psi}_3$, $\bar{\psi}_4$ by means of the Znajek condition at infinity, that is
\begin{equation}
\label{zc_inf_5}
    \sin\theta\partial_\theta \bar{\psi}_{4}^\infty(\theta)-2\cos\theta \bar{\psi}_{4}^\infty(\theta)=-8\left[\sin^2\theta\omega_{5}(\theta)-\frac{k_1 \Theta_1(\theta)+k_3\Theta_3(\theta)+k_5\Theta_5(\theta)}{2}\right].
\end{equation}

\begin{figure}[ht!]
    \centering
    \includegraphics[width=1\textwidth]{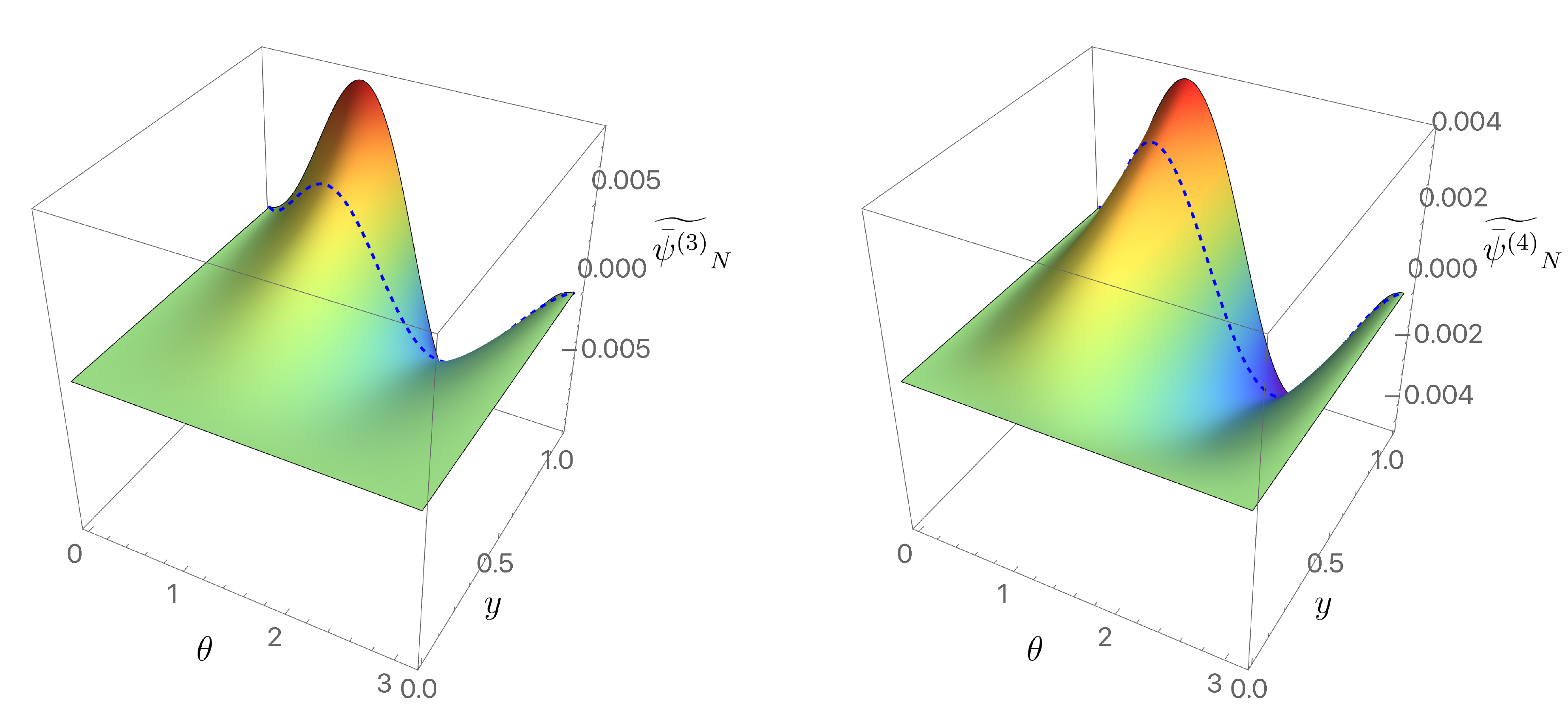}
    \caption{The numerical solution for $\widetilde{\bar{\psi}^{(3)}}_N$ and $\widetilde{\bar{\psi}^{(4)}}_N$ shown as a function of $\theta$ and $y=\bar{r}/(r_0+\bar{r})$. Note the smoothness at the OLS (represented as a dashed blue line). Both $\widetilde{\bar{\psi}^{(3)}}_N$ and $\widetilde{\bar{\psi}^{(4)}}_N$ are defined in Appendix \ref{App:Numerics} and in these plots they were generated with $N=10$ and $N_y=250$.}
    \label{fig:plot_barpsi4}
\end{figure}

To find $\bar{\psi}_4(\bar{r},\theta)$ one can follow the same procedure as for $\bar{\psi}_3(\bar{r},\theta)$, solving Eq.~\eqref{eq_barpsi4} numerically and in the process fixing the asymptotic function $\bar{\psi}^\infty_{4}(\theta)$ by demanding continuity of 
$\bar{\psi}_4(\bar{r},\theta)$ across the OLS. As part of this, one as usual decomposes  $\bar{\psi}^\infty_{4}(\theta)$ in even harmonics
\begin{equation}
    \label{ansatz_barpsi4_inf}
        \bar{\psi}^\infty_{4}(\theta)=\bar{c}^{(4)}_2\Theta_2(\theta)+\bar{c}^{(4)}_4\Theta_4(\theta)+\bar{c}^{(4)}_6\Theta_6(\theta)+\bar{c}^{(4)}_8\Theta_8(\theta)+\dots\,.
\end{equation}
The coefficients $\bar{c}^{(4)}_n$ are then determined numerically by adopting the procedure described in Appendix \ref{App:Numerics}, which enforces the continuity of $\psi$ across the OLS.
The numerical result for $\Bar{\psi}_4$ is presented in Fig.~\ref{fig:plot_barpsi4} and below we list the values we found for the four coefficients in Eq.~\eqref{ansatz_barpsi4_inf}
\begin{equation}
\begin{split}
\label{c_coeff4}
        \bar{c}^{(4)}_2=0.000945488(3)\;, \quad \bar{c}^{(4)}_4=0.00010879(4)\;, \quad \\ \bar{c}^{(4)}_6=-0.00001918(4), \quad \bar{c}^{(4)}_8=0.0000039633(2)\;.
\end{split}
\end{equation}

Considering Eqs.~\eqref{zc_inf_5} and \eqref{ansatz_barpsi4_inf} one can also argue that $\omega_5$ is expressed as an infinite superposition of $\Theta_{2n+1}(\theta)$, via 
\begin{equation}
\label{w5_num_ansatz}
        \omega_5(\theta)=\frac{1}{2}(k_1-4 k_3+8 k_5)+b^{(5)}_1\Theta_1(\theta)+b^{(5)}_3\Theta_3(\theta)+b^{(5)}_5\Theta_5(\theta)+b^{(5)}_7\Theta_7(\theta)+\dots\,,
\end{equation}
with the coefficients $b^{(5)}_n$ related to $\bar{c}^{(4)}_n$ by means of Eq.~\eqref{zc_inf_5}. It turns out that
\begin{equation}
\frac{1}{2}\left(k_1-4 k_3+8 k_5\right)=\frac{1}{32}\,.
\end{equation}

The coefficients $b^{(5)}_n$ play a fundamental role in the expressions for the energy and angular momentum fluxes, which are discussed in Sec.~\ref{Sec: ELoutflows}. For this purpose we collect their numerical values up to $n=7$
\begin{equation}
\begin{split}
\label{b_coeff5}
        b^{(5)}_1&=0.002790808(3)\;, \quad b^{(5)}_3=0.000985432(8)\;, \quad\\ 
        b^{(5)}_5&= -0.00000447(0)
        \;, \quad  
        b^{(5)}_7=0.00000111(6)
        \;.
\end{split}
\end{equation}

\subsection{Matching the \texorpdfstring{$r$}{}-region and the \texorpdfstring{$\bar{r}$}{}-region}
\label{Sec: Matching2}

In this section we show the consistency between the expansions \eqref{EXP_r_reg} in the $r$-region and the expansions \eqref{EXP_rbar_reg} in the $\bar{r}$-region. As explained in Sec. \ref{Sec: MAE}, the two expansions have to match in the overlap region defined in Eq.~\eqref{match_reg}.\footnote{We remind that the two radial coordinates are related by $\bar{r}=\alpha r$.} In what follows only the matching regarding the flux function $\psi$ will be shown; the matching concerning $I(\psi)$ and $\Omega(\psi)$, indeed, follows directly from the integrability conditions, which have been already used. To facilitate the reader in visualizing the matching, terms which translate from one expansion to the other will be highlighted with the same colour.

In the $r$-region we use Eqs.~\eqref{R2greatr}-\eqref{R42greatr}-\eqref{R44greatr} to write the large $r$ expansion of the coefficients
\begin{equation}
    \label{greatr}
    \begin{split}
        &\psi_2(r,\theta)=\left[{\color{red}\frac{1}{8}}\frac{r_0}{r}{\color{red}-\frac{11}{800}}\frac{r_0^2}{r^2}+{\color{red}\frac{1}{40}}\frac{r_0^2}{r^2}\log\frac{r}{r_0}\right]\Theta_2(\theta)+\mathcal{O}\left(\frac{r_0^3}{r^3}\log\frac{r}{r_0}\right),
        \\
        \\
        &\psi_4(r,\theta)=\left[{\color{blue}\frac{1}{224}}\frac{r}{r_0}+{\color{blue}\frac{227}{100800}}+{\color{blue}\frac{1}{1680}}\log\frac{r}{r_0}\right]\Theta_2(\theta)+
        \\
        &\qquad\qquad+\left[{\color{violet}\frac{9}{8960}}\frac{r}{r_0}+{\color{violet}\frac{363}{896000}}+{\color{violet}\frac{3}{22400}}\log\frac{r}{r_0}\right]\Theta_4(\theta)+\mathcal{O}\left(\frac{r_0^3}{r^3}\log\frac{r}{r_0}\right),
        \\
        &\psi_5(r,\theta)=\frac{r^2}{r_0^2}\mathcal{F}(\theta)+\mathcal{O}\left(\frac{r}{r_0}\right),
    \end{split}
\end{equation}
where we recall that the function $\mathcal{F}(\theta)$ is a solution of Eq.~\eqref{eqF}, which contains only $\Theta_{2n}(\theta)$ functions.

In the $\bar{r}$-region, 
using the Frobenius method and the matching with \eqref{greatr}, we get for $\bar{r}/r_0\ll 1$
\begin{equation}
    \label{smallrbar}
	\begin{split}
	    &\bar{\psi}_3(\bar{r},\theta)={\color{red}\frac{1}{8}}\frac{r_0}{\bar{r}}\Theta_2(\theta)+\frac{\bar{r}}{r_0}\left[{\color{blue}\frac{1}{224}}\Theta_2(\theta)+{\color{violet}\frac{9}{8960}}\Theta_4(\theta)\right]+\frac{\bar{r}^2}{r_0^2}\mathcal{F}(\theta)+\mathcal{O}\left(\frac{\bar{r}^3}{r_0^3}\right),
	    \\
	    \\
		&\bar{\psi}_4(\bar{r},\theta) = {\color{red}-\frac{11}{800}}\frac{r_0^2}{\bar{r}^2}\Theta_2(\theta )+{\color{red}\frac{1}{ 40 }}\frac{r_0^2}{\bar{r}^2}\log \frac{\bar{r}}{r_0}\Theta_2(\theta )+ {\color{blue} \frac{227 }{100800} }\Theta_2(\theta )+{\color{violet}\frac{363 }{896000}}\Theta_4(\theta )+
		\\
		&\hspace{55mm}+\left[{\color{blue} \frac{1}{1680}}\Theta_2(\theta )+{\color{violet}\frac{3 }{22400}}\Theta_4(\theta )\right]\log \frac{\bar{r}}{r_0} + \mathcal{O}\left(\frac{\bar{r}}{r_0}\right).
	\end{split}
\end{equation}
For $\bar{\psi}_{4_L}$, using Eq.~\eqref{presol_barpsi4L} and the matching with \eqref{greatr}, we get the constant $K=-1/1680$ and hence $\bar{\psi}_{4_L}$ is given 
by%
\begin{equation}
    \bar{\psi}_{4_L}(\bar{r},\theta) =- {\color{red}\frac{1}{40}}\frac{r_0^2}{\bar{r}^2}\Theta_2(\theta )-\left[{\color{blue}\frac{1}{1680}}\Theta_2(\theta )+{\color{violet}\frac{3}{22400}} \Theta_4(\theta )\right],
\end{equation}
which is Eq.~\eqref{sol_barpsi4L} where we highlighted the matching coefficients with colors.

This reveals that the structure of the matching is quite  intricate: whenever a term such as $\alpha^{m+n} \psi_{m+n}\sim\alpha^{m+n}(r/r_0)^n\log(r/r_0)$ enters the expansion of $\psi$ in the $r$-region, this can only match with a term akin to $\alpha^{n}(\bar{\psi}_n+\log \alpha\; \bar{\psi}_{n_L})$ in the $\bar{r}$-region. Furthermore, the leading term of the expansion for $\bar{\psi}_n$ and $\bar{\psi}_{n_L}$ when $\bar{r}/r_0\ll1$ is always related to the coefficients in the expansion of $\psi_2$. The subleading terms in the $\bar{r}$-region, instead, in general mix coefficients from both $\psi_2$ and $\psi_4$.
This allows one to infer the structure of the expansion in the $\bar{r}$-region and motivates \emph{a posteriori} our ansatz for the two expansions \eqref{EXP_r_reg} and \eqref{EXP_rbar_reg}. 

It is finally important to notice that 
by means of the expansion \eqref{EXP_rbar_reg} 
there is no need to introduce terms proportional to $\log\alpha$ up to the fifth order in the $r$-region; as a consequence, our higher order corrections are consistent with the lower-order corrections derived in \cite{Armas:2020mio}.

\subsection{Consistency check with \texorpdfstring{$\psi_6$}{} and \texorpdfstring{$\psi_{6_L}$}{}}
\label{sec: psi6&6L}

In the rest of the paper, we push the expansion of the magnetic flux function $\psi$ to the sixth order. The first equation in Eq.~\eqref{EXP_r_reg} reads now as 
\begin{equation}
\label{psiEXP_66L}
\psi(r,\theta)=\psi_0(\theta)+\alpha^2 \, \psi_2(r,\theta)+ \alpha^4 \, \psi_4(r,\theta)+\alpha^5 \, \psi_5(r,\theta)+\alpha^6 \, \left[\psi_6(r,\theta)+ \log\alpha ~\psi_{6_L}(r,\theta)\right] + \mathcal{O}(\alpha^7 \log\alpha).
\end{equation}
There are two main reasons why we extend the perturbation of the magnetic flux function to this order. The first is to make a very non-trivial consistency check of the matching between the $r$-region and the $\bar{r}$-region; see Eqs.~\eqref{greatr}-\eqref{smallrbar}. The second reason is to compute the eighth-order term as well as the first non-vanishing logarithmic contribution in the energy flux at the horizon. These results are reported in the next section \ref{Sec: ELoutflows}. 
Notice that for both purposes the knowledge of the sixth order for $\psi$ is sufficient. Pushing the expansions of $I$ and $\Omega$ at higher orders is not necessary.

The function $\psi_{6_L}(r,\theta)$ obeys the stream equation
\begin{equation}\label{SEpsi6L}
    \mathcal{L}\psi_{6_L}(r,\theta) = -\frac{\sin\theta  \left(r^2+r r_0+r_0^2\right) }{2 r^2 r_0^2}\left[\sin\theta~\omega_{5L}'(\theta )+4 \omega_{5L}(\theta ) \cos \theta \right].
\end{equation}
Upon using $\omega_{5_L}$ as given in Eq.~\eqref{sol_omgea5L}, the general solution can be written by separation of variables as follows
\begin{equation}
    \psi_{6_L}(r,\theta) = R_2^{(6_L)}(r)\Theta_2(\theta)+R_4^{(6_L)}(r)\Theta_4(\theta)+R_6^{(6_L)}(r)\Theta_6(\theta).
\end{equation}
The stream equation \eqref{SEpsi6L} can at this point be solved analytically for the three radial functions. The solutions read 
\begin{subequations}
\begin{align}
    R_2^{(6L)}(r) &= -\frac{1}{21} \left(b^{(5_L)}_1+\frac{4}{9} b^{(5_L)}_3\right) \left(344\frac{r^3}{r_0^3}-249\frac{r^2}{r_0^2}+6\frac{ r}{r_0}+7\right)-\frac{19}{13440}\frac{r^3}{r_0^3}+\frac{19}{17920}\frac{r^2}{r_0^2}, 
    \\
    R_4^{(6L)}(r) &= -\frac{ 1}{2800}\left( b^{(5_L)}_1+\frac{36}{11} b^{(5_L)}_3\right)\left(200 \frac{r^2}{r_0^2}+180\frac{ r}{r_0}+189\right),
    \\
    R_6^{(6L)}(r) &= -\frac{5   }{29106}b^{(5_L)}_3\left(441\frac{ r^2}{r_0^2}+420\frac{ r}{r_0}+430\right).
\end{align}
\end{subequations}
We recall that $ b^{(5_L)}_1$ and $ b^{(5_L)}_3$ have been found analytically in Eq.~\eqref{sol_omgea5L_2}, and in the solutions above the integration constants are already fixed. More specifically, for each radial function it is possible to choose one of the integration constants demanding regularity at the horizon. The remaining integration constant is fixed so as to not introduce lower order terms in the expansion \eqref{EXP_rbar_reg} for the $\bar{r}$-region.
These non-trivial cancellations can be compared to what occurs in the case of $\psi_5$, that leads to Eq.~\eqref{R5sol}.
The corresponding limits for $r/r_0 \ll1$ are
\begin{subequations}
\begin{align}
    \label{R6L2_r0}
    R_2^{(6L)}(r) &= -\frac{108}{21}\left( b^{(5_L)}_1+\frac{4}{9} b^{(5_L)}_3\right) - \frac{19}{53760} + \mathcal{O}(r-r_0), \\
    R_4^{(6L)}(r) &= -\frac{569}{2800}\left( b^{(5_L)}_1+\frac{36}{11} b^{(5_L)}_3\right) + \mathcal{O}(r-r_0), \\
    \label{R6L_r0}
    R_6^{(6L)}(r) &= -\frac{6455 }{29106}b^{(5_L)}_3 + \mathcal{O}(r-r_0),
\end{align}
\end{subequations}
while for $r/r_0 \gg 1$ we have
\begin{subequations}
\begin{align}
    R_2^{(6_L)}(r) &= -\textcolor{red}{\frac{3 }{22400}}\frac{r^3}{r_0^3}+\textcolor{blue}{\frac{3 }{22400}}\frac{r^2}{r_0^2}+\mathcal{O}\left(\frac{r}{r_0}\right),\\
    R_4^{(6_L)}(r) &={\color{violet}\frac{61 }{5913600 }}\frac{r^2}{r_0^2}+\mathcal{O}\left(\frac{r}{r_0}\right), \\
    R_6^{(6_L)}(r) &={\color{OliveGreen}\frac{1}{563200 }}\frac{r^2}{r_0^2}+\mathcal{O}\left(\frac{r}{r_0}\right).
\end{align}
\end{subequations}

Turning our attention to the function $\psi_6(r,\theta)$, the stream equation schematically reads as
\begin{equation}\label{SEpsi6}
    \mathcal{L}\psi_{6}(r,\theta) = S_6(r,\theta),
\end{equation}
where the explicit expression of the source $S_6$ is written in Appendix \ref{App:psi6}. Since, among other functions, the stream equation is sourced by $\omega_5$, the general solution for $\psi_6(r,\theta)$ is expressed as the  following infinite superposition of angular harmonics and radial functions
\begin{equation}
    \psi_{6}(r,\theta) = R_2^{(6)}(r)\Theta_2(\theta)+R_4^{(6)}(r)\Theta_4(\theta)+R_6^{(6)}(r)\Theta_6(\theta) + \dots .
\end{equation}
As we will see in the next section, the only radial function whose $r/r_0\ll1$ limit is useful in the computation of the energy extraction is the first one, and this reads
\begin{subequations}
\begin{align}
    \label{R6L2}
    R_2^{(6)}(r) &= -\frac{3555}{28} \zeta (3)^2+\frac{254238098500 \pi }{10169178137}+\frac{1027 \pi ^6}{9408}-\frac{22}{21}b_1^{(5)}-\frac{88}{189}b_3^{(5)}-\frac{16}{189}b_5^{(5)}+\mathcal{O}(r-r_0).
\end{align}
\end{subequations}
For $r/r_0 \gg1$ the behaviour of the first three radial functions is given by
\begin{subequations}
\begin{align}
     \label{R62_r0}
    R_2^{(6)}(r) &= \frac{r^3 }{r_0^3}\bigg[C_0+\textcolor{red}{\frac{3 }{22400}}\log \frac{r_0}{r}\bigg]-\frac{r^2 }{r_0^2}\bigg[\frac{3}{7} b_1^{(5)}+\frac{4}{21} b_3^{(5)}+\frac{8}{231} b_5^{(5)}+\frac{3 }{4}C_0+\frac{V_0}{84} \cr 
    &\quad+\frac{3 W_0}{56}-\frac{\pi ^4}{24192}-\frac{169 \pi ^2}{36288}+\frac{39533771}{870912000}+\textcolor{blue}{\frac{3 }{22400}}\log \frac{r_0}{r}\bigg]+\mathcal{O}\left(\frac{r}{r_0}\log \frac{r}{r_0}\right)  ,\\
    R_4^{(6)}(r) &= \frac{39}{501760}\frac{r^3 }{r_0^3}-\frac{r^2 }{r_0^2} \bigg[\frac{1}{14} b_1^{(5)}+\frac{18}{77} b_3^{(5)}+\frac{32}{715} b_7^{(5)}+\frac{180 }{1001}b_5^{(5)}+\textcolor{violet}{\frac{61 }{5913600}}\log \frac{r_0}{r}\cr 
    &\quad+\frac{23 V_0}{616}+\frac{W_0}{112}-\frac{367 \pi ^2}{88704}+\frac{\pi ^4}{5376}+\frac{29794297}{1354752000}\bigg]+\mathcal{O}\left(\frac{r}{r_0}\log \frac{r}{r_0}\right) ,
    \\
    \label{R66_r0}
    R_6^{(6)}(r) &= \frac{1}{73728 }\frac{r^3 }{r_0^3}-\frac{r^2 }{r_0^2}\bigg[\frac{5}{66} b_3^{(5)}+\frac{7}{33} b_5^{(5)}+\frac{168}{935} b_7^{(5)}+\frac{16}{323} b_9^{(5)}+\textcolor{OliveGreen}{\frac{1}{563200}}\log \frac{r_0}{r}+\frac{7 V_0}{528}
    \\ \nonumber
    &\quad-\frac{575 \pi ^2}{456192}+\frac{\pi ^4}{13824}+\frac{10733797}{1990656000}\bigg] +\mathcal{O}\left(\frac{r}{r_0}\log \frac{r}{r_0}\right),
\end{align}
\end{subequations}
where $C_0$ is an unspecified constant that could be fixed by the matching with the small $\Bar{r}$ expansions of $\Bar{\psi}_3$ and $\Bar{\psi}_4$, specifically at orders in $\Bar{r}$ beyond those included in Eq.~\eqref{smallrbar}.

Our interest here is to investigate how the large-$r$ behaviour of $\psi_6$ and $\psi_{6_L}$ matches the small-$\bar{r}$ behaviour of the magnetic flux function. To this aim we consider the regime $r/r_0 \gg1$ of the sum $\alpha^6\left[\psi_6+ \log\alpha ~\psi_{6_L}\right]$ and we use the relation $r=\alpha^{-1}\bar{r}$. It is readily seen that
%
\begin{equation}
\begin{split}
    \alpha^6\big[\psi_6+ \log\alpha ~\psi_{6_L}\big]&\xrightarrow[]{\;\;r\to\alpha^{-1} \bar{r}\;\;}\alpha^3\bar{\psi}_3^{(\bar{r}^3)}+\alpha^4\bar{\psi}_{4}^{(\bar{r}^2)}+\mathcal{O}\left(\alpha^5 \log \alpha\right),
\end{split}
\end{equation}
where $\bar{\psi}_3^{(\bar{r}^3)}$ and $\bar{\psi}_4^{(\bar{r}^2)}$ are respectively the contributions at order $\mathcal{O}\left(\Bar{r}^3/r_0^3\right)$ of $\bar{\psi}_3$ and  $\mathcal{O}\left(\Bar{r}^2/r_0^2\right)$ of $\Bar{\psi}_4$ in the expansion for $\bar{r}/r_0\ll1$.
This is a very non-trivial consistency check for our expansions \eqref{EXP_r_reg} and \eqref{EXP_rbar_reg}, as it proves that the addition of higher order terms in the $r$-region does not affect the $\bar{r}$-region expansion structure laid out in Eq.~\eqref{EXP_rbar_reg}.
In particular the solution for $\bar{\psi}_{4_L}$ that we found in Eq.~\eqref{sol_barpsi4L} does not receive any new contribution coming from subleading orders of the MAE scheme due to systematic cancellations between $\psi_6$ and $\bar{\psi}_{6_L}$: as highlighted with colors (blue, violet, green) all the $r^2 \log r$ terms in $\psi_6$ cancel with the $r^2$ terms in $\psi_{6 L}$, so that no additional contributions appear in $\bar{\psi}_{4 L}$. Similarly (see the red coefficients) the $r^3 \log r$ terms in $\psi_6$ cancel out the $r^3$ terms in $\psi_{6 L}$, upholding the absence of a contribution akin to $\alpha^3 \log \alpha$ in the $\Bar{r}$-region expansion of $\psi$. In conclusion, the presence of $\psi_6$ and $\psi_{6_L}$ in the MAE scheme totally confirms all of our results at the previous orders.

Moreover, as we show in the next section the solution derived here for $\psi_{6_L}$ is sufficient for computing the first logarithmic contribution in the energy extraction rate of the BZ mechanism. Similarly, the near-horizon solution for $\psi_6(r,\theta)$ will be exploited to compute the eighth-order contribution to the energy flux.

\section{Energy and angular momentum fluxes}
\label{Sec: ELoutflows}

In Sec.~\ref{Sec: Exp} we obtained new higher order contributions in the $\alpha$-expansion of the magnetic flux $\psi$, the angular velocity $\Omega$ and the electric current $I$. These included novel non-analytic terms in $\alpha$. In this section we consider the consequences for the energy and angular momentum fluxes, which are physically measurable quantities. In particular, the energy flux corresponds to the power emitted by the BZ mechanism. As we shall see below, we find for the first time non-analytic contributions to these fluxes.

The Kerr solution with the magnetosphere is a stationary axisymmetric system so we can define conserved flux vectors for energy and angular momentum about the axis of symmetry.
As we have checked previously, the Maxwell energy momentum tensor of the FFE magnetosphere on the Kerr background is conserved: $\nabla_\mu T^{\mu\nu}=0$ with $T_{\mu\nu}=F_\mu^{\:\:\alpha}F_{\nu\alpha}-\frac{1}{4}g_{\mu\nu}F^2$. It follows that if $\xi$ is a Killing vector field then $\nabla_\nu\left( \xi_\mu T^{\mu\nu}\right)=0$, \emph{i.e.} $\xi_\mu T^{\mu\nu}$ is a conserved quantity, where $\xi=\partial_t$ or  $\xi=\partial_\phi$.
Integrating these quantities across a sphere $S$ of constant radius with normal $n_\nu$ we find the associated flux of energy and angular momentum, 
$\int_S \sqrt{-g} \xi_\mu T^{\mu\nu} n_\nu d\theta d\phi$.
The general expression for the flux of energy, $\dot{E}=dE/dt$, and the flux of angular momentum, $\dot{L}=dL/dt$, through a sphere of constant radius $r_c$ is thus given by 
\begin{align}
    \label{dEdt}
    \dot{E}\big\vert_{r_c}&=2\pi\int_0^\pi \Omega(r_c,\theta) I(r_c,\theta) \partial_\theta\psi(r_c,\theta)d\theta,
    \\
    \label{dLdt}
    \dot{L}\big\vert_{r_c}&=2\pi\int_0^\pi I(r_c,\theta) \partial_\theta\psi(r_c,\theta)d\theta.
\end{align}
Since the field variables are expanded in $\alpha$ according to Eqs.~\eqref{EXP_r_reg} and \eqref{EXP_rbar_reg}, then the fluxes can be expanded in the parameter $\alpha$, also with logarithmic terms as $\alpha^n \log \alpha$. We denote by $\dot{E}\big\vert_{r_c}^{(n)}$ the coefficient of $\alpha^n$ and by $\dot{E}\big\vert_{r_c}^{(n_L)}$ the coefficient of $\alpha^n \log \alpha$. 
Similar notation is used for the flux of the angular momentum $\dot{L}$. 

We consider here first the energy flux. Typically, one computes it either at the event horizon $r_c=r_+$ or in the asymptotic region $r_c=\infty$. As we show below, both choices of $r_c$ yield the same result, as expected from energy conservation.
For the energy flux at the event horizon, one finds in general
\begin{equation}
    \label{Eflux_H}
  \dot{E}\big\vert_{r_+} = 2\pi \int_0^\pi \Omega(r_+,\theta)\left[\Omega_{H} - \Omega(r_+,\theta)\right] \left(\partial_{\theta}\psi(r_+,\theta)\right)^2 \sqrt{\frac{g_{\phi\phi}(r_+,\theta)}{g_{\theta\theta}(r_+,\theta)}}d\theta.
\end{equation}
This is obtained by supplementing Eq.~\eqref{dEdt} with the Znajek condition at the horizon \eqref{ZC}.
Substituting the expressions \eqref{EXP_r_reg} of the field variables in the $r$-region, we get the expansion
\begin{equation}
\begin{split}
\label{Eflux_exp}
    \dot{E}\big\vert_{r_+}&=\alpha^2\dot{E}\big\vert_{r_+}^{(2)}+\alpha^4\dot{E}\big\vert_{r_+}^{(4)}+\alpha^6\dot{E}\big\vert_{r_+}^{(6)}+\alpha^7\dot{E}\big\vert_{r_+}^{(7)}+\mathcal{O}(\alpha^8 \log \alpha).
\end{split}
\end{equation}
For the second and fourth order, Eq.~\eqref{Eflux_H} gives
\begin{align}
\alpha^2\dot{E}\big\vert_{r_+}^{(2)}&=\dfrac{\pi}{6} \frac{\alpha^2}{r_0^2}\approx 0.52359878\frac{\alpha^2}{r_0^2},\\
\alpha^4\dot{E}\big\vert_{r_+}^{(4)}
&= 
\dfrac{(56-3\pi^2)\pi}{270}\frac{\alpha^4}{r_0^2}\approx 0.30707540\frac{\alpha^4}{r_0^2},
\end{align}
as originally found in \cite{1977MNRAS.179..433B} and \cite{Tanabe_2008}, respectively. These results rely on Eqs.~\eqref{monopole_config} and \eqref{omega1and3} for $\psi_0$, $\omega_1$ and $\omega_3$. Furthermore, one needs $\psi_2(r_0,\theta)$ which is obtained from Eqs.~\eqref{psi2psi4}, \eqref{R2smallr} and \eqref{Us}.
As shown in \cite{Armas:2020mio}, the fourth order contribution computed in~\cite{Tanabe_2008} is correct, despite the magnetic flux computed in~\cite{Tanabe_2008} having an apparent asymptotic divergence. 
We note that we computed the fifth order contribution to the energy flux to be zero, as previously concluded in \cite{Armas:2020mio}.

From Eq.~\eqref{Eflux_H} one gets the sixth order contribution
\begin{equation}
\label{Edot_sixth}
    \alpha^6\dot{E}\big\vert_{r_+}^{(6)} =
    \frac{\pi(54684\pi^4+421995\pi^2-8912812-12636\zeta(3))}{9525600}\frac{\alpha^6}{r_0^2}\approx 0.18589183\frac{\alpha^6}{r_0^2},
\end{equation}
again using Eqs.~\eqref{monopole_config} and \eqref{omega1and3} for $\psi_0$, $\omega_1$ and $\omega_3$. This expression involves $U_0$ and $U_1$ from $\psi_2$ and $W_0$ from $\psi_4$ written in Eqs.~\eqref{Us} and \eqref{Ws}
 and it does not depend on $\omega_4$ or $\Omega_5$. This is the first time  an exact expression at the sixth order has been derived. However, it is in numerical agreement with the previous result of
\cite{Pan:2015iaa} (with the convention $r_0=2$), despite the convergence assumptions of that paper were proved to be not valid \cite{Grignani_2018,Grignani:2019dqc,Armas:2020mio}. The numerical agreement can be explained by the lack of dependence of Eq.~\eqref{Edot_sixth} on $\omega_4$ and $\Omega_5$, neither of which were obtained correctly in \cite{Pan:2015iaa}.\footnote{Note also that while the computation of $W_0$ was originally carried out in Ref.~\cite{Pan:2015iaa} only an approximate result was found. Our computation is instead fully analytic; see Eq.~\eqref{Ws} for the result.}
 
Turning to the seventh order correction to the energy flux, we compute from Eq.~\eqref{Eflux_H} the novel term
\begin{equation}
    \alpha^7\dot{E}\big\vert_{r_+}^{(7)}=\frac{\alpha^7}{r_0^2}\frac{ \pi}{8} \int_0^{\pi} \sin^2 \theta \left[(4U_0-1)\sin^3 \theta~ \omega_4(\theta) + 2 \partial_{\theta} \psi_5(r_0,\theta) \right]d\theta,
\end{equation}
employing again Eqs.~\eqref{monopole_config} and \eqref{omega1and3}.
Using the decompositions given in Eqs.~\eqref{w4_num_ansatz} and \eqref{psi5}
we obtain

\begin{equation}
    \label{P7_a}
    \alpha^7\dot{E}\big\vert_{r_+}^{(7)}=\frac{\alpha^7}{r_0^2}\frac{2\pi(2\pi^2-15)}{10395}\left(99b_1^{(4)}+44b_3^{(4)}+8b_5^{(4)}\right)\approx0.0007907\frac{\alpha^7}{r_0^2}.
\end{equation}
Here the first equality is an analytic result, whereas the second equality makes use of the coefficients in Eq.~\eqref{b_coeff4} determined numerically.\footnote{We notice that only the coefficient of $\Theta_2$ in $\psi_5(r_0,\theta)$ as well as the coefficients of $\Theta_1$, $\Theta_3$ and $\Theta_5$ of $\omega_4(\theta)$ contribute to $\dot{E}\big\vert_{r_+}^{(7)}$.}
The uncertainty is on the 5th significant digit, and it is due to the numerical uncertainty of the coefficients $b^{(4)}$.%

The seventh order energy flux \eqref{P7_a} is one of the main results of this paper. 
This is the first contribution to the energy flux which is an odd power in the spin. This arises as a consequence of the perturbative analysis, including the matched asymptotic expansions of the $r$-region and $\bar{r}$-regions. Indeed, we find that the coefficients $b_n^{(4)}$, that were obtained by demanding continuity of $\bar{\psi}_3(\bar{r},\theta)$ across the OLS together with the correct asymptotics of the split-monopole solution, directly show up in this contribution to the energy flux.
Notice that since the energy flux $\dot{E}$ has to be even under $\alpha\rightarrow -\alpha$, the seventh order term is in general $|\alpha|^7 \dot{E}\big\vert_{r_+}^{(7)}$. Hence it is also the first example of a term that is non-analytic in the $\alpha$-expansion. 
Below we report on non-analytic terms in the expansion of the angular momentum flux as well.

As a side remark, the computation of the flux of energy up  to  $\mathcal{O}(\alpha^6)$ can be repeated at infinity. Since this computation is performed in the $\bar{r}$-region, one can use the Znajek condition at infinity \eqref{inf_ZC} to simplify the integral as follows
\begin{equation}
    \dot{E}\big\vert_{\infty}=2\pi\int_0^\pi \sin\theta \left[\Omega^\infty(\theta)\partial_\theta\psi^\infty(\theta)\right]^2d\theta.
\end{equation}
Substituting the expressions of the field variables for the $\bar{r}$-region, as given in Eq.~\eqref{EXP_rbar_reg}, one exactly reproduces the flux of energy through the event horizon \eqref{Eflux_exp}. This is obviously an important check to make, since it states that the energy emitted is conserved from the horizon to infinity. Notice that the use of Eq.~\eqref{EXP_rbar_reg} allows one to confirm the computation of the sixth order, made in \cite{Pan:2015iaa} at the horizon, and extends its consistency with the complementary computation at infinity.
Unfortunately the order $\mathcal{O}(\alpha^7)$ cannot be determined at infinity, since we lack the required $\bar{\psi}_5$ and $\bar{\Omega}_6$. 
It is worth mentioning how the cancellations of the fifth  order and of the logarithmic term occur in this computation. 
From Eq.~\eqref{dEdt}, one could in principle have contributions in $\alpha^5$ and $\alpha^6\log\alpha$.
Both these contributions are identically zero since they can be written as the integral of the total derivative $d\left(\sin^2\theta \psi^\infty\right)$, upon using the Znajek condition at the horizon $i_4=-\sin^2 \theta~\omega_4$ and $i_{5_L}=-\sin^2 \theta~\omega_{5_L}$
The vanishing of the fifth order contribution $\dot{E}\big\vert_{\infty}^{(5)}$ and of the logarithmic contribution $\dot{E}\big\vert_{\infty}^{(6_L)}$ is again consistent with the conservation of energy as $\dot{E}\big\vert_{\infty}^{(5)}= \dot{E}\big\vert_{r_+}^{(5)}=0$ together with $\dot{E}^{(6_L)}\big\vert_{\infty} = \dot{E}^{(6_L)}\big\vert_{r_+} =0$. Finally, it is immediate to check that $\dot{E}^{(7_L)}\big\vert_{r_+}$ is identically vanishing.

In order to find logarithmic contributions in the expression for the energy extraction rate one needs to push the perturbation theory at even higher orders.
We notice that by taking into account the sixth order expressions $\psi_6$ and $\psi_{6_L}$ in the field variables expansions (see Eq.~\eqref{psiEXP_66L}), it is sufficient to compute the contributions $\alpha^8$ and $\alpha^8 \log\alpha$ in the energy flux at the horizon \eqref{Eflux_H}. More specifically
\begin{align}
\label{preP8L}
    &\alpha^8 \log \alpha~ \dot{E}\big\vert_{r_+}^{(8_L)}=\frac{\alpha^8 \log \alpha}{r_0^2}\frac{ \pi}{8} \int_0^{\pi} \sin^2 \theta \left[(4U_0-1)\sin^3 \theta~ \omega_{5_L}(\theta) + 2 \partial_{\theta} \psi_{6_L}(r_0,\theta) \right]d\theta,
    \\
\label{preP8}
    &\alpha^8 ~ \dot{E}\big\vert_{r_+}^{(8)}=\frac{\alpha^8}{r_0^2}\frac{\pi}{8}\Bigg\{
   \int_0^{\pi} \sin^2\theta\left[-16\sin\theta ~\omega_4(\theta)^2+(4U_0-1) \sin^3 \theta ~\omega_5(\theta)+ 2 \partial_{\theta} \psi_{6}(r_0,\theta)\right]d\theta\cr
    &  +\frac{40897268365-54\pi^2[91384477-280\pi^2\left(135+587\pi^2\right)]+58320(6605-678\pi^2)\zeta(3)}{2571912000}\Bigg\}.
\end{align}
The second line in $ \dot{E}\big\vert_{r_+}^{(8)}$ is computed by means of Eqs.~\eqref{Us}, \eqref{Ws} and  \eqref{Vs} and comes from contributions proportional to $\psi_4$ and $\psi_2$. Notice also that there is no possible dependence on $I_6$ and $\Omega_6$ by virtue of the Znajek condition at the horizon, already implemented at the level of Eq.~\eqref{Eflux_H}.
The equations for $\psi_6$ and $\psi_{6_L}$ and their solutions in a neighbourhood of the horizon are given in Sec.~\ref{sec: psi6&6L}. Therein we also comment about their role in the matching with the $\Bar{r}$-region of Sec.~\ref{Sec: Matching2}.
Upon computing $\psi_6$ and $\psi_{6_L}$ at $r_0$ one gets explicit numbers for these contributions
\begin{align} \label{Eight}
    &\alpha^8 \log \alpha~ \dot{E}\big\vert_{r_+}^{(8_L)}=\frac{\alpha^8 \log \alpha}{r_0^2}\frac{2\pi(2\pi^2-15)}{105}\left(b^{(5_L)}_1 +\frac{4}{9}b^{(5_L)}_3 \right)\approx-0.00002216\frac{\alpha^8 \log \alpha}{r_0^2},
    \\
   &\alpha^8~ \dot{E}\big\vert_{r_+}^{(8)}\approx0.12115911 \frac{\alpha^8}{r_0^2},
\end{align}
where, in the integrals of Eqs.~\eqref{preP8L} and \eqref{preP8}, the only contribution coming from $\psi_{6_L}$ and $\psi_6$ are proportional respectively to $R^{(6_L)}_2(r_0)$ and $R^{(6)}_2(r_0)$, whose analytic expressions are given in Eqs.~\eqref{R6L2_r0} and \eqref{R62_r0}. Notice that because of the presence of the term $\omega_4^2$ in Eq.~\eqref{preP8} all the harmonics $\Theta_{2n+1}(\theta)$ appear. The result above is obtained by truncating the series at $\Theta_7(\theta)$ included. We emphasize that, while the coefficients $b^{(5_L)}$ are exact and so is the logarithmic contribution at the eight-order, the coefficient of $\alpha^8$ has the same inaccuracy of the coefficients $b^{(5)}$. We approximate the numerical value to the 8th significant digit.

We now turn to the angular momentum flux. Using the Znajek condition at the horizon \eqref{ZC} one gets 
\begin{equation}
   \dot{L}\big\vert_{r_+}= 2\pi \int_0^\pi \left[\Omega_{H} - \Omega(r_+,\theta)\right] \left(\partial_{\theta}\psi(r_+,\theta)\right)^2 \sqrt{\frac{g_{\phi\phi}(r_+,\theta)}{g_{\theta\theta}(r_+,\theta)}}d\theta.
\end{equation}
Substituting the expressions of the field variables for the $r$-region \eqref{EXP_r_reg}, employing Eqs.~\eqref{w4_num_ansatz}, \eqref{w5_num_ansatz} and \eqref{sol_omgea5L} to rewrite respectively $\omega_4$,  $\omega_5$ and $\omega_{5_L}$, we find\footnote{ Here we notice that, if $\omega_n(\theta)=b^{(n)}_1\Theta_1(\theta)+b^{(n)}_3\Theta_3(\theta)+b^{(n)}_5 \Theta_5(\theta)+\dots$, this implies 
\begin{equation*}
    \int_0^\pi \sin^3\theta\, \omega_n(\theta)d \theta=\frac{32}{105}\left(b^{(n)}_3+\frac{7}{2}b^{(n)}_1\right).
\end{equation*}
}
\begin{equation}
\label{Ldot_alpha}
    \begin{split}
        \dot{L}\big\vert_{r_+}&= \dfrac{2\pi}{3}\frac{\alpha}{r_0}  + \dfrac{(56-3\pi^2)\pi}{135}\frac{\alpha^3}{r_0} - \frac{64 \pi}{105}\left(b^{(4)}_3+\frac{7}{2}b^{(4)}_1\right)\frac{\alpha^4}{r_0}
        \\
        &\quad- \left[ \frac{\pi  \big(25272 \zeta (3)-105588 \pi ^4-981330 \pi ^2+19483259\big)}{4762800}+\frac{64 \pi}{105}\left(b^{(5)}_3+\frac{7}{2}b^{(5)}_1\right)\right]\frac{\alpha ^5}{r_0}
        \\
        &\quad- \frac{64 \pi}{105}\left(b^{(5_L)}_3+\frac{7}{2}b^{(5_L)}_1\right)\frac{\alpha^5 \log{\alpha}}{r_0} + \mathcal{O}(\alpha^6)\\
        &\approx 2.09439510 \frac{\alpha}{r_0} + 0.61415080 \frac{\alpha^3}{r_0} - 0.0181836(6) \frac{\alpha^4}{r_0}+ 0.28080637 \frac{\alpha^5}{r_0}\\ 
        &\quad
        +0.00049866 \frac{\alpha^5 \log{\alpha}}{r_0} + \mathcal{O}(\alpha^6).
    \end{split}
\end{equation}
This is another relevant result of our paper: odd powers in $\alpha$ and logs appear in the angular momentum flux, and they do so at lower orders than in the energy flux. The explanation of this follows immediately once we notice that the integrands of Eqs.~\eqref{dEdt} and \eqref{dLdt} only differ by a factor $\Omega$. Its absence in the angular momentum flux integrand prevents the cancellation between $I$ and $\Omega$ that instead takes place in the energy flux integral because of the Znajek condition at the horizon.

\section{Conclusions}
\label{sec: conclusions}

In this work we found new higher-order terms in the perturbative expansion of the BZ split-monopole magnetosphere around a slowly spinning Kerr black hole.
We employed the method of matched asymptotics expansions, as reviewed in Sec.~\ref{sec: MAE_BZ} (see also Appendix \ref{App: previous_orders}), building on the recent work \cite{Armas:2020mio}.
In this approach one employs the critical surfaces to determine the field variables together with the stream equation, as envisioned by \cite{2013ApJ...765..113C}. 
In Sec.~\ref{Sec: Exp} we obtained an expansion in the low spin  regime both in the $r$-region, that contains the event horizon and the ILS, and in the $\bar{r}$-region, that contains the OLS and the asymptotic region.
We imposed that the magnetic flux is continuous across these critical surfaces, this is non-trivial at the OLS where we employed numerical methods to solve the stream equation (see Appendix \ref{App:Numerics}). Moreover, in Sec.~\ref{Sec: Matching2} we matched the two expansions in the overlap region.
Exploring the higher order corrections reveals the ineluctability of terms containing $\log \alpha$, {\emph{i.e.}}~terms in the perturbative expansion for $\alpha \ll 1$ that are proportional to $\alpha^n (\log \alpha)^m$ with $m>0$. This feature was briefly anticipated in \cite{Armas:2020mio}. 
This is one of the main results of this paper, as it reveals the low spin expansion to be non-analytic in the expansion parameter. 
This is a novel aspect of the low spin expansion of the BZ split-monopole that was not present in earlier works
\cite{Tanabe_2008,Pan_2015,Pan:2015iaa,Pan:2015imp,Grignani_2018,Grignani:2019dqc}. 

In Sec.~\ref{Sec: ELoutflows} we computed the energy and angular momentum fluxes that result from our new higher order corrections. 
For the energy flux, we computed three new terms in the low spin expansion, namely the seventh order correction written in Eq.~\eqref{P7_a} 
and the eighth order corrections in Eq.~\eqref{Eight}, including a logarithmic contribution. 
The seventh order and the logarithmic contribution at the eighth order are non-analytic in $\alpha$. It is important to emphasize that, because of the evenness of the energy flux under $\alpha\rightarrow -\alpha$, these new contributions are, respectively, proportional to $|\alpha|^7$ and $\alpha^8\log( |\alpha|)$. This is another of our main results.
Furthermore, we compute for the first time the exact sixth order correction to the energy flux Eq.~\eqref{Edot_sixth}
that previously was found only numerically \cite{Pan:2015iaa}.%
\footnote{As noted in Sec.~\ref{Sec: ELoutflows}, this is only the case when $\dot{E}$ is expressed as a function of $\alpha$. For the sixth order correction to $\dot{E}(\Omega_H)$ the result of \cite{Pan:2015iaa} is inaccurate, as one can see by comparing to Eq.~\eqref{Power_OmegaH}.}
For the angular momentum flux we found three new higher order corrections, that also include terms proportional to $\log |\alpha|$, as recorded in Eq.~\eqref{Ldot_alpha}. Again, this demonstrates the non-analytic dependence on the expansion parameter.

In addition to the above-mentioned main results, we have several other important results worth advertising here: 
\begin{itemize}
\item In Appendix \ref{App:Numerics} we present a novel numerical method, with higher accuracy and faster convergence, that does not rely on a minimization procedure, to solve the stream equation in the $\bar{r}$-region and impose continuity across the OLS. 
Note that the numerical coefficients for the expansions of $\omega_4$ in Eq.~\eqref{b_coeff4} and $\omega_5$ in Eq.~\eqref{b_coeff5} are obtained from this numerical computation. These coefficients enter the fluxes of energy and angular momentum 
and hence it is important to obtain them with increased precision.
\item We check in Sec.~\ref{Sec: ELoutflows} that the energy flux computed at the horizon matches the energy flux computed at infinity. This is another consistency check for our matched asymptotic expansions, since that computation involves both the perturbative expansions in the $r$ and $\bar{r}$ regions.
This also serves as a confirmation of the fact that the asymptotic behaviour of $\psi_4$ for large $r$ in \cite{Pan:2015iaa} is in fact not valid, since it would violate energy conservation. 
%
%
%
\item Regarding the perturbative expansion of $\psi(\bar{r},\theta)$ in $\alpha$ for the $\bar{r}$-region, we show in Appendix \ref{App: previous_orders} why one cannot have  corrections of order $\alpha$ and $\alpha^2$. This is an important result as it ensures that there cannot be higher order corrections in $\alpha$ in the $r$-region that potentially can alter the perturbative expansion in the $\bar{r}$-region.
\item In Appendix \ref{app:explicit} we present for the first time the full exact solution for $\psi_4(r,\theta)$, \emph{i.e.}~the fourth order magnetic flux in the $r$-region. 
\item In Section \ref{Subsec: SE_cs} we find a new form for the stream equation in Eq.~\eqref{stream3} that can ease the study of light surfaces. 
Using this, we show that, at each perturbative order in $\alpha$, the ILS condition obtained from Eq.~\eqref{stream_LS2} is equivalent to the Znajek condition at the horizon \eqref{ZC}.
\end{itemize}

As already concluded, the BZ split-monopole magnetosphere expanded for low spin has non-analytic terms in the expansion parameter both in the magnetic flux as well as in the energy and angular momentum fluxes. It would be highly interesting to understand the origin of this non-analytic behavior. For finite $\alpha$, the Kerr metric is obviously an analytic function of $\alpha$. Thus, this suggests that also the magnetosphere should be an analytic function of $\alpha$. Our take on this, is that this might very well be the case, and that the non-analyticity is an artifact of taking the $\alpha \rightarrow 0$ limit. Therefore, if one could obtain the exact solution of the split-monopole magnetosphere for finite $\alpha$, it should reveal a function that is analytic in $\alpha$, but notably a function that exhibits non-analytic terms in the $\alpha \rightarrow 0$ limit. This points to an interesting non-perturbative structure of the BZ split-monopole that might indicate how to approach a finite $\alpha$ solution.
To illustrate this point, consider the following example of a function expanded for small $\alpha$
\begin{equation}
\frac{1}{\alpha   ^2+1}+\alpha ^2 \sqrt{\left(\alpha ^2\right)^{\alpha ^4}-\frac{1}{\alpha ^2+1}}  = 1 -\alpha^2 +  |\alpha|^3 + \alpha^4 - \frac{1}{2}  |\alpha|^5 + |\alpha|^5 \log |\alpha|  + \mathcal{O} ( \alpha^6 ).
\end{equation}
On the LHS we have clearly an even function of $\alpha$ which would give an expansion that is analytic around non-zero values of $\alpha$. However,  when expanded around $\alpha=0$ the function exhibits non-analytic terms in $\alpha$, both in the form of odd powers of $|\alpha|$ as well as terms involving $\log |\alpha|$. 

\begin{figure}[ht]
    \centering
    \includegraphics[width=0.8\textwidth]{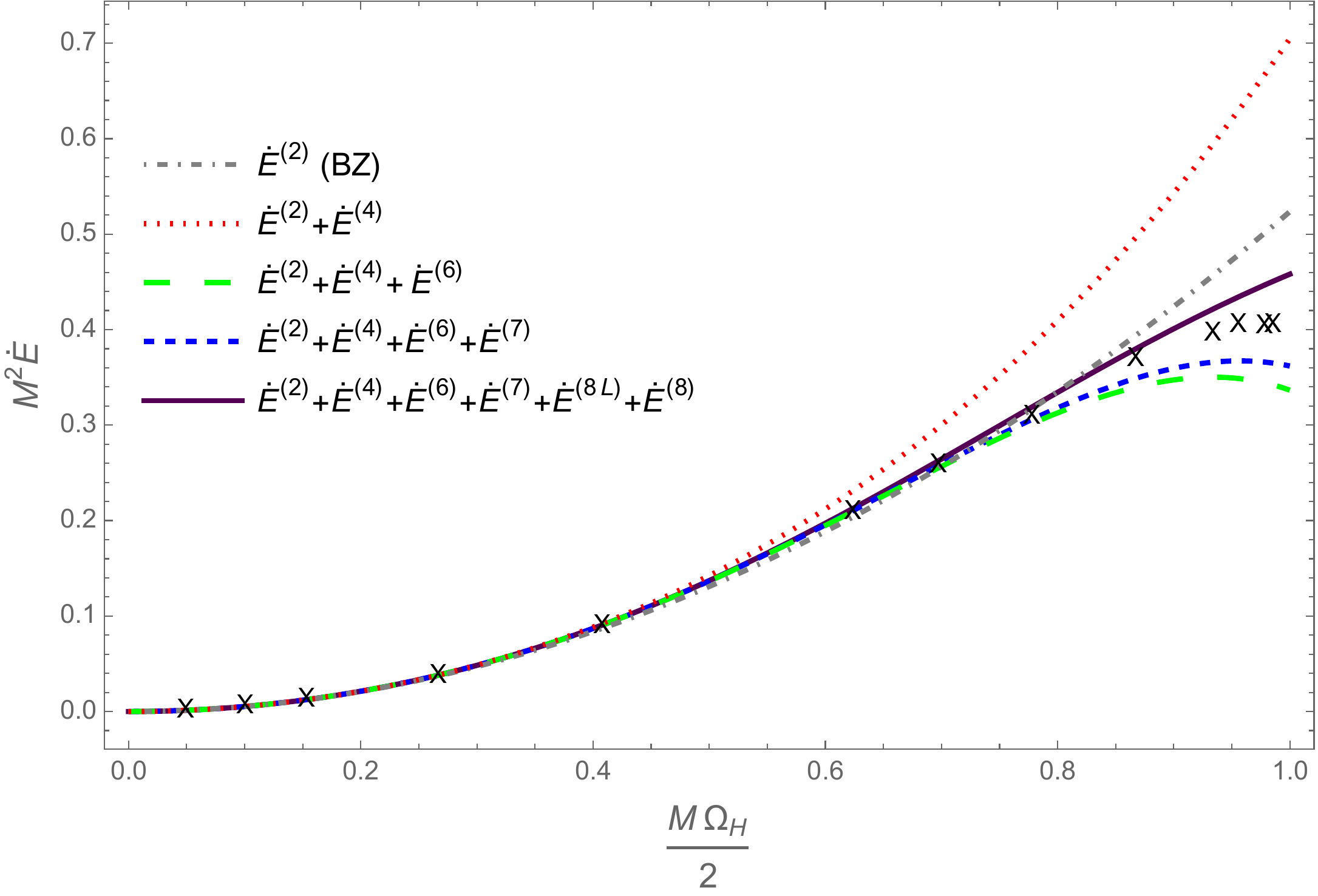}
    \caption{We depict here the energy flux $M^2 \dot{E}$ as function of $M\Omega_H/2$. 
    We plot the curves for second (original BZ result), fourth, sixth, seventh and eighth orders (log included).
    The data points of \cite{Tchekhovskoy:2009ba} are marked with crosses.}
    \label{fig:perturbative_comparison}
\end{figure}

It is interesting to compare our new analytical results for the energy flux $\dot{E}$ to the numerical data obtained for finite $\alpha$. We shall consider the energy flux $\dot{E}(\Omega_H)$ as a function of the angular velocity $\Omega_H$ of the Kerr black hole.
The approximate numerical values for the perturbative coefficients are written in Eq.~\eqref{Power_OmegaH}.
For the numerical comparison, we employ the simulation data of \cite{Tchekhovskoy:2009ba}. 
In Fig.~\ref{fig:perturbative_comparison} we have drawn the curves obtained by adding more and more perturbative corrections to $\dot{E}(\Omega_H)$ and compared these curves to the simulation of \cite{Tchekhovskoy:2009ba}.
In the range from $0.5$ to $1$, only the fourth order curve, where one truncates the powers of $\Omega_H$ after the fourth order term, is deviating considerably from the simulation data. Among the others, the curve that includes all perturbative terms up to the eighth order in $\Omega_H$ is the one that globally best approximates the simulation data of \cite{Tchekhovskoy:2009ba}. 

\begin{figure}[ht]
    \centering
    \hspace{-1mm}\includegraphics[width=0.8\textwidth]{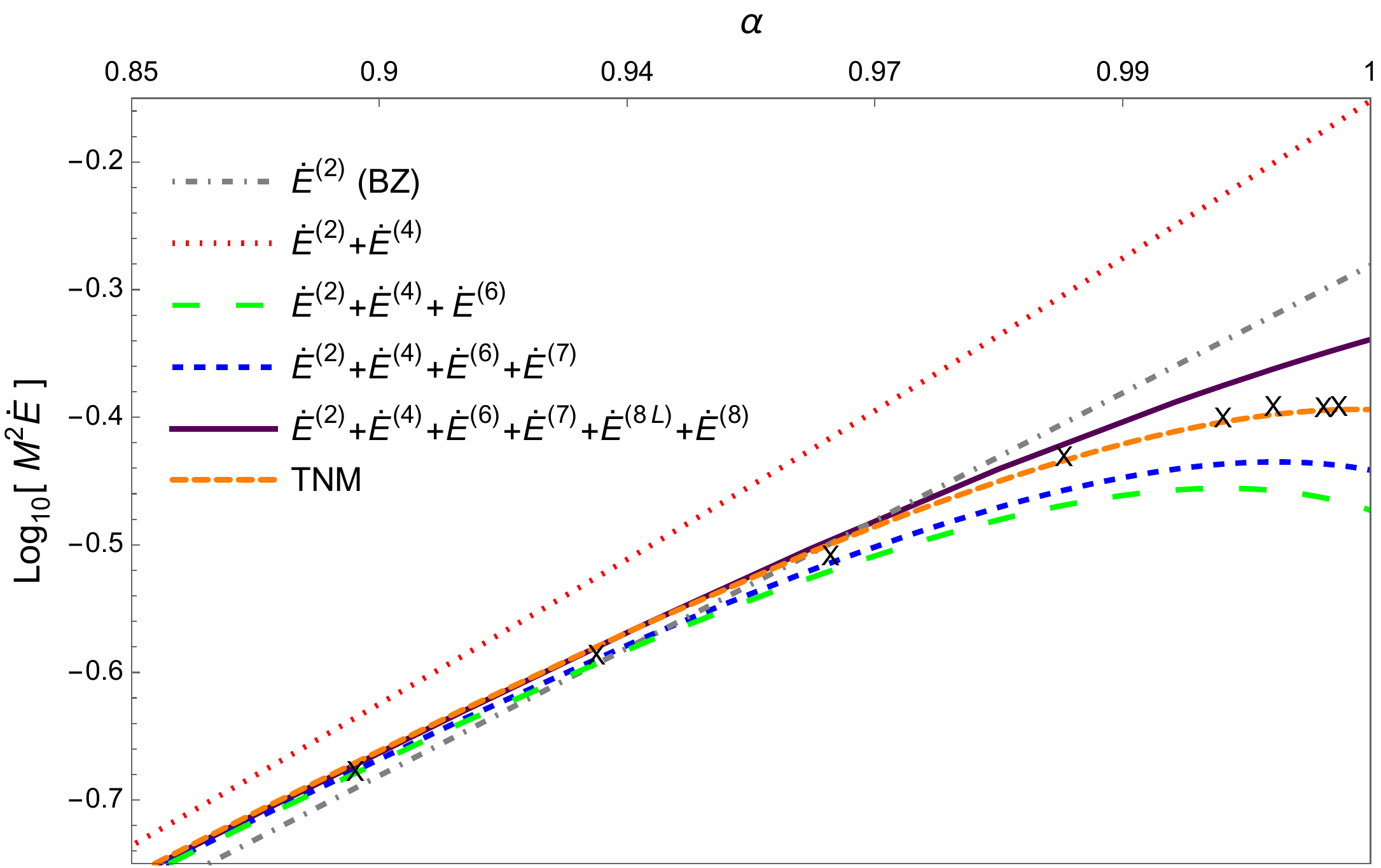}
    \includegraphics[height=0.215\textwidth, width=0.795\textwidth]{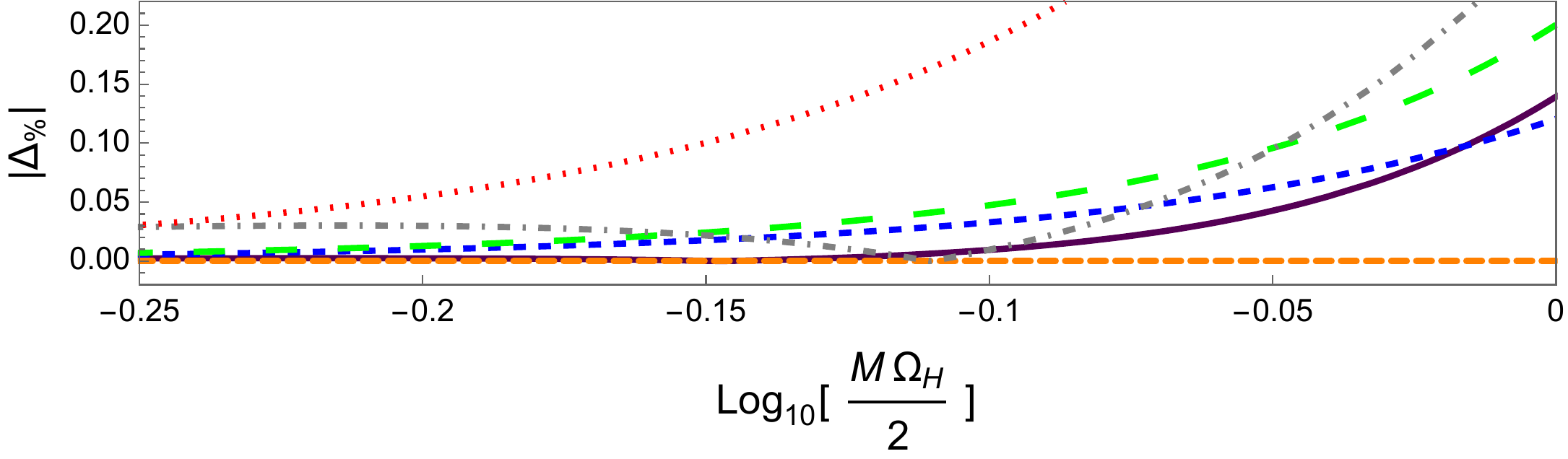}
    \caption{We depict in the top diagram the energy flux $\dot{E}$
    in terms of $\log_{10}(M^2\dot{E})$ as function of $\log_{10}(M\Omega_H/2)$. We plot the curves for second (original BZ result), fourth, sixth, seventh and eighth orders (log included). The TNM curve is that of Eq.~\eqref{TMNcurve}. The data points of \cite{Tchekhovskoy:2009ba} are marked with crosses. In the bottom diagram we compare the absolute fractional deviation $|\Delta_\%|$ between the perturbative curves and the curve of Eq.~\eqref{TMNcurve}.}
    \label{fig:full_comparison}
\end{figure}

We explore this comparison further in Fig.~\ref{fig:full_comparison}. 
In this figure we have used that the simulation data of \cite{Tchekhovskoy:2009ba} is well approximated by the following formula by Tchekhovsky, Narayan and McKinney (TNM) \footnote{See Eq.~(9) of \cite{Tchekhovskoy:2009ba} with $\alpha_{\rm(TNM)} \approx 0.346 r_0^2$ and $\beta_{\rm(TNM)}\approx -0.575 r_0^4 $; for $r_0=2$ one has $\alpha_{\rm(TNM)}\approx 1.38$ and $\beta_{\rm(TNM)}\approx -9.2$.}
\begin{equation}
\label{TMNcurve}
         \dot{E}_{\rm (TNM)}=\frac{2\pi}{3} \Omega_H^2 \Big[1+ 0.346 ~ r_0^2\Omega_H^2 -0.575 ~r_0^4\Omega_H^4 \Big],
     \end{equation}
which has the same second and fourth order corrections as our perturbative energy flux, but where the sixth order term is obtained by fitting the curve to $\alpha=0.9999$. Henceforth we will refer to Eq.~\eqref{TMNcurve} as the TNM curve.

One can make the comparison of curves more quantitative by considering the fractional deviations $|\Delta_\%|$. This is done in the bottom diagram of Fig.~\ref{fig:full_comparison}. For instance, the original quadratic BZ model depicted with a  grey  dashed line  is  unable  to  reproduce  numerical  data in the high spin regime as it deviates from the TNM curve with $|\Delta_\%|\approx28.7\%$ when $\alpha\simeq 1$.
Similarly, the sixth order approximation depicted with a green dashed line can be seen to have deviations with $|\Delta_\%|\approx20.0\%$. Our seventh order approximation instead behaves significantly better, with a maximal $|\Delta_\%|$ around $12\%$. The inclusion of the eighth order correction increases the accuracy in the nearly-extreme regime, and still attains a deviation of only $|\Delta_\%|\approx 13.9\%$ when $\alpha\simeq1$. 
In the sub-extreme regime, for instance at the Thorne limit $\alpha\simeq0.998$ \cite{Thorne:1974ve}, our result at the eighth order deviates from the TNM curve of only $|\Delta_\%|\approx7.5\%$, the seventh order deviates of $|\Delta_\%|\approx8.4\%$, and sixth order of $|\Delta_\%|\approx13.3\%$.

There are many directions one could pursue from the results of this work:
\begin{itemize}
\item
It would be interesting to push our perturbative results to higher orders in $\alpha$. Computing such higher terms could reveal hints about a possible non-perturbative structure. 
\item The numerical data at finite $\alpha$ are currently not precise enough to give a satisfying comparison with the perturbative results. It would be important to obtain the full curve $\dot{E}(\alpha)$ with higher precision, and with a range all the way from 0 to 1. Currently, the numerical data is lacking for values of $\alpha$ close to 0 in which a comparison to the perturbatively obtained $\dot{E}(\alpha)$ curve would be appropriate.
\item One could use our improved higher-order understanding of the BZ split-monopole to study the BZ mechanism. Indeed, in the recent paper \cite{Kimura:2021dsa} the authors study the back-reaction of the BZ process on black holes and its relation to the black hole mechanics. It would be interesting to understand the interplay between the BZ mechanism and the black hole mechanics for exact solutions describing black holes immersed into a back-reacting electromagnetic fields \cite{Gibbons:2013dna,Booth:2015nwa,Astorino:2016hls}.
Another interesting direction for future work is to study the propagation of the Alfv\'en waves and investigate resonant or superradiance effects that might occur in the magnetosphere \cite{Noda:2021ivs}.
\item It would be highly interesting to approach the BZ split-monopole with analytical methods from the extreme case $\alpha=1$ and compute its deviation in the near-extreme spin regime. However, a magnetically-dominated analytic solution to FFE around extreme Kerr background is still missing so far. A first attempt in this direction has been already taken by identifying the attractor solutions to FFE in the near-horizon region of extreme Kerr \cite{Gralla:2016jfc}, and near-extreme Kerr spacetime \cite{Camilloni:2020qah}, and constructing perturbatively a solution away from the near-horizon region \cite{Camilloni:2020hns}.

\item Finally, it would be interesting to consider other profiles. For example, force-free magnetospheres with field lines with vertical topology received some attention in the literature \cite{Nathanail:2014aua,Pan:2014bja,Yang:2015ata,Pan:2017npg,East:2018ayf}. A careful investigation should be performed to understand whether the MAE scheme can be applied in the vertical case to have a consistent perturbation scheme.
\end{itemize}

We hope to tackle some of these open problems in the future.


\section*{Acknowledgment}
We thank Jay Armas and Geoffrey Comp\`ere for useful correspondence, and the anonymous referee for suggestions to improve the manuscript.
We thank Tetyana Pitik for helping us with the numerical data in Figs.~\ref{fig:perturbative_comparison} and \ref{fig:full_comparison}.
 O.~C.~D. acknowledges financial support from the STFC ``Particle Physics Grants Panel (PPGP) 2018" Grant No.~ST/T000775/1. T.H. acknowledges support by the Independent Research Fund Denmark grant number DFF-6108-00340 “Towards a deeper understanding of black holes with non-relativistic holography”.
The research of R.O. is funded by the European Structural and Investment Funds (ESIF) and the Czech Ministry of Education, Youth and Sports (MSMT), Project CoGraDS - CZ.02.1.01/0.0/
0.0/15003/0000437. R.O. would like to thank the Institut d'Astrophysique de Paris for the hospitality at different stages of this project.
M.O. and G.G. acknowledge support from the project “MOSAICO” financed by Fondo Ricerca di Base 2020 of the University of Perugia. F.C. G.G., R.O., M.O. and A.P. thank Niels Bohr Institute for hospitality. T.H. and R.O. thank the University of Perugia for the hospitality. J.~E.~S has been partially supported by STFC consolidated grant ST/T000694/1.


\appendix

\section{Perturbation of the Blandford-Znajek split-monopole: review of past results
}
\label{App: previous_orders}

In this appendix we offer a streamlined review of the BZ perturbative approach applied to the split-monopole solution.
Even though the main results that follow are not new and just confirm those of Refs.~\cite{Armas:2020mio,1977MNRAS.179..433B,Tanabe_2008,Pan_2015,Pan:2015iaa,Pan:2015imp,Grignani_2018,Grignani:2019dqc} our aim here is to make our analysis as thorough and self-contained as possible.
For the sake of a clear notation we recall the general expansions of the field variables at the $\alpha$ orders taken into account in the following: in the $r$-region they are
\begin{equation}
    \label{Exp_r_general}
    \begin{split}
        \psi(r,\theta)&=\psi_0(r,\theta)+\alpha \; \psi_1(r,\theta)+\alpha^2\;\psi_2(r,\theta)+\alpha^3\;\psi_3(r,\theta)+\alpha^4\;\psi_4(r,\theta)+\mathcal{O}(\alpha^5),
        \\
        r_0I(r,\theta)&=I_0(r,\theta)+\alpha\; I_1(r,\theta)+\alpha^2\; I_2(r,\theta)+\alpha^3\;I_3(r,\theta)
        +\alpha^4\;I_4(r,\theta)+\mathcal{O}(\alpha^5),
        \\
        r_0\Omega(r,\theta)&=\Omega_0(r,\theta)+\alpha\; \Omega_1(r,\theta)+\alpha^2\; \Omega_2(r,\theta)+\alpha^3\;\Omega_3(r,\theta)+\alpha^4\;\Omega_4(r,\theta)+\mathcal{O}(\alpha^5),
    \end{split}
\end{equation}
while in the $\bar{r}$-region 
\begin{equation}
    \label{Exp_rbar_general}
    \begin{split}
        \psi(\Bar{r},\theta)&=\Bar{\psi}_0(r,\theta)+\alpha\; \Bar{\psi}_1(\bar{r},\theta)+\alpha^2\;\Bar{\psi}_2(\bar{r},\theta)+\alpha^3\;\Bar{\psi}_3(\bar{r},\theta)+\mathcal{O}(\alpha^4),
        \\
        r_0I(\bar{r},\theta)&=\bar{I}_0(\bar{r},\theta)+\alpha\; \bar{I}_1(\bar{r},\theta)+\alpha^2\; \bar{I}_2(\bar{r},\theta)+\alpha^3\;\bar{I}_3(\bar{r},\theta)+\alpha^4\;\bar{I}_4(\bar{r},\theta)+\mathcal{O}(\alpha^5),
        \\
        r_0\Omega(\bar{r},\theta)&=\bar{\Omega}_0(\bar{r},\theta)+\alpha\; \bar{\Omega}_1(\bar{r},\theta)+\alpha^2\; \bar{\Omega}_2(\bar{r},\theta)+\alpha^3\;\bar{\Omega}_3(\bar{r},\theta)+\alpha^4\;\bar{\Omega}_4(\bar{r},\theta)+\mathcal{O}(\alpha^5).
    \end{split}
\end{equation}
%
\subsection*{Leading order}
\label{Subsec:4.1}
%
\subsubsection*{$r$-region}
\label{Subsec:r0}
Given that in this work we consider the split-monopole configuration, our perturbative scheme is built around
\begin{equation}
    \psi_0(r,\theta) = 1-\cos\theta, \quad I_0(r,\theta) = \Omega_0(r,\theta) = 0.
\end{equation}
where indeed, in term of the Schwarzschild operator \eqref{Lr}, we have
\begin{equation}
   \mathcal{L} \psi_0(r,\theta) = 0.
\end{equation}
In such a configuration all the magnetic field lines escape from the horizon of the black hole and reach infinity.

It is possible to determine the location of the ILS and the OLS by equating to zero respectively the two invariants \eqref{chi2} and \eqref{eta2}.
Accordingly, we see that in the limit $\alpha \to 0$ the ILS approaches the Schwarzschild event horizon $r_{\rm ILS}\to r_0$, whereas the OLS approaches infinity $r_{\rm OLS}\to\infty$. This means that the OLS position in terms of the $r$ coordinate has a negative power of $\alpha$ as the leading perturbative contribution and thus suggests the necessity of a radial coordinate like $\Bar{r}= \alpha r$ for a proper perturbative description, as first noted in Ref.~\cite{Armas:2020mio}.
%
%
\subsubsection*{$\Bar{r}$-region}
\label{Subsec:rbar0}
In compliance with what we did in the other region, we assume that at leading order we have
\begin{equation}
    \Bar{\psi}_0(\Bar{r},\theta) = 1-\cos\theta, \quad \Bar{I}_0(\Bar{r},\theta) = \Bar{\Omega}_0(\Bar{r},\theta) = 0.
\end{equation}
This is consistent with the class of solutions found in \cite{1973ApJ...180L.133M} to the stream equation in flat spacetime\footnote{At leading order in the $\Bar{r}$-dependent expansion of the field variables this is what the stream equation \eqref{stream2} reduces to} and in \cite{Armas:2020mio} it has been argued that other possible solutions satisfying all the boundary conditions at hand probably do not exist.
Indeed in terms of the operator \eqref{Lbarr} we have
\begin{equation}
   \bar{\mathcal{L}} \Bar{\psi}_0(\Bar{r},\theta) = 0.
\end{equation}
%
%
%
%
\subsection*{First subleading order}
\label{Subsec:4.2}
\subsubsection*{$r$-region}
\label{Subsec:r1}
The first subleading contribution of the $\alpha$-expanded stream equation in the $r$-region is simply
\begin{equation}
    \mathcal{L} \psi_1(r,\theta)=0.
\end{equation}
The most general solution to this equation satisfying the split-monopole boundary conditions \eqref{BC_split-monopole} is \cite{Gralla:2015vta}
\begin{equation}
    \label{psione}
    \psi_1(r,\theta)= \sum^{\infty}_{l = 1} \left[ B^{<}_{2l}R^{<}_{2l}(r)+B^{>}_{2l}R^{>}_{2l}(r)\right] \, \Theta_{2l} (\theta),
\end{equation}
where $B^{<}_{2l}$, $B^{>}_{2l}$ are free constants and the radial functions are given by
\begin{align}
    R^{<}_{l}(r) & = \frac{2 r^2}{ r_0^2}\frac{\Gamma(l+2)^2}{\Gamma(2l+1)} \, {}_{2}F_1 \left[l+2,1-l,3,\frac{r}{ r_0}\right],\\
    R^{>}_{l}(r) & = -\frac{2}{\sqrt{\pi}} \left(\frac{r}{2 r_0}\right)^{-l} \left\{{}_{2}F_1 \left[l+2,l;1;1-\frac{r_0}{ r}\right] \log\left(1-\frac{r_0}{r}\right)+ P_l\left(\frac{r}{r_0}\right)\right\}.
\end{align}
Here ${}_{2}F_1$ is the Gaussian hypergeometric function and the polynomials $P_l$ have the recursive definition
\begin{align}
    P_1(x) & = x^2 + \frac{x}{2}, \cr
    P_2(x) & = 4 x^4 - x^3 -\frac{x^2}{6}, \\
    P_l(x) & = \frac{(2l-1)\left[l(l-1)(2x -1)-1\right]x P_{l-1}(x)-l^2(l-2)x^2 P_{l-2}(x)}{(l+1)(l-1)^2}.\nonumber
\end{align}
In general one has that $R^{>}_{l}(r)$ diverges logarithmically for $r\to r_0$ and
\begin{align}
\label{Rlarger<}
   R^{<}_{l}(r) & \sim \left(\frac{r}{r_0}\right)^{l+1},\\
\label{Rlarger>}
   R^{>}_{l}(r) & \sim \left(\frac{r}{r_0}\right)^{-l},
\end{align}
for $r\to \infty$. While a divergence for $r\to\infty$ can be in general acceptable when it leads to a matching with the $\Bar{r}$-region as discussed in Sec.~\ref{Sec: Matching2}, one has to choose $B^{>}_{2l}=0$ $\forall l$ so as to avoid a diverging solution for $r\to r_0$\footnote{In Ref.~\cite{Armas:2020mio} the authors proved that even by resolving an additional region around the ILS one can't find any new matching condition that could make divergences for $r\to r_0$ acceptable.}. However we can also argue that $B^{<}_{2l}=0$ $\forall l$: let us focus on the case $l=1$ and suppose that $B^{<}_{2}\neq0$, then considering Eq.~\eqref{Rlarger>} the quantity $\alpha \psi_1 (r,\theta)$ for $r\to \infty$ would go at least like $\alpha r^3/r_0^3$, which should match an impossible $\mathcal{O}(\alpha^{-2})$ term in the $\bar{r}$-region. For higher $l$ the situation is the same, with just higher powers in $r$ in the large $r$-behaviour of $\alpha \psi_1 (r,\theta)$. Therefore overall we must have
\begin{equation}
    \psi_1 (r,\theta) = 0
\end{equation}
In addition, from the integrability conditions \eqref{intcond1} and \eqref{intcond2}
we get
\begin{equation}
    \partial_r \Omega_1 (r,\theta) = 0,\quad \partial_r I_1 (r,\theta) = 0.
\end{equation}
Thus we can write
\begin{equation}
    \label{Om_first}
    \Omega_1 (r,\theta)= \omega_1(\theta),\quad
I _1(r,\theta)= i_1(\theta),
\end{equation}
in terms of two yet unspecified function of $\theta$, $i_1$ and $\omega_1$.

From the Znajek condition at the horizon \eqref{ZC} expanded to the
lowest order in $\alpha$ follows
\begin{equation}
    \label{zc_1}
    i_1(\theta) =  \frac{1}{2}\bigg(1-2\omega_1(\theta) \bigg) \sin^2 \theta .
\end{equation}
We can check that the latter result can also be inferred by imposing the regularity condition at the ILS.
Assuming $0\leq\Omega\leq\Omega_H$, the ILS lies between the horizon radius $r_+$ and the static limit $r_{\rm{ergo}}(\theta)$ \cite{Kom2004}. 
Solving Eq.~\eqref{chi2} to lowest order in $\alpha$ yields
\begin{equation}
    r_{\rm ILS} (\theta) = r_0 - \frac{r_0}{4} \left[ \cos^2 \theta+ 4\omega_1(\theta)\bigg(1-\omega_1(\theta)\bigg)\sin^2 \theta    \right]\alpha^2 + \mathcal{O}(\alpha^3).
\end{equation}
By inserting this in the regularity condition \eqref{stream_LS2} and considering only the lowest order in $\alpha$ one gets
\begin{equation}
     \partial_{\theta}i_1^2(\theta)=\partial_\theta\left[\frac{1}{2}\bigg(1-2\omega_1(\theta)\bigg)\sin^2\theta\right]^2.
\end{equation}
We observe that the Znajek condition \eqref{zc_1} is the solution of this equation after the regularity on the axis has been imposed. In other words, we confirm that regularity at the ILS, supported by the regularity at the axis, is equivalent to regularity at the horizon: both of them can be used to derive the expression \eqref{zc_1} for $i_1(\theta)$.

\subsubsection*{$\Bar{r}$-region}
At this order the integrability conditions \eqref{intcond1} and \eqref{intcond2} entail
\begin{equation}
    \partial_{\bar{r}} \bar{\Omega}_1 (\bar{r},\theta) = 0,\quad \partial_{\bar{r}} \bar{I}_1 (\bar{r},\theta) = 0.
\end{equation}
Taking into account the necessary matching with the $r$-region this means that
\begin{equation}
   \bar{\Omega}_1 (\bar{r},\theta) = \omega_1(\theta),\quad
    \bar{I}_1 (\bar{r},\theta) = i_1(\theta),
\end{equation}
where $\omega_1(\theta)$ and $i_1(\theta)$ are the same functions introduced in Eq.~\eqref{Om_first}.
Then, from the first non-zero contribution of the $\alpha$-expanded stream equation we find
\begin{equation}
    \label{streamrbar1}
    \partial_\theta  i_1^2(\theta) = \partial_\theta \big[ \omega_1^2(\theta) \sin^4 \theta \big].
\end{equation}
By combining this with Eq. \eqref{zc_1} we get
\begin{equation}
    \partial_\theta \left[ \sin^4 \theta \left( \omega_1(\theta)- \frac{1}{4} \right) \right] = 0,
\end{equation}
with the only solution which is also regular on the axis being
\begin{equation}
    \label{omega_1}
    \omega_1 (\theta)= \frac{1}{4}.
\end{equation}
The same conclusion could be reached by looking at the first $\alpha$ order of the Znajek condition at infinity \eqref{inf_ZC}, which gives the relation
\begin{equation}
    \label{zc_inf_1}
    i_1(\theta)=\omega_1(\theta) \sin^2\theta,
\end{equation}
that in fact inserted in Eq. \eqref{zc_1} gives again $\omega_1 (\theta)= 1/4$.

We can now compute the position of the OLS at leading order. From Eq.~\eqref{eta2} expanded to zeroth order in $\alpha$ we get
\begin{equation}
    \Bar{r}_{\rm OLS}(\theta) =  \frac{r_0}{ \omega_1 (\theta) \sin \theta} + \mathcal{O}( \alpha) = \frac{4 r_0}{ \sin \theta} + \mathcal{O}( \alpha) .
\end{equation}
\subsection*{Second subleading order}

\subsubsection*{$r$-region}

At this order the $\alpha$-expansion of the stream equation \eqref{stream3} leads to
\begin{equation}
    \label{stream_2nd}
    \mathcal{L} \psi_2(r,\theta) = - \dfrac{r_0 (r+r_0)}{2r^4}\dfrac{\Theta_2(\theta)}{\sin{\theta}}.
\end{equation}
The corresponding particular solution for $\psi_2$ is given by \cite{1977MNRAS.179..433B}
\begin{equation}
    \label{psi2r}
    \psi_2 (r,\theta)=  R_2^{(2)}(r) \Theta_2(\theta),
\end{equation}
where
\begin{equation}
    \label{R2}
    \begin{aligned}
        R_2^{(2)}(r) =&\frac{r_0^2+6r_0r-24r^2}{12r_0^2} \log \left(\frac{r}{r_0}\right)+\frac{11}{72}+\frac{r_0}{6r}+\frac{r}{r_0}-\frac{2r^2}{r_0^2}
        \\ \vspace{1mm}
        &  +\left[\mbox{Li}_2\left(\frac{r_0}{r}\right)-\log \left(\frac{r}{r_0}\right) \log \left(1-\frac{r_0}{r}\right)\right]\frac{r^2(4r-3r_0)}{2r_0^3}\,.
    \end{aligned}
\end{equation}
The general solution is found by adding to Eq.~\eqref{stream_2nd} an homogeneous term akin to the RHS of Eq.~\eqref{psione}. However the same considerations that imply $\psi_1=0$ can be repeated here to argue that such an homogeneous term must be fixed to zero for a consistent perturbative scheme.\\
Coming to the integrability conditions, at this order we have
\begin{equation}
\Omega_2 (r,\theta) = \omega_2(\theta), \quad I_2(r,\theta) = i_2(\theta).
\end{equation}
From the Znajek condition at the horizon \eqref{ZC} we thus get
\begin{equation}
    \label{zc_2}
    i_2 (\theta) = - \omega_2 (\theta)  \sin^2 \theta.
\end{equation}
The ILS is now located at
\begin{equation}
    r_{\rm ILS} (\theta) = r_0 -  \frac{ \alpha^2 r_0}{4} \left( 1- \frac{1}{4} \sin^2 \theta    \right) - \frac{\alpha^3 r_0}{2}   \omega_2(\theta) \sin^2 \theta +\mathcal{O}(\alpha^4).
\end{equation}
By inserting this in the reduced stream equation \eqref{stream_LS2} we get
\begin{equation}
    i_2(\theta) + \frac{1}{2}\tan \theta \partial_\theta i_2(\theta) = -2 \omega_2(\theta)  \sin^2 \theta - \frac{1}{2} \sin^2\theta \tan\theta \partial_\theta \omega_2(\theta).
\end{equation}
Again, the solution to this equation, upon demanding regularity on the axis, corresponds to $i_2$ as expressed in \eqref{zc_2}.
%
%
\subsubsection*{$\Bar{r}$-region}
%
%
At this order the integrability conditions give
\begin{equation}
    \partial_{\Bar{r}}\Omega_2(\Bar{r},\theta) = 0, \quad \partial_{\Bar{r}}I_2(\Bar{r},\theta) = \frac{1}{2} \cos{\theta} \partial_{\Bar{r}}\Bar{\psi}_1(\Bar{r},\theta),
\end{equation}
that is, considering the matching to the $r$-region,
\begin{equation}
    \Omega_2(\Bar{r},\theta) = \omega_2(\theta), \quad  I_2(\Bar{r},\theta) = i_2(\theta) + \frac{1}{2} \cos{\theta}  \Bar{\psi}_1(\Bar{r},\theta).
\end{equation}
Using this in the $\alpha$ expansion of the stream equation yields
\begin{equation}
\label{barpsieq1}
    \Bar{\mathcal{L}}\Bar{\psi}_1(\Bar{r},\theta) = \frac{\partial_\theta\left[\sin^2{\theta}\left(\sin^2\theta \, \omega_{2}(\theta)-i_{2}(\theta)\right)\right]}{4r_0^2\sin{\theta}} = \frac{\partial_\theta\left[\sin^4{\theta} \, \omega_{2}(\theta)\right]}{2r_0^2\sin{\theta}},
\end{equation}
where the last equality holds thanks to Eq.~\eqref{zc_2}.
A simple solution to this equation that satisfies the boundary conditions \eqref{BC_split-monopole} is
\begin{equation}
\label{solpsi1omega2}
    \Bar{\psi}_1(\Bar{r},\theta) = 0, \quad \omega_{2}(\theta) = 0,
\end{equation}
which also implies $i_{2}(\theta) = 0$ via Eq.~\eqref{zc_2}. This is compatible with the Znajek condition at infinity, that at this order reads
\begin{equation}
    \sin\theta\partial_\theta \bar{\psi}_{1}^\infty(\theta)-2\cos\theta \bar{\psi}_{1}^\infty(\theta)=-4 \left(\sin^2\theta \, \omega_{2}(\theta)-i_{2}(\theta)\right).
\end{equation}
However, it is conceivable that $\bar{\psi}_1$ can have a non-zero boundary condition in the overlap region $\bar{r}\ll r_0$. To examine this possibility we consider the leading behavior of $\bar{\psi_1}$ for small $\bar{r}$ as
\begin{equation}
    \bar{\psi}_1(r,\theta) = \frac{\bar{r}^n}{r_0^n} A(\theta) + \cdots ,
\end{equation}
where the dots are subleading terms. For $n \geq 3$ this gives the equation for $A(\theta)$
\begin{equation}
\label{Aeq}
\left[ 4  \cos^2 \theta - (2+n(n+1))\sin^2 \theta \right] A    - \cos \theta \sin\theta A' - \sin^2 \theta A'' = 0,
\end{equation}
which is solved  as
\begin{equation}
\label{Asol}
  A(\theta)= C \sin^2\theta P_{\frac{\sqrt{25+4n+4n^2}-1}{2}}(\cos \theta),
\end{equation}
where $C$ is an arbitrary constant and $P_n(x)$ is a Legendre Polynomials of the first kind, since the second kind is divergent for $\theta=0$. However, for $n\geq 3$ this is non-zero at the equator $\theta=\pi/2$ which violates the split-monopole boundary conditions \eqref{BC_split-monopole}.
Therefore, we can conclude that non-zero terms with $n \geq 3$ in $\bar{\psi}_1$ are forbidden for small $\bar{r}$. Comparing with the $r$-region, this excludes that $\bar{\psi}_1$ could be turned on by perturbative corrections in the $r$-region with $\alpha^{n+1}$ with $n\geq 3$.
As we shall see below, this fixes in particular $\psi_4$ in the $r$-region. Moreover, there are no non-zero boundary conditions from the $r$-region with lower powers of $\alpha$. Thus, we can conclude that \eqref{solpsi1omega2} is the correct solution.

Note also that it is a simple exercise to see that if one starts with a leading term 
\begin{equation}
    \bar{\psi}_1(r,\theta) = \left( \log \frac{\bar{r}}{r_0}\right)^m \frac{\bar{r}^n}{r_0^n} A(\theta) + \cdots ,
\end{equation}
with $m$ an integer,
then $A(\theta)$ obeys the same equation \eqref{Aeq} and hence has the same solution \eqref{Asol} for $n \geq 3$. Thus, also contributions of this type seems excluded.

As for the position of the OLS, from Eqs.~\eqref{eta2} and \eqref{solpsi1omega2} we find
\begin{equation}
    \Bar{r}_{\rm OLS}(\theta)  = \frac{4 r_0}{ \sin \theta} -\frac{\alpha}{2}+ \mathcal{O}( \alpha^2).
\end{equation}
\subsection*{Third subleading order}

\subsubsection*{$r$-region}

From the stream equation it follows that
\begin{equation}
   \mathcal{L} \psi_3(r, \theta) = 0.
\end{equation}
Similarly to what happens for $\psi_1$, the only acceptable solution to this equation is
\begin{equation}
  \psi_3(r, \theta) = 0.
\end{equation}
The integrability conditions tell us 
\begin{equation}
    \Omega_3(r,\theta) = \omega_3(\theta), \quad I_3(r,\theta) =   i_3(\theta)+\frac{1}{2} \cos{\theta} \psi_2(r,\theta),
\end{equation}
with $\psi_2$ given in Eq.~\eqref{psi2r}.\\
From the Znajek condition \eqref{ZC} and the near horizon behaviour of $R_2^{(2)}(r)$ given in Eq.~\eqref{R2smallr}, we get
\begin{equation}
    \label{zc_3}
    i_3(\theta) =  \sin^2 \theta\left[\frac{1}{8}-\omega_3(\theta)\right] + \frac{\sin^4 \theta}{4}\left(\frac{1}{4}-U_0\right).
\end{equation}
Then, with Eq.~\eqref{chi2} we can update the ILS position to
\begin{equation}
    r_{\rm ILS} (\theta) = r_0 -  \frac{ \alpha^2 r_0}{4} \left( 1- \frac{1}{4} \sin^2 \theta    \right) + \alpha^4 r_0 \, r^{(4)}_{\rm ILS}(\theta) +\mathcal{O}(\alpha^5),
\end{equation}
with 
\begin{equation}
    \label{rILS4}
    r^{(4)}_{\rm ILS}(\theta) = - \frac{\cos^2 \theta}{16} - \frac{\sin^2\theta}{2}\left[\omega_3(\theta)+\frac{ \sin^2 \theta}{128}\right].
\end{equation}
If we insert the latter into the reduced stream equation \eqref{stream_LS2} we get
\begin{equation}
    i_3(\theta) + \frac{1}{2} \tan \theta\, \partial_\theta i_3(\theta) = - \sin^2\theta\left[2\omega_3(\theta)+\frac{\tan\theta}{2}\partial_\theta\omega_3(\theta)-\frac{1}{4}-\frac{(1-2U_1)}{16}\sin^2\theta\right].
\end{equation}
Then, by substituting $i_3(\theta)$ from the Znajek condition at the horizon \eqref{zc_3} we find
\begin{equation}
    1  - 6 U_0 +  U_1 = 0.
\end{equation}
Recalling Eqs.~\eqref{Us}, we see that the above condition is automatically satisfied. Therefore again we can assert the equivalence between the regularity at the ILS and the Znajek condition at the horizon.
\subsubsection*{$\Bar{r}$-region}

The integrability conditions \eqref{intcond1} and \eqref{intcond2} supplemented by the matching condition to the $r$-region now give
\begin{equation}
   \Bar{\Omega}_3(\Bar{r}, \theta) = \omega_3(\theta), \quad \Bar{I}_3(\Bar{r}, \theta) =i_3(\theta) + \frac{1}{2} \cos{\theta} \Bar{\psi}_2(\Bar{r},\theta). 
\end{equation}
Considering this, the expansion in $\alpha$ of the stream equation yields
\begin{equation}
\label{stremrbar3}
    \Bar{\mathcal{L}} \Bar{\psi}_2 (\Bar{r}, \theta) = \frac{\partial_\theta\left[\sin^2{\theta} \left(\sin^2\theta \, \omega_{3}(\theta)-i_{3}(\theta)\right)\right]}{4r_0^2\sin{\theta}}.
\end{equation}
One can perform a similar analysis as for $\bar{\psi}_1$. Writing the leading term in a small $\bar{r}$ expansion as 
\begin{equation}
    \bar{\psi}_2(r,\theta) = \left( \log \frac{\bar{r}}{r_0}\right)^m \frac{\bar{r}^n}{r_0^n} A(\theta) + \cdots ,
\end{equation}
with integers $n\geq 1$ and $m \geq 0$, one finds that $A(\theta)$ obeys Eq.~\eqref{Aeq} which is solved by \eqref{Asol}. Since $\bar{\psi}_2$ should be zero at the equator $\theta=\pi/2$
 this excludes that $A(\theta)$ can be non-zero. A non-zero boundary condition for $\bar{\psi}_2$ can thus only arise from perturbative corrections to the $\alpha$-expansion in the $r$-region that dominates over $\alpha^2$. Since no such terms are found in the above, we can conclude that 
\begin{equation}
\label{psi2barr}
    \Bar{\psi}_2 (\Bar{r}, \theta)=0.
\end{equation}
Equipped with the Znajek condition \eqref{zc_3}, we solve for the $\omega_{3}(\theta)$ that ensures the RHS of Eq.~\eqref{stremrbar3} to vanish. In practice we obtain a simple first order differential equation whose unique solution regular at the axis turns out 
\begin{equation}
    \label{omega_3}
    \omega_3(\theta)=\frac{1}{16}+(1-4U_0)\frac{\sin^2\theta}{32}.
\end{equation}
Notice that this result is in accordance with the Znajek condition at infinity \eqref{inf_ZC}, namely
\begin{equation}
    \sin\theta\partial_\theta \bar{\psi}_{2}^\infty(\theta)-2\cos\theta \bar{\psi}_{2}^\infty(\theta)=-4 \left[\sin^2\theta \, \omega_{3}(\theta)-i_{3}(\theta)\right],
\end{equation}
once Eqs.~\eqref{psi2barr} and \eqref{zc_3} are used.

Finally, from Eqs.~\eqref{eta2} and \eqref{omega_3} we get an additional contribution to the position of the OLS:
\begin{equation}
    \Bar{r}_{\rm OLS}(\theta)  = \frac{4 r_0}{ \sin \theta} -\frac{\alpha}{2}- \alpha^2 r_0\,\left[ \frac{r_0}{\sin \theta} +  \left(\frac{19}{32} -2U_0 \right) r_0 \sin \theta\right] + \mathcal{O}( \alpha^3).
\end{equation}

\subsection*{Fourth subleading order}
\label{Subapp: 4subleading}

\subsubsection*{$r$-region}
At this order the stream equation reads
\begin{equation}
    \label{stream_r_4}
    \mathcal{L} \psi_4 = S^{(4)}_{2}(r) \frac{\Theta_2(\theta)}{\sin\theta} + S^{(4)}_{4}(r) \frac{\Theta_4(\theta)}{\sin\theta},
\end{equation}
where the explicit expressions for the sources are
\begin{align}
    \label{S42}
  &  S_2^{(4)}(r)=\frac{1}{56 r^6 r_0^2 (r-r_0)} \bigg\{
  2 r^2 \bigg[(r-r_0) \bigg(r \left(r^5-4 r^2 r_0^3-3 r_0^5\right)\partial_r^2 R_2^{(2)}(r)\cr
  &+ \left(2 r^5+4 r^2 r_0^3+9 r_0^5\right) \partial_r R_2^{(2)}(r)\bigg)-2 R_2^{(2)}(r) \left(3 r^5-10 r^2 r_0^3+21 r r_0^4-11 r_0^5\right)\bigg]\cr 
  &+r_0^2 \bigg(r^5+4 r^4 r_0 (3 U_0-2)+6 r^2 r_0^3+7 r r_0^4-6 r_0^5\bigg)\bigg\},\\
  \label{S44}
  &  S_4^{(4)}(r)= \frac{1}{224 r^6 r_0^2 (r-r_0)} \bigg\{ 6 r^2 \bigg[(r-r_0) \bigg(r \left(r^5-4 r^2 r_0^3+4 r_0^5\right) \partial_r^2 R_2^{(2)}(r)\cr 
  &-2 \left(r^5+2 r^2 r_0^3-6 r_0^5\right) \partial_r R_2^{(2)}(r)\bigg)-6 R_2^{(2)}(r) \left(r^5-8 r^2 r_0^3+8 r_0^5\right)\bigg]\cr 
  &+3 r_0^2 \bigg(r^5+r^4 r_0 (12 U_0-1)-8 r^2 r_0^3+8 r_0^5\bigg)\bigg\},
\end{align}
in terms of the $R_2^{(2)}(r)$ given in Eq.~\eqref{R2} and the $U_0$ in Eq.~\eqref{Us}.
By just looking at the form of the equation we can write its particular solution as a decomposition of the kind
\begin{equation}
\label{ansatzpsi4}
    \psi_4(r,\theta) = R_4^{(2)}(r) \Theta_2(\theta) + R_4^{(4)}(r) \Theta_4(\theta).
\end{equation}
From here the idea is to insert Eq.~\eqref{ansatzpsi4} into Eq.~\eqref{stream_r_4} and project the latter on $\Theta_2(\theta)$ and $\Theta_4(\theta)$ so as to solve separately for the radial components $R_4^{(2)}(r) $ and $R_4^{(4)}(r)$. The analytical solutions for these functions are rather long and we give them explicitly in Appendix \ref{app:explicit} . Nevertheless, from them we can easily infer the behaviour of $\psi_4(r,\theta)$ for $r\to r_0$ and $r \to \infty$ by means of either the Green function method described in Appendix~\ref{App:Green} or the Frobenius method.
The results of this operation can be found in Eqs.~\eqref{R42smallr}, \eqref{R44smallr} and in Eqs.~\eqref{R42greatr}, \eqref{R44greatr}.

In addition to the particular solution for $\psi_4(r,\theta)$ described above, one can furthermore add the homogeneous terms 
\begin{equation}
    c_1 \left(\frac{r^3}{r_0^3}-\frac{3r^2}{4r_0^2} \right) \Theta_2(\theta) + c_2 \left( \frac{r^5}{6r_0^5}-\frac{5r^4}{16r_0^4}+\frac{5r^3}{28r_0^3}-\frac{5r^2}{168r_0^2}\right) \Theta_4(\theta).
\end{equation}
One can immediately see that $c_2=0$, since in the overlap region $r \gg r_0$ a leading term that goes like $\alpha^4 r^5$ would need to match with $\alpha^{-1} \bar{r}^5$ in the $\bar{r}$-region which is not allowed in our perturbative expansion. Regarding $c_1$, having this non-zero would require the matching with a term of the form $ \alpha c_1 \bar{r}^3 r_0^{-3} \Theta_2(\theta)$ which is not allowed since, as seen above, $\bar{\psi}_1=0$. Thus, also $c_1=0$.

As for the other field variables, the integrability conditions \eqref{intcond1} and \eqref{intcond2} impose
\begin{equation}
\Omega_4(r,\theta) = \omega_4(\theta) \quad I_4(r,\theta) = i_4(\theta).
\end{equation}
From the Znajek condition at the horizon \eqref{ZC} we thus get
\begin{equation}
    \label{zc_4}
    i_4 (\theta) = - \omega_4(\theta)  \sin^2 \theta.
\end{equation}
This result also follows by using the condition of regularity at the ILS, whose position given by Eq.~\eqref{chi2} now reads
\begin{equation}
    \frac{r_{\rm ILS} (\theta)}{r_0} = 1 + \alpha^2 r_{\rm ILS}^{(2)} + \alpha^4 r_{\rm ILS}^{(4)}+ \alpha^5 r_{\rm ILS}^{(5)}
    +\mathcal{O}(\alpha^6),
\end{equation}
with the new term being
\begin{equation}
\label{rils4}
r_{\rm ILS}^{(5)}= - \frac{\omega_4}{2} \sin^2 \theta. 
\end{equation}
In fact, by evaluating the reduced stream equation \eqref{stream_LS2} at this point and considering only the terms at order $\alpha^5$ one gets
\begin{equation}
    \label{ils4equation}
    i_4 + \frac{1}{2}\tan \theta\, \partial_\theta i_4 = -\sin^2\theta\left(2 \omega_4 - \frac{1}{2} \, \partial_\theta \omega_4\, \tan\theta\right).
\end{equation}
As usual, the solution of this ODE preserving regularity at the rotational axis is given by Eq.~\eqref{zc_4}. This establishes once again the equivalence between the horizon and the ILS at the level of regularity conditions.

\subsubsection*{$\Bar{r}$-region}
We start again from the integrability conditions \eqref{intcond1} and \eqref{intcond2} plus the matching condition to the $r$-region, which ultimately give
\begin{equation}
    \Bar{\Omega}_4 (\Bar{r}, \theta) = \omega_4(\theta), \quad  \Bar{I}_4 (\Bar{r}, \theta) = i_4(\theta) + \frac{1}{2} \cos{\theta}  \Bar{\psi}_3(\Bar{r},\theta).
\end{equation}
The corresponding $\alpha$ order in the stream equation expansion gives Equation \eqref{eq_psibar3}.
Notice that this time we cannot just set $\Bar{\psi}_3(\Bar{r},\theta)=0$ since the source term is $\Bar{r}$-dependent. This equation can only be tackled numerically and this is precisely what has been done in \cite{Armas:2020mio}. Taking into account the new-found importance of the numerical parameters that result from this computation (see the discussion below Eq.~\eqref{P7_a}), we repeated it in an effort to fix such parameters with greater accuracy. More details about the numerics can be found in Appendix~\ref{App:Numerics}, while a plot for the numerical solution for $\Bar{\psi}_3$ is shown on the left panel of Fig.~\ref{fig:plot_barpsi4}.


\section{Green's function method for \texorpdfstring{$\psi_n$}{} in the \texorpdfstring{$r$}{}-region }
\label{App:Green}

\subsection*{Homogeneous solutions in the $r$-region}
The source-free stream equation in the Schwarzschild spacetime reads (see Eq.~\eqref{L_schw})
\begin{equation}
    \label{L_schw_app}
    \mathcal{L}\psi=\frac{1}{\sin\theta}\partial_r\left[\left(1-\frac{r_0}{r}\right)\partial_r\psi\right]+\frac{1}{r^2}\partial_\theta\left(\frac{1}{\sin\theta}\partial_\theta\psi\right)=0.
\end{equation}
The equation is separable thus we can look for a solution of the form
\begin{equation}\label{schwfactor}
\psi(r,\theta)=R(r)\Theta(\theta).
\end{equation}
Since we would like to get a symmetric solution about $\theta=\pi/2$, we can limit ourselves to the region $0\le\theta\le\pi/2$. We also look for solution that are zero at $\theta= 0$. 
Eq.\eqref{schwfactor} separates into 
\begin{align}
 &\mathcal{L}_\ell^\theta [\Theta]=\frac{d}{d\theta} \left(\frac{1}{\sin\theta}\frac{d\Theta}{d\theta}  \right)+\frac{\ell(\ell+1)\Theta}{\sin\theta}=0,\label{Thetaeq}\\
 &\mathcal{L}_\ell^r[ R]=\frac{d}{dr} \left[\left(1-\frac{r_0}{r}\right)\frac{d R}{dr} \right]-\frac{\ell(\ell+1)R}{r^2}=0.\label{Req}
\end{align}
Suitably normalized solutions of $\mathcal{L}_\ell^\theta\left[\Theta_{\ell}(\theta)\right]=0$, vanishing at both poles, are given by the hypergeometric functions,
\begin{align}
	\label{eq:ThetaOdd}
	\Theta_{2k-1}(\theta)&={_2F_1}\!\br{-k,k-\frac{1}{2};\frac{1}{2};\cos^2{\theta}},\\
	\label{eq:ThetaEven}
	\Theta_{2k}(\theta)&={_2F_1}\!\br{-k,k+\frac{1}{2};\frac{3}{2};\cos^2{\theta}}\cos{\theta},
\end{align}
where $k$ (and hence $\ell$) is a positive integer. Note that $\Theta_{2k}({\pi}/{2})=0$. The $\Theta_\ell$ are proportional to the Gegenbauer polynomials $C_{\ell-1}^{(3/2)} (x)$ \cite{Gralla:2015vta}
\begin{numcases}
	{\Theta_\ell(\theta)=}
		{_2F_1}\!\br{\frac{\ell}{2},-\frac{\ell+1}{2};\frac{1}{2};\cos^2{\theta}}
		=-\frac{\Gamma\!\pa{-\frac{\ell}{2}}\Gamma\!\pa{\frac{\ell+1}{2}}}{2\sqrt{\pi}}
		\sin^2{\theta}\, C^{(3/2)}_{\ell-1}(\cos{\theta})
		& $\ell$ odd,\nonumber\\
		&\\
		{_2F_1}\!\br{-\frac{\ell}{2},\frac{\ell+1}{2};\frac{3}{2};\cos^2{\theta}}\cos{\theta}
		=-(-1)^{\ell/2}\frac{\sqrt{\pi}\,\Gamma\!\pa{\frac{\ell}{2}}}{4\,\Gamma\!\pa{\frac{\ell+3}{2}}}
		\sin^2{\theta}\,C^{(3/2)}_{\ell-1}(\cos{\theta})
		&$\ell$ even.\nonumber
\end{numcases}
 The $C_\ell^\alpha (x)$ are orthogonal polynomials in the interval $[-1,1]$ with respect to the weight function $(1-x^2)^{\alpha - 1/2}$, with $\alpha>1/2$. In particular, we will make use of the equation
\begin{equation}\label{ortogonality}
\int_0^\pi d\theta\, C_\ell^{(3/2)}(\cos\theta) C_{\ell'}^{(3/2)}(\cos\theta) \sin^3\theta=\frac{(\ell+2)(\ell+1)}{\ell+3/2}\delta_{\ell\ell'}.
\end{equation}

As we are only interested in what happens outside the Schwarzschild horizon, we solve Eq.~\eqref{Req} in the region $r\ge 2 M=r_0$. The equation has two independent solutions, the first one, when $\ell\to \ell +1$ in Eq.~\eqref{Req}, is given in terms of Jacobi Polynomials,  $P_\ell^{(\alpha,\beta)}(x)$,
\begin{equation}
U_{\ell}(r)= r_0^\ell r^2 P_\ell^{(2,0)}\left(1-\frac{2 r}{r_0}\right),
\end{equation}
while the second solution is obtained in terms of the first as
\begin{equation}
\label{V_l}
V_\ell(r)=U_\ell(r)\int_r^\infty\frac{ x dx}{(r_0- x)U_\ell^2(x)}.
\end{equation}
We  note that all the $U_\ell(r)$ solutions are regular at $r=r_0$ but diverge as $r^{\ell+2}$ for $r \to\infty$, while the $V_\ell(r)$ solutions are divergent in $r=r_0$ but go as $r^{-\ell-1}$ for $r\to\infty$.\footnote{The functions $U_\ell(r)$ and  $V_\ell(r)$ defined here corresponds respectively to $R^{<}_{l+1}(r)$ and $R^{>}_{l+1}(r)$ in the notation of Appendix \ref{App: previous_orders}. The notation used here follows the one of the original paper by Blandford and Znajek \cite{1977MNRAS.179..433B}.}

Before writing the total solution we need to consider separately the case $\ell=0$  that corresponds to the monopole solution. For the angular part we have the solution
$\Theta_0(\theta) = C(1- \cos \theta).$ The corresponding radial solution is
$R_0(r) = a + b[r + r_0 \ln(r - r_0)]$.
This solution is divergent both at $r = r_0$ and for $r\to\infty$.
The general solution can now be written as a linear combination of the solutions described above
\begin{equation}
\psi_0(r,\theta)=(1-\cos\theta)\left[\alpha+\beta(r+r_0\log(r-r_0))\right]+\sum_{\ell=0}^\infty\left[A_\ell U_\ell(r)+B_\ell V_\ell(r)\right]\Theta_{\ell+1}(\theta).
\end{equation}

\subsection*{Green's function}
The Green's function $G(r,\theta;r',\theta')$ for the operator $\mathcal{L}$ defined in Eq.~\eqref{L_schw_app} is the solution to the equation
\begin{equation}
 \mathcal{L} G(r,\theta ; r',\theta')=\delta(r-r')\delta(\theta -\theta')=\sin\theta\delta(r-r')\delta(\cos\theta -\cos\theta').
\end{equation}
The Green function $G(r,\theta;r',\theta')$ was derived by Blandford and Znajek in \cite{1977MNRAS.179..433B} as
\begin{equation}
G(r,\theta;r',\theta')=\sum_{\ell=0}^\infty \frac{\ell+3/2}{(\ell+1)(\ell+2)}\sin^2\theta\sin^2 \theta' C_\ell^{(3/2)}(\cos\theta) C_\ell^{(3/2)}(\cos\theta')R_\ell(r, r'),
\end{equation}
where $R_\ell(r, r')$ is given by
\begin{equation}
R_\ell(r, r')=
\begin{cases}
U_\ell(r)V_\ell(r')~\text{if}~ r<r',
\\
V_\ell(r)U_\ell(r') ~\text{if}~ r>r'.
\end{cases}
\end{equation}
The function $G(r,\theta;r',\theta')$ satisfies the boundary conditions to be finite at $r = r_0$ and to approach zero at $r\to\infty$.

In the $r$-region, at $\mathcal{O}(\alpha^n)$, in general we have to solve equations of the form
\begin{equation}\label{inhom_n}
    \mathcal{L}\psi_n=S_n(\omega_{n-1}, i_{n-1}; r, \theta),
\end{equation}
where the  $S_n$ are the sources of the equations.  Given Eq.~\eqref{inhom_n}, we can then construct a particular solution as
\begin{align}\label{particularsol}
\psi_n(r,\theta)&=\int_0^\pi d\theta' \int_{r_0}^\infty dr' G(r,\theta;r',\theta') S_n(r', \theta')\cr
&=\sum_{\ell=0}^\infty \frac{\ell+3/2}{(\ell+1)(\ell+2)}\sin^2\theta C_\ell^{(3/2)}(\cos\theta)\int_0^\pi d\theta'\sin^2 \theta'C_\ell^{(3/2)}(\cos\theta')\times
\cr
&\quad\times \left[V_\ell(r) \int_{r_0}^r dr' U_\ell(r')+U_\ell(r) \int_{r}^\infty dr' V_\ell(r')\right] S_n(r', \theta').
\end{align}
In particular, the value of this solution at $r=r_0$ is given by:
\begin{align}\label{psin0}
\psi_n(r_0,\theta)&=\sum_{\ell=0}^\infty \frac{\ell+3/2}{(\ell+1)(\ell+2)}\sin^2\theta C_\ell^{(3/2)}(\cos\theta)\times\cr
&\quad\times U_\ell(r_0)\int_0^\pi d\theta' \int_{r_0}^\infty dr' \sin^2 \theta'C_\ell^{(3/2)}(\cos\theta')
V_\ell(r') S_n(r', \theta').
\end{align}

\subsubsection*{Second order}

At the second order in the $\alpha$ expansion the stream equation in the $r$-region reads~\eqref{stream_2nd}
\begin{equation}
    \mathcal{L} \psi_2(r,\theta) = S_2(r,\theta)= - \dfrac{r_0 (r+r_0)}{2r^4}\dfrac{\Theta_2(\theta)}{\sin{\theta}}= - \frac{r_0 (r+r_0)}{6 r^4} C_1^{(3/2)}(\cos\theta)\sin\theta.
\end{equation}
Inserting this into Eq.~\eqref{particularsol}, thanks to the orthogonality condition \eqref{ortogonality}, the only term in the series in $\ell$ that survives the integration on $\theta'$ is $\ell=1$
\begin{equation}\label{psi2green}
\psi_2(r,\theta) =-\sin^2\theta C_1^{(3/2)}(\cos\theta)
 \left[V_1(r) \int_{r_0}^r dr' U_1(r')+U_1(r) \int_{r}^\infty dr' V_1(r')\right]  \frac{r_0 (r'+r_0)}{6 r'^4},
\end{equation}
where
\begin{align}\label{U1V1}
U_1(r)&=r^2 (3r_0-4 r),\cr
V_1(r)&=\frac{1}{6 r_0^5}\left[
r_0 \left(-24 r^2+6 r r_0+r_0^2\right)+6
   r^2 (4 r-3 r_0) \log
   \left(\frac{r}{r-r_0}\right)\right].
\end{align}
   
At $r=r_0$, we have:
\begin{align}\label{psi2greenhor}
\psi_2(r_0,\theta) &=-\sin^2\theta C_1^{(3/2)}(\cos\theta)
U_1(r_0) \int_{r_0}^\infty dr' V_1(r')  \frac{r_0 (r'+r_0)}{6 r'^4}\cr
&=\frac{6\pi^2-49}{72} \sin^2\theta\cos\theta\equiv U_0 \sin^2\theta\cos\theta.
\end{align}

\subsubsection*{Fourth order}

At the fourth-order, the stream equation becomes
\begin{align}\label{stream_4rd}
    \mathcal{L} \psi_4(r,\theta) =S_4(r,\theta)&= S^{(4)}_{2}(r) \frac{\Theta_2(\theta)}{\sin\theta} + S^{(4)}_{4}(r) \frac{\Theta_4(\theta)}{\sin\theta} \nonumber\\
    &= \frac{1}{3}S^{(4)}_{2}(r) C_1^{(3/2)}(\cos\theta)\sin\theta -\frac{2}{15}S^{(4)}_{4}(r)C_3^{(3/2)}(\cos\theta)\sin\theta,
\end{align}
where the explicit expressions for $S^{(4)}_{2}(r)$ and $S^{(4)}_{4}(r)$ are given in Eqs.~\eqref{S42} and \eqref{S44}.

Inserting the source expression into Eq.~\eqref{particularsol}, thanks to the orthogonality condition \eqref{ortogonality}, the only terms in the series in $\ell$ that survive the integration on $\theta'$ are $\ell=1$ and 
$\ell=3$. A particular solution of Eq.~\eqref{stream_4rd} is
\begin{align}\label{psi4green}
\psi_4(r,\theta) =\frac{1}{3}\sin^2\theta C_1^{(3/2)}(\cos\theta)
 \left[V_1(r) \int_{r_0}^r dr' U_1(r')+U_1(r) \int_{r}^\infty dr' V_1(r')\right] S_2^{(4)}(r')
 \cr
 -\frac{2}{15}\sin^2\theta C_3^{(3/2)}(\cos\theta)
 \left[V_3(r) \int_{r_0}^r dr' U_3(r')+U_3(r) \int_{r}^\infty dr' V_3(r')\right] S_4^{(4)}(r'),
\end{align}
where $U_1(r)$ and $V_1(r)$ are given in Eq.~\eqref{U1V1} and 
\begin{align}
&U_3(r)=r^2 \left(-56 r^3+105 r^2 r_0-60 r r_0^2+10
   r_0^3\right),\cr
   &V_3(r)=-\frac{56 r^4}{r_0^8}+\frac{77 r^3}{r_0^7}-\frac{157
   r^2}{6 r_0^6}+\frac{r}
   {r_0^5}+\frac{1}{20 r_0^4}+\frac{r^2}{r_0^9} \left(-56 r^3+105 r^2
   r_0-60 r r_0^2+10 r_0^3\right) \log
   \left(1-\frac{r_0}{r}\right).\cr&
\end{align}
At $r_0$ one has
\begin{align}\label{psi4greenhorizon}
\psi_4(r_0,\theta) &=\frac{1}{3}\sin^2\theta C_1^{(3/2)}(\cos\theta)
U_1(r_0) \int_{r_0}^\infty drV_1(r) S_2^{(4)}(r)\nonumber\\
 &-\frac{2}{15}\sin^2\theta C_3^{(3/2)}(\cos\theta)
U_3(r_0) \int_{r_0}^\infty dr V_3(r) S_4^{(4)}(r).
\end{align}
The function $\psi_4(r,\theta)$ is dimensionless; it is then possible to scale out $r_0$ everywhere in Eq.~\eqref{psi4greenhorizon} and to perform the integrals analytically defining $x=r/r_0$. We get
\begin{align}
W_0 &\equiv \frac{U_1(r_0)}{r_0^3}\int_{1}^\infty dxV_1(x) S_2^{(4)}(x)\nonumber\\
&=\frac{39 \zeta (3)}{3920}+\frac{17929399}{2540160}-\frac{3877 \pi ^2}{12096}-\frac{19 \pi ^4}{480} \simeq  0.051159~,\\
V_0&\equiv \frac{U_3(r_0)}{r_0^5}\int_{1}^\infty dxV_3(x) S_4^{(4)}(x)\nonumber\\
&=\frac{79 \pi ^4}{640}-\frac{963 \zeta (3)}{3920}-\frac{2012505017}{67737600}+\frac{9791 \pi ^2}{5376} \simeq -0.006732~,
\end{align}
in agreement with the numerical values computed in \cite{Pan:2015iaa}.

\subsubsection*{Fifth order}
At the fifth-order, the stream equation becomes (see Eq.~\eqref{eqpsi5})
\begin{equation}
    \mathcal{L}\psi_5(r,\theta)=S_5(r,\theta)=\frac{r^2+r_0 r+r_0^2}{2r_0^2 r^2\sin^2\theta}\partial_\theta\left[\omega_4(\theta)\sin^4\theta\right].
\end{equation}
Substituting the numerical ansatz for $\omega_4(\theta)$ as in Eq.~\eqref{w4_num_ansatz} and performing the integral in Eq.~\eqref{psin0}, one gets the numerical value of the coefficients of $\Theta_2(\theta)$ obtained in Eq.~\eqref{psi5}, namely
\begin{equation}
    -\frac{2}{189}\left(99b_1^{(4)}+44b_3^{(4)}+8b_5^{(4)} \right) = -0.0029210(6).
\end{equation}

\section{Numerical methods}
\label{App:Numerics}

In this section we detail the numerical method we used to tackle this problem. The aim is to solve both Eq.~(\ref{SE_rbar_reg}) for $n=3$ and $n=4$, subject to the relevant boundary conditions.

We first note that the mode functions $P_k(\theta)\equiv \Theta_{2k}(\theta)$ appearing in Eq.~(\ref{eq:ThetaOdd}) form a basis for the angular dependence of $\bar{\psi}_n(\bar{r},\theta)$. A \emph{priori} these functions might not look too useful for solving the problem at hand, because they are not associated to separable solutions of the operator (\ref{Lbarr}). In fact, we strongly suspect that the operator (\ref{Lbarr}) does not admit separable solutions. Nevertheless, the functions $P_{k}(\theta)$ do form a basis. As such, we can expand $\bar{\psi}_n$ as
\begin{subequations}
\begin{equation}
\bar{\psi}_n(\bar{r},\theta)=\frac{\sin ^2\theta \cos \theta }{8}\frac{r_0}{\bar{r}}\delta_{n\,,3}+\sum_{k=1}^{+\infty}P_{k}(\theta) \bar{Q}_k^{(n)}(\bar{r})\,.
\label{eq:expansion}
\end{equation}
Note that (\ref{SE_rbar_reg}) also implicitly depends on $\bar{\psi}^{\infty}_n(\theta)$ (see for instance Eq.~(\ref{eq_barpsi4})). As such, we introduce
\begin{equation}
\bar{\psi}^{\infty}_n(\theta)=\sum_{k=1}^{+\infty}P_{k}(\theta) \bar{c}_k^{(n)}\,.
\label{eq:cs}
\end{equation}
with the understanding that
\begin{equation}
\lim_{\bar{r}\to+\infty}\bar{Q}_k^{(n)}(\bar{r})=\bar{c}_k^{(n)}\,.
\end{equation}
\end{subequations}

Since for each value of $n$, Eq.~(\ref{SE_rbar_reg}) comprises a linear equation for $\bar{\psi}_n(\theta,\bar{r})$ with a source term determined from the previous $(n-1)$ orders, we can hope to find an (infinite) system of coupled ordinary differential equations for the $Q_n^{(k)}(\bar{r})$. In order to derive these, we will need to tabulate a couple of integrals:
\begin{subequations}
\begin{equation}
\int_0^{\pi} \frac{P_{n_1}(\theta)\,P_{n_2}(\theta)}{\sin \theta}\,\mathrm{d}\theta = A(n_1)\,\delta_{n_1,n_2}
\end{equation}
and
\begin{multline}
\int_0^{\pi} \sin \theta \, P_{n_1}(\theta)\,P_{n_2}(\theta) \mathrm{d}\theta=A(n_1)\Bigg[\frac{(2 n_1-1)(2 n_1+1)}{(4 n_1-3)(4 n_1-1)}\delta _{n_1,n_2+1}+\frac{4 n_1 \left(n_1+1\right)}{\left(4 n_1+3\right) \left(4 n_1+5\right)} \delta _{n_1+1,n_2}
\\
+\frac{4 n_1 \left(2n_1+1\right)}{(4 n_1-1)(4 n_1+3)}\delta _{n_1,n_2}\Bigg]\,,
\end{multline}
with
\begin{equation}
A(n_1)\equiv \frac{\pi  \Gamma (n_1) \Gamma (n_1+1)}{2 (4 n_1+1) \Gamma \left(n_1+\frac{1}{2}\right) \Gamma \left(n_1+\frac{3}{2}\right)}\,.
\end{equation}
\label{eq:integrals}
\end{subequations}

The \emph{R\`egle du jeux} are now simple: we take the expansion (\ref{eq:expansion}) into (\ref{SE_rbar_reg}). Let us call the resulting equation $\mathcal{B}(\bar{\psi}_n)=0$. We then project the equation as
\begin{equation}
\mathcal{B}_k(\bar{Q}_k^{(n)}(\bar{r}))\equiv \int_0^{\pi} \frac{P_{k}(\theta)}{\sin \theta}\mathcal{B}(\bar{\psi}_n)\,\mathrm{d}\theta
\end{equation}
and find the resulting equations for $\bar{Q}_k^{(n)}(\bar{r})$ by using the integrals (\ref{eq:integrals}) and integration by parts. For $\bar{\psi}_3$ we find
\begin{multline}
\partial^2_{z}\bar{Q}_k^{(3)}(z)-\frac{1}{16}\Bigg[\frac{(2 k-1)(2 k+1)}{(4k-3)(4k-1)}\partial_{z}(z^2\partial_z \bar{Q}_{k-1}^{(3)}(z))+\frac{4 k \left(k+1\right)}{\left(4 k+3\right) \left(4 k+5\right)} \partial_{z}(z^2\partial_z \bar{Q}_{k+1}^{(3)}(z))
\\
+\frac{4 k \left(2k+1\right)}{(4k-1)(4k+3)}\partial_{z}(z^2\partial_ z\bar{Q}_k^{(3)}(z))\Bigg]-\frac{2k(2k+1)}{z^2}\bar{Q}_k^{(3)}(z)+\frac{1}{4}\bar{Q}_k^{(3)}(z)\\
+ \frac{(k+1) (2 k-3) (2 k-1) (2 k+1)}{8 (4 k-3) (4 k-1)}\bar{Q}_{k-1}^{(3)}(z)+\frac{k (k+1) (k+2) (2 k-1)}{2 (4 k+3) (4 k+5)}\bar{Q}_{k+1}^{(3)}(z)
\\
+\frac{k (2 k+1) \left(4 k^2+2 k-7\right)}{4 (4 k-1) (4 k+3)}\bar{Q}_k^{(3)}(z)+\frac{3}{112 z}\delta_{k,\,1}+\frac{9}{448 z}\delta_{k,\,2}\\
-\Bigg[\frac{(k+1) (2 k-3) (2 k-1) (2 k+1)}{8 (4 k-3) (4 k-1)} \bar{c}^{(3)}_{k-1}+\frac{k (k+1) (k+2) (2 k-1)}{2 (4 k+3) (4 k+5)}\bar{c}^{(3)}_{k+1}
\\
+\frac{(k+1) (2 k-1) \left(4 k^2+2 k+3\right)}{4 (4 k-1) (4 k+3)}\bar{c}^{(3)}_{k}\Bigg]=0\,,\qquad\text{for}\qquad k=1,2,3,\ldots
\label{eq:nuts1}
\end{multline}
with $z\equiv \bar{r}/r_0$, $\bar{Q}_0^{(n)}(z)=0$, $\bar{c}_0^{(n)}=0$. For $\bar{\psi}_4$ we find
\begin{multline}
\partial^2_{z}\bar{Q}_k^{(4)}(z)-\frac{1}{16}\Bigg[\frac{(2 k-1)(2 k+1)}{(4k-3)(4k-1)}\partial_{z}(z^2\partial_z \bar{Q}_{k-1}^{(4)}(z))+\frac{4 k \left(k+1\right)}{\left(4 k+3\right) \left(4 k+5\right)} \partial_{z}(z^2\partial_z \bar{Q}_{k+1}^{(4)}(z))
\\
+\frac{4 k \left(2k+1\right)}{(4k-1)(4k+3)}\partial_{z}(z^2\partial_z \bar{Q}_k^{(4)}(z))\Bigg]-\frac{2k(2k+1)}{z^2}\bar{Q}_k^{(4)}(z)+\frac{1}{4}\bar{Q}_k^{(4)}(z)\\
+ \frac{(k+1) (2 k-3) (2 k-1) (2 k+1)}{8 (4 k-3) (4 k-1)}\bar{Q}_{k-1}^{(4)}(z)+\frac{k (k+1) (k+2) (2 k-1)}{2 (4 k+3) (4 k+5)}\bar{Q}_{k+1}^{(4)}(z)
\\
+\frac{k (2 k+1) \left(4 k^2+2 k-7\right)}{4 (4 k-1) (4 k+3)}\bar{Q}_k^{(4)}(z)+\frac{3}{112 z}\delta_{k,\,1}+\frac{9}{448 z}\delta_{k,\,2}\\
-\Bigg[\frac{(k+1) (2 k-3) (2 k-1) (2 k+1)}{8 (4 k-3) (4 k-1)} \bar{c}^{(4)}_{k-1}+\frac{k (k+1) (k+2) (2 k-1)}{2 (4 k+3) (4 k+5)}\bar{c}^{(4)}_{k+1}
\\
+\frac{(k+1) (2 k-1) \left(4 k^2+2 k+3\right)}{4 (4 k-1) (4 k+3)}\bar{c}^{(4)}_{k}\Bigg]-\Bigg[2\partial_z\left(\frac{1}{z}\partial_z \bar{Q}_k^{(3)}(z)\right)+\frac{\partial_z \bar{Q}_k^{(3)}(z)}{z^2}\Bigg]
\\
+\frac{1}{16}\Bigg\{\frac{(2 k-1)(2 k+1)}{(4k-3)(4k-1)}\left[\partial_{z}(z \partial_ z\bar{Q}_{k-1}^{(3)}(z))+\partial_ z\bar{Q}_{k-1}^{(3)}(z)\right]
\\
+\frac{4 k \left(k+1\right)}{\left(4 k+3\right) \left(4 k+5\right)} \left[\partial_{z}(z \partial_ z\bar{Q}_{k+1}^{(3)}(z))+\partial_ z\bar{Q}_{k+1}^{(3)}(z)\right]
\\
+\frac{4 k \left(2k+1\right)}{(4k-1)(4k+3)}\left[\partial_{z}(z \partial_ z\bar{Q}_{k}^{(3)}(z))+\partial_ z\bar{Q}_{k}^{(3)}(z)\right]\Bigg\}
+\frac{2k(2k+1)}{z^2}\bar{Q}_{k}^{(3)}(z)=0\,
\label{eq:nuts2}
\end{multline}
for $k=1,2,3,\dots$ . Numerically, we will not be able to solve the equations for $\bar{Q}_k^{(3)}(z)$ and $\bar{Q}_k^{(4)}(z)$ for infinite values of $k$. Therefore, we introduce a regulator $N$ which accounts for the maximum of harmonics that we intend to use in our approximation scheme. The equations remain as above, except that we take $\bar{Q}_k^{(n)}(z)=\bar{c}_k^{(n)}=0$ for $k\geq N+1$ and $n=3,4$. In the end we should estimate the numerical error in reading physical quantities by changing $N$. For this reason, all our quantities have an additional index $N$, which indicates how many harmonics we have used: $\bar{Q}_{k\;N}^{(n)}(z)$ and $\bar{c}_{k\;N}^{(n)}$.

We are then left with a system of $2 N$ second order differential equations to solve for $\bar{Q}_{k\;N}^{(3)}(z)$ and $\bar{Q}_{k\;N}^{(4)}(z)$. Note that, as expected, we can solve first for $\bar{Q}_{k\;N}^{(3)}(z)$ and only afterwards determine $\bar{Q}_{k\;N}^{(4)}(z)$. In this sense, we have $N$ second order differential equation to solve for each order $n=3,4$. However, it is not hard to see that the ordinary differential equations above have themselves outer light surfaces that are dependent on $k$. These comes from the first and fourth terms in both Eq.~(\ref{eq:nuts1}) and Eq.~(\ref{eq:nuts2}) and are given by
\begin{equation}
z_{\mathrm{OLS}\;k\;N}=\frac{2\sqrt{(4 k-1) (4 k+3)}}{\sqrt{k (2 k+1)}}\,.
\end{equation}
What remains is then how one can use a numerical method to solve these second order ODEs, while determining all the $\bar{c}_k^{n}$ in the process by requiring continuity at $z=z_{\mathrm{OLS}\;k\;N}$. We do this in a series of steps. First, we promote all $\bar{c}_k^{n}$ to functions of $z$ and supplement the $2N$ second order differential equations for $\bar{Q}_{k\;N}^{(n)}(z)$ with additional $2N$ second order differential equations of the form
\begin{equation}
\partial_{z}^2\bar{c}_{k\;N}^{n}(z)=0\,.
\label{eq:csartificial}
\end{equation}
The generic solution to these equations is simply
\begin{equation}
\bar{c}_{k\;N}^{n}(z)=\bar{c}_{k\;N}^{n,\,0}+z\,\bar{c}_{k\;N}^{n,\,1},
\end{equation}
where $\bar{c}_{k\;N}^{n,\,0}$ and $\bar{c}_{k\;N}^{n,\,1}$ are constants. We thus demand that Eqs.~(\ref{eq:csartificial}) are solved restricted to the boundary condition $\left.\partial_z \bar{c}_{k\;N}^{n}(z)\right|_{z=0}$. It seems we have not gained much by doing this procedure yet. However, we now perform a couple of tricks.

We first introduce a compact coordinate $y$ defined as
\begin{equation}
y=\frac{z}{1+z}\,,
\end{equation}
so that $z=0$ and $z\to+\infty$ are mapped to $y=0$ and $y=1$, respectively. The $k$ dependent OLS are mapped using the same rule, \emph{i.e.}
\begin{equation}
y_{\mathrm{OLS}\;k\;N}=\frac{z_{\mathrm{OLS}\;k\;N}}{1+z_{\mathrm{OLS}\;k\;N}}.
\end{equation}

Next, we note that the generic Frobenius behaviour around each of the $y_{\mathrm{OLS}\;k\;N}$ predicts a logarithmic behaviour in $y-y_{\mathrm{OLS}\;k\;N}$ (which we want to discard) and a smooth component that we want to keep. Motivated by this, we write all ordinary differential equations above in first order form for the $\bar{Q}_{k\;N}^{(n)}(y)$ and use a spectral collocation scheme (such as the ones reviewed in \cite{Dias:2015nua}) on a Chebyshev-Gauss-Lobatto grid to solve the resulting $4N+2N$ ordinary differential equations. This scheme assumes that all functions can be decomposed as a sum of (shifted) Chebyshev polynomials of the first kind in $y\in(0,1)$, which in particular are continuous and smooth across $y=y_{\mathrm{OLS}\;k\;N}$. we thus expect a unique smooth solution for $\bar{Q}_{k\;N}^{(n)}(y)$ and $\bar{c}_{k\;N}^{n}$. Indeed, we find this to be the case. The advantage of this method over the one detailed in \cite{Armas:2020mio} is that we do not need to minimise any functional to determine $\bar{c}_{k\;N}^{n}$: they are determined via a simple ordinary differential solver.

There is one additional technical remark we will make. The functions $\bar{Q}_{k\;N}^{(3)}(y)$ have regular boundary conditions at $y=0$ and $y=1$ (indeed, this is the reason why we introduced the first factor in Eq.~(\ref{eq:expansion}):
\begin{equation}
\bar{Q}_{k\;N}^{(3)}(0)=0\quad\text{and}\quad \left.\frac{\partial \bar{Q}_{k\;N}^{(3)}(y)}{\partial y}\right|_{y=1}=\frac{1}{8} \delta_{k,\,1}.
\end{equation}
However, this is not the case for $\bar{Q}_{1\;N}^{(4)}(y)$ and $\bar{Q}_{1\;N}^{(4)}(y)$ whose small $y$ behaviour has to be matched with Eq.~(\ref{smallrbar}). In order to achieve this, we do not solve for $\bar{Q}_{1\;N}^{(4)}(y)$ and $\bar{Q}_{2\;N}^{(4)}(y)$, but instead make an additional functional redefinition of the form
\begin{subequations}
\begin{align}
\bar{Q}_{1\;N}^{(4)}(z)=&-\frac{11}{800 z^2}+\frac{1}{40 z^2}\log \left(\frac{z}{1+z}\right)+\frac{1}{40 z}-\frac{1}{80}
\nonumber\\
&+\frac{1}{1680 \left(1+z^4\right)}\log \left(\frac{z}{1+z}\right)+\frac{227}{100800}+\tilde{\bar{Q}}_{1\;N}^{(4)}(z)\,,
\end{align}
\begin{equation}
\bar{Q}_{2\;N}^{(4)}(z)=\frac{3}{22400 \left(1+z^4\right)} \log \left(\frac{z}{1+z}\right)+\frac{363}{896000}+\tilde{\bar{Q}}_{2\;N}^{(4)}(z)\,.
\end{equation}
\end{subequations}
It is relatively easy to check that the required boundary conditions are simply $\tilde{\bar{Q}}_{1\;N}^{(4)}(z)=\tilde{\bar{Q}}_{2\;N}^{(4)}(z)=\bar{Q}_{k\;N}^{(4)}(z)=0$, for $k\geq3$ and
\begin{equation}
\left.\frac{\partial \bar{Q}_{k\;N}^{(4)}(y)}{\partial y}\right|_{y=1}=\frac{1}{8} \delta_{k,\,1}+\frac{1}{40} \delta_{k,\,2}\,.
\end{equation}

The functions we plot in Fig.~\ref{fig:plot_barpsi4} are simply given by $\bar{\psi}^{(n)}$ with its singular behaviour removed at $\bar{r}=0$. To wit
\begin{subequations}
\begin{equation}
\widetilde{\bar{\psi}^{(3)}}(\bar{r},\theta)=\bar{\psi}^{(3)}(\bar{r},\theta)-\frac{\sin ^2\theta \cos \theta }{8}\frac{r_0}{\bar{r}}=\sum_{k=1}^{N}P_{k}(\theta) \bar{Q}_{k\;N}^{(n)}(\bar{r})
\end{equation}
and
\begin{equation}
\widetilde{\bar{\psi}^{(4)}}(\bar{r},\theta)=\tilde{\bar{Q}}_{1\;N}^{(4)}(\bar{r})P_1(\theta)+\tilde{\bar{Q}}_{2\;N}^{(4)}(\bar{r})P_2(\theta)+\sum_{k=3}^{N}P_{k}(\theta) \bar{Q}_k^{(4)}(\bar{r})\,.
\end{equation}
\end{subequations}

We now make a parenthetical remark: we have also solved this problem using a Chebyshev collocation scheme. In such an approach, the partial differential equations are discretized on a tensor product grid of two Chebyshev-Gauss-Lobatto grids having $N_z$ and $N_{\theta}$ collocation points along the $z$  and $\theta$ directions, respectively. Again, the system is found to be dependent on two regulators: the number of grid points along each integration directions. In this sense, the numerical method we outline above is no different. One might wonder why we did not use such collocation methods to tackle the problem at hand. Indeed, we found that using Chebyshev-Gauss-Lobatto collocation methods also demanded that we introduce \emph{additional regularisation parameters} besides $N_{z}$ and $N_{\theta}$, which were difficult to control numerically. Such additional regularisation parameters were also needed in \cite{Armas:2020mio}, but are \emph{not} required in the numerical scheme that intrinsically uses the $P_k(\theta)$ harmonics detailed above.

Perhaps interestingly, one can solve the equations exactly for $N=1$. This turns out to be a good test of our numerical procedures. For $Q_{1\;N=1}^{(3)}(z)$ we find
\begin{subequations}
\begin{align}
&\bar{Q}_{1\;N=1}^{(3)}(z)=-\frac{3}{200704 z^2}\left[4704 z+3 z^3 \left(\sqrt{7} \pi ^2 z-56\right)\right]
\nonumber\\
&-\frac{3 \left(2352+56 z^2+3 z^4\right)}{28672 \sqrt{7} z^2} \left[4\, \mathrm{arctanh}\left(\frac{z}{2 \sqrt{7}}\right) \log
   \left(\frac{z}{\sqrt{28}}\right)-4\,\text{Li}_2\left(\frac{z}{2 \sqrt{7}}\right)+\text{Li}_2\left(\frac{z^2}{28}\right)\right]
\nonumber\\
   &+\frac{126 z \left(28+z^2\right)}{50176 z^2} \log
   \left(\frac{z}{\sqrt{28}}\right)
   \label{eq:explicit}
\end{align}
with
\begin{equation}
\bar{c}_{1\;N=1}^{(3)}=\frac{3 \pi ^2}{512 \sqrt{7}}\,.
\end{equation}
\end{subequations}
The closed form expression for $\bar{Q}_{1\;N=1}^{(4)}(z)$ occupies a few pages and is not particularly enlightening, but $\bar{c}_1^{(4)}$ turns out to be given by
\begin{equation}
\bar{c}_{1\;N=1}^{(4)}=\frac{379+120 \log 2+60 \log 7}{201600}-\frac{9 \pi ^4}{458752}\,.
\end{equation}
\subsection*{Convergence of the numerical method to the continuum limit}
\begin{figure}[ht!]
    \centering
    \includegraphics[width=0.8\textwidth]{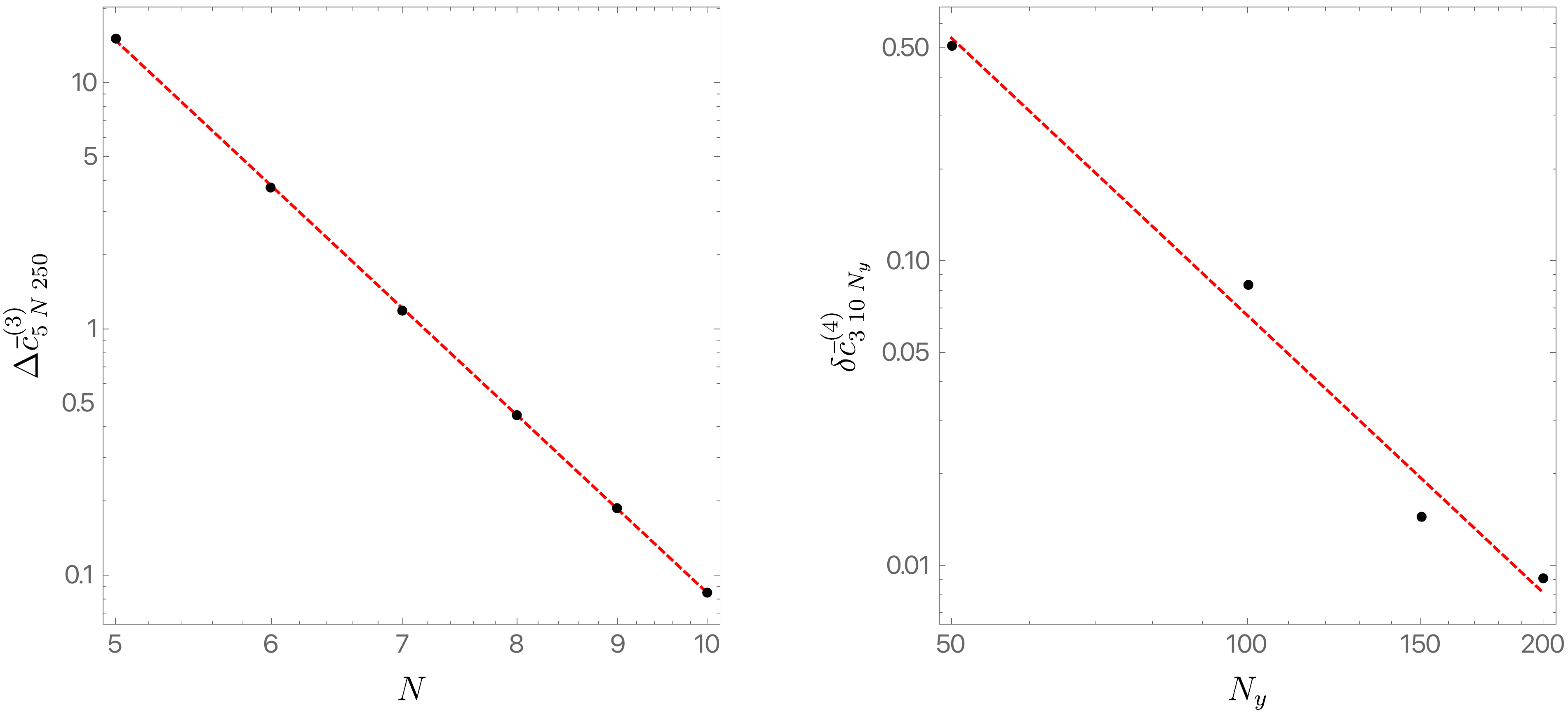}
    \caption{On the left panel we plot $\Delta \bar{c}_ {5\;N \;250}^{(3)}$ as a function of $N$ finding a consistent scaling with $N^{-13/2}$ (given by the best linear fit represented as a dashed red line), while on the right panel we plot $\delta \bar{c}_ {3\;10\;N_y}^{(4)}$ as a function of $N_y$ and find a consistent scaling with $N_y^{-3}$ (given by the best linear fit represented as a dashed red line).}
    \label{fig:convergence}
\end{figure}

A complete convergence study involves changing both $N$ (the number of harmonics used) and $N_y$ ( the number of grid points in the Chebyshev-Gauss-Lobatto grid). Because the functions we solve for exhibit explicit logarithmic dependence (see for instance the analytic solution for $N=1$ (\ref{eq:explicit})) we do not expect exponential convergence in $N_y$ or $N$, but instead power law. To monitor the convergence, let us denote $\bar{c}_{k\;N\;N_y}^{(n)}$ with $k=1,\ldots,5$ and $n=3,4$, as computing $\bar{c}_{k\;N}^{(n)}$ with $N_y$ collocation points in the $y$ direction. For each of these we define
\begin{equation}
\delta \bar{c}_{k\;N\;N_y}^{(n)}\equiv 100\left|1-\frac{\bar{c}_{k\;N\;N_y}^{(n)}}{\bar{c}_{k\;N\;N_y+50}^{(n)}}\right|\,.
\end{equation}
Similarly, we can define
\begin{equation}
\Delta \bar{c}_{k\;N\;N_y}^{(n)}\equiv 100\left|1-\frac{\bar{c}_{k\;N\;N_y}^{(n)}}{\bar{c}_{k\;N+1\;N_y}^{(n)}}\right|\,.
\end{equation}
The quantity $\delta \bar{c}_{k\;N\;N_y}^{(n)}$ measures the convergence in $N_y$, whereas $\Delta \bar{c}_{k\;N\;N_y}^{(n)}$ measures the convergence in $N$. As expected, we find polynomial convergence in $N$ and $N_y$. For $\Delta \bar{c}_{k\;N\;N_y}^{(n)}$ we find a convergence in the continuum limit compatible with $N^{-13/2}$ while for $\delta \bar{c}_{k\;N\;N_y}^{(n)}$ we find a convergence compatible with $N_y^{-3}$. Examples of the studies we performed are detailed in Fig.~\ref{fig:convergence} where we plot $\Delta \bar{c}_ {5\;N \;250}^{(3)}$ (on the left panel) and $\delta \bar{c}_ {3\;10\;N_y}^{(4)}$ (on the right panel). Different choices of $n$ and $k$ yield similar results. Note that both $\delta \bar{c}_{k\;N\;N_y}^{(n)}$ and $\Delta \bar{c}_{k\;N\;N_y}^{(n)}$ are measured already in percentage level. All plots in the manuscript were generated with $N_y=250$ and $N=11$ and all quantities reported have an error estimated to be well under the $10^{-1}\%$ level.

\section{\label{app:explicit}Explicit expressions for \texorpdfstring{$R^{(4)}_2(r)$}{} and \texorpdfstring{$R^{(4)}_4(r)$}{}}
In this section we take $y\equiv r_0/r$, so that $y\in[0,1]$. The function $R^{(4)}_2(y)$ is given by
\begin{multline}
\label{R42}
R^{(4)}_2(y)=\log y \Bigg[\left(\frac{14319}{1960 y^2}-\frac{13}{490 y^3}-\frac{51}{28 y}-\frac{17}{56}\right) \text{Li}_2(y)+\left(\frac{153}{14 y^2}-\frac{102}{7 y^3}\right) \text{Li}_3(1-y)
\\
   +\frac{2 y^2}{63 (y-1)}+\pi ^2 \left(\frac{1}{56 y}-\frac{1}{14 y^2}+\frac{1}{336}\right)+\frac{758063}{35280 (y-1) y^2}+\frac{1}{98 (y-1) y^3}-\frac{3}{98 (y-1)
   y^4}
\\
   +\frac{2119 y}{1512 (y-1)}+\frac{1562143}{211680 (y-1)}-\frac{2144087}{70560 (y-1) y}+\left(\frac{102}{7
   y^3}-\frac{153}{14 y^2}\right) \zeta (3)\Bigg]
\\
   +\left(\frac{274241}{11760y^2}-\frac{758729}{35280 y^3}+\frac{1}{196 y^4}+\frac{3}{98 y^5}-\frac{40009}{11760 y}-\frac{57859}{70560}\right) \text{Li}_2(y)
\\
   +\left(\frac{153}{56 y^2}-\frac{51}{14 y^3}\right) \text{Li}_2(y){}^2+\pi ^2 \Bigg[\left(\frac{3}{56 y^2}-\frac{1}{14 y^3}\right)
   \text{Li}_2(y)+\frac{1}{14 y^2}+\frac{1}{288}\Bigg]
\\
   +\left(\frac{39}{980 y^3}-\frac{57237}{3920 y^2}+\frac{51}{14
   y}+\frac{17}{28}\right) \text{Li}_3(y)+\left(\frac{15}{y^3}-\frac{45}{4 y^2}\right) \text{Li}_4(1-y)+\left(\frac{153}{14 y^2}-\frac{102}{7 y^3}\right) \text{Li}_4(y)
\\
   +\left(\frac{153}{14 y^2}-\frac{102}{7
   y^3}\right) \text{Li}_4\left(\frac{y}{y-1}\right)+\left(\frac{3}{7 y^2}-\frac{3}{28 y}-\frac{1}{56}\right) \text{Li}_3(1-y)
\\
   +\log (1-y) \Bigg\{\Bigg[\left(\frac{153}{28 y^2}-\frac{51}{7 y^3}\right) \text{Li}_2(y)+\pi ^2 \left(\frac{33}{14 y^3}-\frac{99}{56
   y^2}\right)+\frac{274241}{11760 y^2}-\frac{758729}{35280 y^3}+\frac{1}{196 y^4}
\\
   +\frac{3}{98 y^5}-\frac{40009}{11760 y}-\frac{57859}{70560}\Bigg] \log y+\left(\frac{39}{7840 y^2}-\frac{13}{1960 y^3}\right)
   \log ^2y\Bigg\}+\left(\frac{102}{7 y^2}-\frac{51}{14 y}-\frac{17}{28}\right) \zeta (3)
\\
   -\frac{y^3}{840}+\frac{89 y^2}{6720}+\pi ^4 \left(\frac{1}{8 y^2}-\frac{1}{6 y^3}\right)+\frac{18929}{882
   y^2}-\frac{5}{392 y^3}-\frac{3}{98 y^4}+\left(\frac{51}{112 y^2}-\frac{17}{28 y^3}\right) \log ^4(1-y)
\\
   +\Bigg[\pi ^2 \left(\frac{51}{56
   y^2}-\frac{17}{14 y^3}\right)+\left(\frac{153}{28 y^2}-\frac{51}{7 y^3}\right) \log ^2y\Bigg] \log ^2(1-y)+\left(\frac{13}{7840 y}-\frac{13}{1960 y^2}+\frac{13}{47040}\right) \log ^2(y)
\\
   +\left(\frac{17}{7 y^3}-\frac{51}{28 y^2}\right) \log y \log ^3(1-y)-\frac{361y}{3024}-\frac{585751}{47040 y}-\frac{122875}{72576}\,,
\end{multline}
while for $R^{(4)}_4(y)$ we find
\begin{multline}
\label{R44}
R^{(4)}_4(y)=\left(\frac{3645}{112 y^3}-\frac{1215}{224 y^2}-\frac{3645}{64 y^4}+\frac{243}{8 y^5}\right) \log ^4(1-y)+\frac{y^3}{840}-\frac{17
   y^2}{1680}+\frac{239y}{2520}
\\
    +\left(\frac{1215}{56 y^2}-\frac{3645}{28 y^3}+\frac{3645}{16 y^4}-\frac{243}{2 y^5}\right) \log (y) \log
   ^3(1-y)+\Bigg[\Bigg(\frac{10935}{28 y^3}-\frac{3645}{56 y^2}-\frac{10935}{16 y^4}
\\
    +\frac{729}{2 y^5}\Bigg) \log ^2y+\pi ^2 \left(\frac{3645}{56 y^3}-\frac{1215}{112 y^2}-\frac{3645}{32 y^4}+\frac{243}{4
   y^5}\right)\Bigg] \log ^2(1-y)+\Bigg\{\Bigg(\frac{7527}{1568 y^3}-\frac{2509}{3136 y^2}
\\-\frac{7527}{896 y^4}+\frac{2509}{560 y^5}\Bigg) \log ^2y +\Bigg[\pi ^2 \left(\frac{2445}{112 y^2}-\frac{7335}{56
   y^3}+\frac{7335}{32 y^4}-\frac{489}{4 y^5}\right)+\Bigg(\frac{10935}{28 y^3}-\frac{3645}{56 y^2}
\\
   -\frac{10935}{16 y^4} +\frac{729}{2 y^5}\Bigg)\text{Li}_2(y) +\frac{80023}{3920 y}-\frac{11191681}{23520
   y^2}+\frac{14805167}{7840 y^3}-\frac{164946493}{62720 y^4}+\frac{70892653}{58800 y^5}
\\
    +\frac{355309}{313600}\Bigg] \log y\Bigg\} \log (1-y)+\left(\frac{393913}{188160 y^2}-\frac{2509}{627200}-\frac{2509}{31360 y}-\frac{27599}{4480 y^3}+\frac{2509}{560 y^4}\right) \log ^2y
\\
    +\left(\frac{10935}{56y^3}-\frac{3645}{112 y^2}-\frac{10935}{32 y^4}+\frac{729}{4 y^5}\right) \text{Li}_2(y){}^2+\pi ^4 \left(\frac{69}{8 y^3}-\frac{23}{16 y^2}-\frac{483}{32 y^4}+\frac{161}{20 y^5}\right)
\\
    +\left(\frac{355309}{313600}+\frac{80023}{3920 y}-\frac{11191681}{23520 y^2}+\frac{14805167}{7840 y^3}-\frac{164946493}{62720 y^4}+\frac{70892653}{58800 y^5}\right) \text{Li}_2(y)
\\
    +\pi ^2 \Bigg[\left(\frac{15}{112 y^2}-\frac{45}{56 y^3}+\frac{45}{32 y^4}-\frac{3}{4 y^5}\right) \text{Li}_2(y)-\frac{9}{640 y}+\frac{15461}{26880 y^2}-\frac{1233}{640 y^3}+\frac{123}{80
   y^4}\Bigg]
   \\
    +\left(\frac{141}{112 y^2}-\frac{9}{2240}-\frac{9}{112 y}-\frac{9}{8 y^3}-\frac{279}{64 y^4}+\frac{189}{40 y^5}\right) \text{Li}_3(1-y)
\\
    +\left(\frac{536643}{1568
   y^2}-\frac{729}{1120}-\frac{729}{56 y}-\frac{793449}{784 y^3}+\frac{334179}{448 y^4}-\frac{2529}{280 y^5}\right) \text{Li}_3(y)
\\
    +\left(\frac{1035}{8 y^2}-\frac{3105}{4 y^3}+\frac{21735}{16 y^4}-\frac{1449}{2 y^5}\right)
   \text{Li}_4(1-y)+\left(\frac{10935}{14 y^3}-\frac{3645}{28 y^2}-\frac{10935}{8 y^4}+\frac{729}{y^5}\right) \text{Li}_4(y)
\\
    +\left(\frac{10935}{14 y^3}-\frac{3645}{28 y^2}-\frac{10935}{8
   y^4}+\frac{729}{y^5}\right) \text{Li}_4\left(\frac{y}{y-1}\right)+\log y \Bigg[\pi ^2 \Bigg(\frac{3}{4480}+\frac{3}{224 y}-\frac{157}{448 y^2}+\frac{33}{32
   y^3}
\\
    -\frac{3}{4 y^4}\Bigg)-\frac{y^2}{56 (y-1)}-\frac{1413 y}{560 (y-1)}+\Bigg(\frac{729}{2240}+\frac{729}{112 y}-\frac{33697}{196 y^2}+\frac{50061}{98 y^3}-\frac{170853}{448 y^4}
\\
    +\frac{2519}{280 y^5}\Bigg) \text{Li}_2(y)+\left(\frac{10935}{14 y^3}-\frac{3645}{28y^2}-\frac{10935}{8 y^4}+\frac{729}{y^5}\right) \text{Li}_3(1-y)-\frac{16648951}{313600 (y-1)}
\\
    +\left(\frac{3645}{28 y^2}-\frac{10935}{14 y^3}+\frac{10935}{8 y^4}-\frac{729}{y^5}\right)
   \zeta (3)+\frac{4537252639}{5644800 (y-1) y}-\frac{14540866027}{5644800 (y-1) y^2}
\\
    +\frac{2845437599}{940800 (y-1) y^3}-\frac{70363663}{58800 (y-1) y^4}\Bigg]+\frac{571610737}{5644800y}-\frac{32993904487}{33868800 y^2}+\frac{1915231103}{940800 y^3}
\\
    -\frac{70361563}{58800 y^4}+\left(\frac{729}{1120}+\frac{729}{56 y}-\frac{76113}{224 y^2}+\frac{15957}{16 y^3}-\frac{46089}{64 y^4}-\frac{189}{40
   y^5}\right) \zeta (3)+\frac{12815693}{3225600}\,.
\end{multline}

\section{Explicit expression for \texorpdfstring{$S_6(r,\theta)$}{}}\label{App:psi6}

The expression of the source $S_6(r,\theta)$ in Eq.~\eqref{SEpsi6} is

\begin{align}
    S_6(r,\theta) \nonumber &= - \frac{\left(r^2+r_0 r+r_0^2\right)}{2 r^2 r_0^2}\left[4 \cos \theta~\omega_5(\theta )+\sin \theta ~\omega_5'(\theta )\right] \sin \theta \\ \nonumber
    &+\bigg(\frac{r_0^6}{32 r^6}+\frac{U_0 r_0^4}{16 r^4}-\frac{5 r_0^4}{128 r^4}+\frac{U_0 r_0^3}{32 r^3}-\frac{5 U_0 r_0^2}{64 r^2}+\frac{15 r_0^2}{256 r^2}+\frac{U_0 r_0}{32 r}-\frac{3 r_0}{128 r}+\frac{7 r U_0^2}{8 r_0}+\frac{3 U_0}{64}\\ \nonumber
    &+\frac{r U_0}{32 r_0}+\frac{3 r^2 U_1}{32 r_0^2}+\frac{5 r V_0}{4 r_0}+\frac{3 r W_0}{8 r_0}-\frac{7}{256}-\frac{r}{8 r_0}-\frac{3 r U_1}{32 r_0}-\frac{5 r^2 V_0}{4 r_0^2}+\frac{r^2}{8 r_0^2}-\frac{7 r^2 U_0^2}{8 r_0^2}\\ \nonumber
    &-\frac{3 r^2 W_0}{8 r_0^2}-\frac{r^2 U_0}{32 r_0^2}\bigg)\frac{ \cos \theta }{(r-r_0)^2}+\bigg(-\frac{r_0^7}{16 r^7}+\frac{r_0^6}{16 r^6}-\frac{7 U_0 r_0^4}{32 r^4}+\frac{11 r_0^4}{128 r^4}+\frac{U_0 r_0^3}{8 r^3}-\frac{5 r_0^3}{64 r^3}\\ \nonumber
    &+\frac{5 U_0 r_0^2}{32 r^2}-\frac{5 r_0^2}{64 r^2}-\frac{U_0 r_0}{16 r}+\frac{r_0}{32 r}+\frac{29 r^2 U_0^2}{8 r_0^2}+\frac{7 r U_0}{32 r_0}+\frac{3 r U_1}{32 r_0}+\frac{19 r^2 V_0}{4 r_0^2}+\frac{3 r^2 W_0}{8 r_0^2}\\ \nonumber
    &-\frac{3 U_0}{32}+\frac{5}{128}-\frac{19 r V_0}{4 r_0}-\frac{29 r U_0^2}{8 r_0}-\frac{3 r W_0}{8 r_0}+\frac{5 r}{64 r_0}-\frac{7 r^2 U_0}{32 r_0^2}-\frac{3 r^2 U_1}{32 r_0^2}-\frac{5 r^2}{64 r_0^2}\bigg)\frac{ \cos ^3\theta}{(r-r_0)^2}\\ \nonumber
    &+\bigg(\frac{3 r_0^8}{32 r^8}-\frac{3 r_0^7}{32 r^7}-\frac{3 r_0^6}{32 r^6}+\frac{3 r_0^5}{32 r^5}+\frac{5 U_0 r_0^4}{32 r^4}-\frac{r_0^4}{64 r^4}-\frac{5 U_0 r_0^3}{32 r^3}+\frac{r_0^3}{64 r^3}-\frac{5 U_0 r_0^2}{64 r^2}\\ \nonumber
    &+\frac{5 r_0^2}{256 r^2}+\frac{U_0 r_0}{32 r}-\frac{r_0}{128 r}+\frac{11 r U_0^2}{4 r_0}+\frac{3 U_0}{64}+\frac{r^2 U_0}{4 r_0^2}+\frac{7 r V_0}{2 r_0}-\frac{3}{256}-\frac{r U_0}{4 r_0}-\frac{r}{64 r_0}\\ \nonumber
    &-\frac{7 r^2 V_0}{2 r_0^2}-\frac{11 r^2 U_0^2}{4 r_0^2}+\frac{r^2}{64 r_0^2}\bigg)\frac{ \cos ^5 \theta }{(r-r_0)^2}\\ \nonumber
    &+\bigg(-\frac{r_0^5}{8 r^5}-\frac{r_0^4}{4 r^4}+\frac{3 r_0^3}{8 r^3}-\frac{11 U_0 r_0^2}{8 r^2}+\frac{11 r_0^2}{16 r^2}+\frac{11 U_0 r_0}{8 r}-\frac{13 r_0}{16 r}+\frac{1}{32}\bigg)R^{(2)}_{2}(r)\frac{ \cos \theta \sin \theta  }{(r-r_0)^2}\\ \nonumber
    &+\bigg(\frac{r_0^6}{4 r^6}+\frac{r_0^5}{2 r^5}-\frac{r_0^4}{r^4}+\frac{r_0^3}{4 r^3}+\frac{17 U_0 r_0^2}{4 r^2}-\frac{11 r_0^2}{8 r^2}-\frac{17 U_0 r_0}{4 r}+\frac{51 r_0}{32 r}-\frac{1}{8}\bigg)R^{(2)}_2(r)\frac{ \cos ^3\theta \sin \theta}{(r-r_0)^2}\\ \nonumber
    &+ \left(\frac{r U_0}{2 r_0}-\frac{r^2 U_0}{2 r_0^2}\right) R^{(2)}_2(r)\frac{\cos \theta \sin ^3\theta}{(r-r_0)^2}+ \bigg[\left(\frac{5 r}{16 r_0}-\frac{5 r^2}{16 r_0^2}\right) R^{(2)}_2(r)+\bigg(-\frac{U_0 r^3}{8 r_0^2}+\frac{3 r^3}{32 r_0^2}\\ \nonumber
    &+\frac{U_0 r^2}{4 r_0}-\frac{3 r^2}{16 r_0}-\frac{U_0 r}{8}+\frac{3 r}{32}+\frac{5 r_0^3}{64 r^2}+\frac{5 r_0}{64}+\frac{r_0^2 U_0}{4 r}-\frac{r_0 U_0}{8}-\frac{5 r_0^2}{32 r}-\frac{r_0^3 U_0}{8 r^2}\bigg) R^{(2)\prime}_{2}(r)\\ \nonumber
    &+\bigg(-\frac{U_0 r^4}{16 r_0^2}+\frac{3 r^4}{64 r_0^2}+\frac{U_0 r^3}{8 r_0}-\frac{3 r^3}{32 r_0}-\frac{U_0 r^2}{16}+\frac{r^2}{16}+\frac{r_0 U_0 r}{8}-\frac{7 r_0 r}{64}+\frac{11 r_0^2}{64}+\frac{r_0^3 U_0}{8 r}\\ \nonumber
    &-\frac{r_0^2 U_0}{4}-\frac{5 r_0^3}{64 r}\bigg) R^{(2)\prime \prime }_{2}(r)\bigg]\frac{\cos \theta \sin ^2\theta}{(r-r_0)^2}+\bigg[\bigg(\frac{3 r^2}{8 r_0^2}-\frac{3 r}{8 r_0}\bigg) R^{(2)}_{2}(r)^2+\bigg(\frac{r^2}{8 r_0^2}-\frac{r}{8 r_0}\bigg) R^{(2)}_{2}(r)\\ \nonumber
    &+\bigg(-\frac{3 r_0^5}{16 r^4}+\frac{3 r_0^4}{8 r^3}+\frac{U_0 r_0^3}{8 r^2}-\frac{13 r_0^3}{64 r^2}-\frac{U_0 r_0^2}{4 r}+\frac{r_0^2}{32 r}+\frac{U_0 r_0}{8}-\frac{r_0}{64}+\frac{r U_0}{8}+\frac{r^3 U_0}{8 r_0^2}-\frac{r}{32}\\ \nonumber
    &-\frac{r^2 U_0}{4 r_0}+\frac{r^2}{16 r_0}-\frac{r^3}{32 r_0^2}\bigg) R^{(2)\prime }_{2}(r)+\bigg(\frac{r_0^5}{16 r^3}-\frac{r_0^4}{8 r^2}-\frac{U_0 r_0^3}{8 r}+\frac{5 r_0^3}{64 r}+\frac{U_0 r_0^2}{4}-\frac{r_0^2}{32}+\frac{r r_0}{64}\\ \nonumber
    &-\frac{r U_0 r_0}{8}+\frac{r^2 U_0}{16}+\frac{r^4 U_0}{16 r_0^2}-\frac{r^2}{64}-\frac{r^3 U_0}{8 r_0}+\frac{r^3}{32 r_0}-\frac{r^4}{64 r_0^2}\bigg) R^{(2)\prime \prime }_{2}(r)\bigg]\frac{\cos ^3\theta \sin ^2\theta }{(r-r_0)^2}\\ \nonumber
    &+\bigg[\bigg(-\frac{9 r_0^6}{8 r^6}+\frac{9 r_0^5}{8 r^5}+\frac{9 r_0^4}{8 r^4}-\frac{9 r_0^3}{8 r^3}-\frac{23 U_0 r_0^2}{8 r^2}+\frac{7 r_0^2}{16 r^2}+\frac{23 U_0 r_0}{8 r}-\frac{17 r_0}{32 r}+\frac{3}{32}\bigg) R^{(2)}_{2}(r)\\ \nonumber
    &+\bigg(\frac{5 r_0^7}{16 r^6}-\frac{5 r_0^6}{8 r^5}+\frac{5 r_0^5}{16 r^4}\bigg) R^{(2)\prime }_{2}(r)+\bigg(-\frac{r_0^7}{16 r^5}+\frac{r_0^6}{8 r^4}-\frac{r_0^5}{16 r^3}\bigg) R^{(2)\prime \prime }_{2}(r)\bigg]\frac{\cos ^5\theta \sin \theta }{(r-r_0)^2}\\ \nonumber
    &+\bigg[-\frac{3 r R^{(4)}_2(r)}{8 r_0^2}+\bigg(\frac{r^2}{8 r_0^2}-\frac{r_0^2}{4 r^2}-\frac{r}{8 r_0}+\frac{r_0}{4 r}\bigg) \left[R^{(4)\prime}_2(r)+ R^{(4)\prime}_4(r)\right]+\bigg(\frac{r^3}{16 r_0^2}-\frac{r^2}{16 r_0}\\ \nonumber
    &+\frac{r_0^2}{4 r}-\frac{r_0}{4}\bigg) \left[R^{(4)\prime \prime}_2(r)+ R^{(4)\prime \prime}_4(r)\right]\bigg]\frac{\sin ^2\theta \cos \theta }{r-r_0}+\bigg[\bigg(-\frac{7 r^2}{24 r_0^2}+\frac{7 r_0^2}{12 r^2}+\frac{7 r}{24 r_0}\\ \nonumber
    &-\frac{7 r_0}{12 r}\bigg) R^{(4)\prime}_2(r)+\left(-\frac{7 r^3}{48 r_0^2}+\frac{7 r^2}{48 r_0}-\frac{7 r_0^2}{12 r}+\frac{7 r_0}{12}\right) R^{(4)\prime \prime}_2(r)\bigg]\frac{\sin ^2\theta \cos ^3\theta }{r-r_0}\\ \nonumber
    &+\bigg[\bigg(\frac{r_0^3}{2 r^4}+\frac{3 r_0^2}{2 r^3}-\frac{2 r_0}{r^2}\bigg) R^{(4)}_2(r)+\left(\frac{r_0^3}{2 r^4}+\frac{5 r_0^2}{r^3}-\frac{11 r_0}{2 r^2}+\frac{5 r}{4 r_0^2}\right) R^{(4)}_4(r)\bigg]\frac{\sin \theta \cos \theta }{r-r_0}\\ \nonumber
    &+\bigg[\bigg(\frac{3 r_0}{r^2}-\frac{3 r_0^3}{r^4}\bigg) R^{(4)}_2(r)+\bigg(-\frac{10 r_0^3}{r^4}-\frac{35 r_0^2}{3 r^3}+\frac{65 r_0}{3 r^2}-\frac{19 r}{4 r_0^2}\bigg) R^{(4)}_4(r)\\ \nonumber
    &+\bigg(\frac{3 r_0^4}{4 r^4}-\frac{3 r_0^3}{4 r^3}\bigg) \left[R^{(4)\prime}_2(r)+ R^{(4)\prime}_4(r)\right]+\left(\frac{r_0^3}{4 r^2}-\frac{r_0^4}{4 r^3}\right) \left[R^{(4)\prime \prime}_2(r)+ R^{(4)\prime \prime}_4(r)\right]\bigg]\frac{\sin \theta \cos ^3\theta }{r-r_0}\\ \nonumber
    &+ \bigg[\bigg(\frac{35 r_0^3}{2 r^4}-\frac{35 r_0}{2 r^2}+\frac{7 r}{2 r_0^2}\bigg) R^{(4)}_4(r)+\bigg(\frac{7 r_0^3}{4 r^3}-\frac{7 r_0^4}{4 r^4}\bigg) R^{(4)\prime}_4(r)\\
    &+\bigg(\frac{7 r_0^4}{12 r^3}-\frac{7 r_0^3}{12 r^2}\bigg) R^{(4)\prime \prime}_2(r)\bigg]\frac{\sin \theta \cos ^5\theta}{r-r_0},
\end{align}
where we remind that the constants $U_0$, $U_1$, $W_0$, and  $V_0$ are given in Eqs.~\eqref{Us}, \eqref{Ws} and \eqref{Vs} while the radial functions $R^{(2)}_2(r)$, $R^{(4)}_2(r)$ and $R^{(4)}_4(r)$ appear respectively in Eqs.~\eqref{R2}, \eqref{R42} and \eqref{R44}.

\bibliography{Bibliography}

\end{document}